\documentclass[aps,twocolumn,showpacs,pre]{revtex4-1}
\usepackage{graphicx,graphics}
\usepackage{amsmath,amssymb,amsfonts}
\usepackage{multirow}
\usepackage{hyperref}
\usepackage{graphicx}

\usepackage{color}

\begin{document}
\title{Random walks on weighted networks: Exploring local and non-local navigation strategies}
\author{A.P. Riascos}
\email{aaappprrr@gmail.com}
\author{Jos\'e L. Mateos}
\email{mateos@fisica.unam.mx}
\affiliation{Instituto de F\'isica,Universidad Nacional Aut\'onoma de M\'exico,
Apartado Postal 20-364, 01000 M\'exico, D.F., M\'exico}

\date{\today}

\begin{abstract}
In this paper, we present an overview of different types of random walk strategies with local and non-local transitions on undirected connected networks. We present a general approach to analyzing these strategies by defining the dynamics as a discrete time Markovian process with probabilities of transition expressed in terms of a symmetric matrix of weights. In the first part, we describe the matrices of weights that define local random walk strategies like the normal random walk, biased random walks, random walks in the context of digital image processing and maximum entropy random walks. In addition, we explore non-local random walks like L\'evy flights on networks, fractional transport and applications in the context of human mobility.  Explicit relations for the stationary probability distribution, the mean first passage time and global times to characterize the random walk strategies are obtained in terms of the elements of the matrix of weights and its respective eigenvalues and eigenvectors. Finally, we apply the results to the analysis of particular local and non-local random walk strategies; we discuss their efficiency and capacity to explore different types of structures. Our results allow to study and compare on the same basis the global dynamics of different types of random walk strategies. 
\end{abstract}

\pacs{89.75.Hc, 05.40.Fb, 02.50.-r, 05.60.Cd}
%     89.75.Hc : Networks and genealogical trees
%     05.40.Fb : Random walks and Levy flights
%     02.50.-r : Probability theory, stochastic processes, and statistics
%     05.60.Cd : Classical transport

\maketitle

\section{Introduction}

Since their introduction as an informal question posted in the journal Nature in 1905 by Rayleigh, random walks have had an important impact in science with applications in a broad range of fields like biology, physics, chemistry, economy, computation, among many others \cite{KlafterSokolov,MasudaPhysRep2017}. The success of random walk models in different applications lies in their simplicity, typically defined as a walker in a particular space that moves randomly without memory of its previous path, making this characteristic a good candidate in the description of processes like the diffusive transport, chemical reactions, fluctuations in the economy and even the foraging of some animal species \cite{KlafterSokolov,MasudaPhysRep2017,SRedner,VKampen,ForeignBook,Viswanathan_Plos2017}. Despite the mentioned simplicity in the definition, the consequences of the dynamics of a random walker are non-trivial and continue to surprise us with new results and with all the complexity that emerges from its simple rules. 
\\[2mm]
In recent years, much of the interest in random walks have migrated to the study of complex systems described through networks \cite{NewmanBook,barabasi2016book,Latora2017}. In this context, the interplay between the topology of the network and the dynamical processes taking place on this structure are of utmost importance \cite{NewmanBook,VespiBook,VanMieghem}. In particular, random walk strategies that allow transitions from one node to one of its nearest neighbors on the network constitutes a paradigmatic case and are the natural framework to study diffusive transport \cite{VespiBook,Hughes,Lovasz1996,MulkenPR502}, navigation and search processes in networks \cite{NohRieger,FronczakPRE2009,TejedorPRE2009,Baronchelli2017}, multiplex networks \cite{Manlio2016}, with applications in a variety of systems like the propagation of epidemics and spreading phenomena \cite{Durrett2010,RomualdoRMP2015}, the dynamics on social networks \cite{Sarkar2011}, the analysis of information \cite{BlanchardBook2011}, human mobility \cite{RiascosMateosPlos2017}, among others \cite{MasudaPhysRep2017,VespiBook}. 
\\[2mm]
\begin{figure}[!b] 
\begin{center}
\includegraphics*[width=0.35\textwidth]{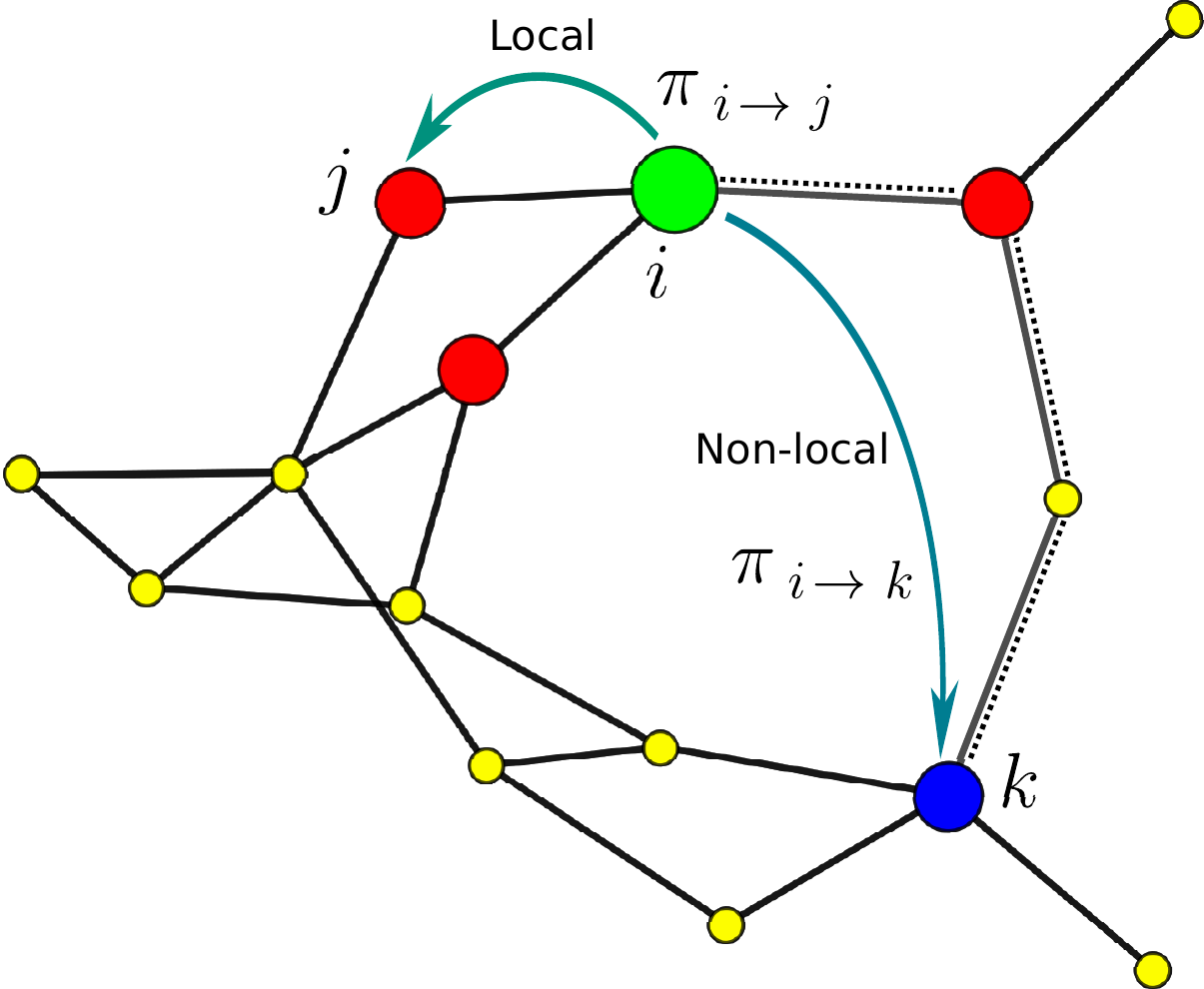}
\end{center}
\vspace{-5mm}
\caption{\label{FigureRW1}(Color online) Two types of transitions of a random walker on a network. The walker can hop from $i$ to the node $j$ that is one of the three nearest-neighbors available for a local transition. Also can make a non-local transition to reach the node $k$. In the non-local displacement, the length of the shortest path is three as indicated by the dashed line.} 
\end{figure}
On the other hand, in different cases full or partial knowledge of the network structure is available to define a random walk capable to use this information to increase the capacity to visit nodes with hops to the nearest neighbors but also long-range transitions beyond this local neighborhood. In Fig. \ref{FigureRW1} we illustrate local and non-local transitions in a network. In this case, the walker can visit one of the nearest-neighbors with a local transition but also there is the option of a non-local transition. By using this long-range dynamics the random walker can contact directly long-distance nodes without the intervention of intermediate nodes and without altering the topology of the network. As we will see, some non-local random walk strategies consider the shortest path connecting two nodes whereas others include quantities that contain all the possible paths between nodes. 
\\[2mm]
In addition, it is important to mention that random walks with a non-local character have been explored in the literature. This is the case of L\'evy flights on networks where random transitions occur to non-nearest neighbors with a probability that decays as a power law of the distance separating two nodes \cite{RiascosMateos2012}; the capacity of this strategy to explore networks has been studied in \cite{RiascosMateos2012,ZhaoPhysA2014,Huang2014132,Weng2015,Weng2016,Guo2016,Zheng2017}. L\'evy flights on networks were generalized by Estrada et. al. by using a random multi-hopper model defined in terms of decaying functions of the shortest-path distances; this approach is explored in detail in \cite{Estrada2017Multihopper}. Furthermore, we also found non-local dynamics in the fractional transport on networks defined in terms of the fractional Laplacian of a graph \cite{RiascosMateosFD2014}. In this case, long-range displacements on the network emerge from a formalism that is introduced as the equivalent of the fractional diffusion equation on networks \cite{RiascosMateosFD2014,RiascosMateosFD2015}.  This strategy is studied in the context of transport in networks and lattices \cite{RiascosMateosFD2014,RiascosMateosFD2015,Michelitsch2016Chaos,Michelitsch2017PhysA,Michelitsch2017PhysARecurrence,deNigris2016,DeNigris2017}, in connection with information analysis \cite{SdeNigris2017} and quantum transport on networks \cite{RiascosMateos2015}.  The fractional transport is a particular case of a series of strategies that can be defined in terms of functions of the Laplacian of a network \cite{RiascosMichelitsch2017_gL}. In general, the study of random walks with long-range displacements on networks opens several questions regarding the way in which these large displacements can appear or be induced in different applications. Moreover, it is necessary the introduction of new methods and quantities that allow us to compare in the same background the efficiency to visit the nodes of a network through random walk strategies.
\\[2mm]
In this paper, we explore different types of local and non-local random walks on networks. We present a general approach to study these processes on the same basis by using the information contained in a symmetric matrix of weights used to define the probability of transition between nodes. We model the dynamics as a discrete time Markovian process. In the first part, we describe the matrices of weights that define local random walk strategies: traditional random walks, biased random walks, random walks in the context of digital image processing and, maximum entropy random walks. In the same way, examples of non-local random walks are described: L\'evy flights on networks, fractional dynamics and applications in the context of human mobility. In all these cases, explicit relations for the stationary probability distribution of the random walker are obtained in terms of the elements of the matrix of weights that defines each strategy. After analyzing the transition matrix for these different processes, in a second part of the paper, a general formalism to calculate the mean first passage time and global times to characterize the dynamics is presented. Analytical expressions in terms of eigenvalues and eigenvectors of the transition matrix are obtained for all these quantities. Finally, we apply the results to the analysis of local and non-local random walk strategies to discuss and compare their efficiency and capacity to explore networks.

\section{Random walks on weighted networks}
\label{RW}
In this section, we introduce different concepts about random walks on weighted networks and the notation implemented to describe this process. We introduce a general random walker with probabilities of transition defined in terms of a network and a matrix of weights, the respective temporal evolution is modeled as a discrete time Markovian process for which we find an analytical result for its stationary probability distribution.
\\[2mm]
We consider undirected weighted networks with $N$ nodes $i=1,\ldots ,N$. The topology of the network is described by an adjacency matrix $\mathbf{A}$ with elements $A_{ij}=A_{ji}=1$ if there is an edge between the nodes $i$ and $j$ and $A_{ij}=0$ otherwise; in particular, $A_{ii}=0$ to avoid lines connecting a node with itself. The degree of the node $i$ is the number of neighbors that this node has and is given by $k_i=\sum_{l=1}^N A_{il}$.  Additionally to the network structure, we have a $N\times N$ symmetric matrix of weights $\mathbf{\Omega}$ with elements $\Omega_{ij}=\Omega_{ji}\geq 0$ and $\Omega_{ii}=0$. The matrix $\mathbf{\Omega}$ can include information of the structure of the network or incorporate additional data describing characteristics of links and nodes. By definition, the strength of the node $i$ is given by $S_i=\sum_{l=1}^N \Omega_{il}$ and represents the total weight of the node $i$.
\\[2mm]  
In the following, we study discrete time random walks on connected weighted networks with transition probabilities between nodes determined by the elements of the matrix of weights $\mathbf{\Omega}$. The occupation probability to find the random walker in the node $j$ at time $t$ starting from $i$ at $t=0$ is given by $P_{ij}(t)$ and obeys  the master equation \cite{Hughes,Weiss}
\begin{equation}\label{master}
P_{ij}(t+1) = \sum_{m=1}^N  P_{im} (t) \pi_{m\rightarrow j} \ ,
\end{equation}
where the transition probability $\pi_{i\rightarrow j}$ between the nodes $i$ and $j$ is given by 
\begin{equation}\label{wijomega}
\pi_{i\rightarrow j}=\frac{\Omega_{ij}}{\sum_{l=1}^N \Omega_{il}}=\frac{\Omega_{ij}}{S_i}.
\end{equation}
The transition matrix $\mathbf{\Pi}$, with elements $\pi_{i\rightarrow j}$, in the general case is not symmetric; however, as a consequence of Eq. (\ref{wijomega}) and the symmetry of the matrix $\mathbf{\Omega}$, we obtain $S_{i} \pi_{i\rightarrow j}=S_{j} \pi_{j\rightarrow i}$, a result that establishes a connection between the transition probabilities $\pi_{i\rightarrow j}$ and $\pi_{j\rightarrow i}$. On the other hand, iterating the master equation (\ref{master}), the probability $P_{ij}(t)$ takes the form
\begin{equation}\label{factors}
P_{ij}(t) = \sum_{j_1,\ldots,j_{t-1}} 
\pi_{i\rightarrow j_1} \cdot \pi_{j_1\rightarrow j_2} 
\cdots  
\pi_{j_{t-1}\rightarrow j} \, 
\end{equation}
and, using Eq. (\ref{factors}), we obtain
\begin{eqnarray}\nonumber
P_{ij}(t) & = \sum_{j_1,\ldots,j_{t-1}}
\frac{S_{j_1}}{S_i}\ldots \frac{S_{j}}{S_{j_{t-1}}}
\pi_{j\to j_{t-1}}  \ldots \pi_{j_1\to i}\\ \label{c2PijPji}
& =\frac{S_{j}}{S_i} P_{ji}(t).
\end{eqnarray}
In this way, the detailed balance condition
\begin{equation}\label{dbalance}
S_{i} P_{ij}(t)=S_{j} P_{ji}(t)
\end{equation} 
is deduced as a direct consequence of the symmetry of $\mathbf{\Omega}$.  The relation in Eq. (\ref{dbalance}) allows to obtain the stationary probability distribution $P_j^{\infty}=\lim_{t\to\infty}P_{ij}(t)$, that gives the probability to find the random walker in the node $j$ when $t\to \infty$. We have
\begin{equation}\label{Pinf}
	P_i^{\infty}= \frac{S_{i}}{\sum_{l=1}^N S_{l}} \, ,
\end{equation}
showing that the stationary distribution $P_i^{\infty}$ of the node $i$ is directly proportional to its strength $S_i$. The stationary distribution $P_{i}^\infty$ in Eq. (\ref{Pinf}) is a general result that characterizes the global behavior of the random walker. As we will see in the next section, this quantity allows to rank and classify the nodes of the network with a measure that combines the topological characteristics of the network structure with their capacity of transport modeled by the master equation (\ref{master}) and the transition matrix $\mathbf{\Pi}$. Furthermore, it is well known in the context of Markovian processes that the value $1/P_{i}^\infty$ is the average number of steps required for the random walker to return for the first time to the node $i$  \cite{ZhangPRE2013,Condamin2007}.
\section{Random Walk Strategies}
\label{SectRWS}
Diverse types of random walk strategies can be explored in terms of the matrix of weights formalism described before. The only restrictions to this approach are the symmetry of  the elements of the matrix of weights $\Omega_{ij}=\Omega_{ji}$, the condition $\Omega_{ij}\geq 0$ and $\Omega_{ii}=0$. In this section, we present particular cases of navigation strategies that can be described by using this method. We divide our discussion into local strategies, for which the transitions of the random walker are restricted to adjacent sites on the network, and long-range strategies, for which the walker can hop with displacements beyond its nearest neighbors.
\subsection{Local random walks}
In local random walk navigation strategies, the random walker always hops from a node to one of its nearest neighbors on the network. As a consequence, the elements of the matrix of weights take the form $\Omega_{ij}=g_{ij} A_{ij}$, where, as we explain in the following part, the value $g_{ij}$ is related to quantities assigned to each node or to the weight of the link that connects the nodes $i$ and $j$.
\subsubsection{Normal random walk}
In this case, the weights coincide with the elements of the adjacency matrix; therefore $\Omega_{ij}= A_{ij}$. As a consequence, from Eq. 
(\ref{wijomega}), the transition matrix is given by \cite{NohRieger}
\begin{equation}\label{wijNRW}
\pi_{i\rightarrow j}=\frac{A_{ij}}{k_i}.
\end{equation}
By definition, the normal random walker hops with equal probability from a node to one of its nearest neighbors in the network. In addition, from Eq. (\ref{Pinf}), the stationary distribution is $P_i^{\infty}= \frac{k_i} {\sum_{l=1}^N k_l}$. Normal random walks have been extensively studied in different contexts with applications in diverse types of networks; in particular, lattices \cite{Hughes,Weiss}, general graphs \cite{Telcs1989,Lovasz1996}, complex networks \cite{YangPRE2005,SandersPRE2009,TejedorPRE2009,KishorePRE2011}, fractal and recursive structures \cite{AgliariBenichouPRE2012}, among others \cite{BlanchardBook2011}.
\subsubsection{Preferential navigation}
In the preferential navigation strategy, a random walker hops with transition probabilities $\pi_{i \to j}$ that depend of the quantity $q_i>0$ assigned to each node $i$ of the network. The value $q_i$ can represent a topological feature of the respective node (e.g., the degree, the betweenness centrality, the eigenvector centrality, the clustering coefficient, among other measures \cite{NewmanBook}) or a value, independent of the network structure, that quantifies an existing resource at each node. We define preferential random walks with local information by means of the weights $\Omega_{ij}=(q_i q_j)^\beta A_{ij}$, where the exponent $\beta$ is a real parameter. Then, from Eq. (\ref{wijomega}), we have
\begin{equation}\label{wijNBRW}
\pi_{i\rightarrow j}=\frac{A_{ij} q_j^{\beta}}{\sum_{l=1}^N A_{il} q_l^{\beta}}.
\end{equation}
In Eq. (\ref{wijNBRW}), $\beta>0$ describes the tendency to hop to neighbor nodes with large values of $q$, whereas for $\beta<0$ this behavior is inverted and the walker tends to hop to nodes with lower values of $q$. On the other hand, for $\beta=0$ the normal random walk is recovered. By means of Eq. (\ref{Pinf}), the stationary distribution for the preferential random walk is
\begin{equation}
	P_i^{\infty}=\frac{\sum_{l=1}^N (q_i q_l)^\beta A_{il}}{\sum_{l,m=1}^N (q_l q_m)^\beta A_{lm}} \, .
\end{equation}
As we will see in the next part, the general preferential strategy defined by Eq. (\ref{wijNBRW}) determines different types of local random walks depending of the election of the quantities $q_i$. 
\subsubsection{Degree biased random walks}
\label{SecDBRW}
\begin{figure*}[!t]
\begin{center}
\includegraphics*[width=0.8\textwidth]{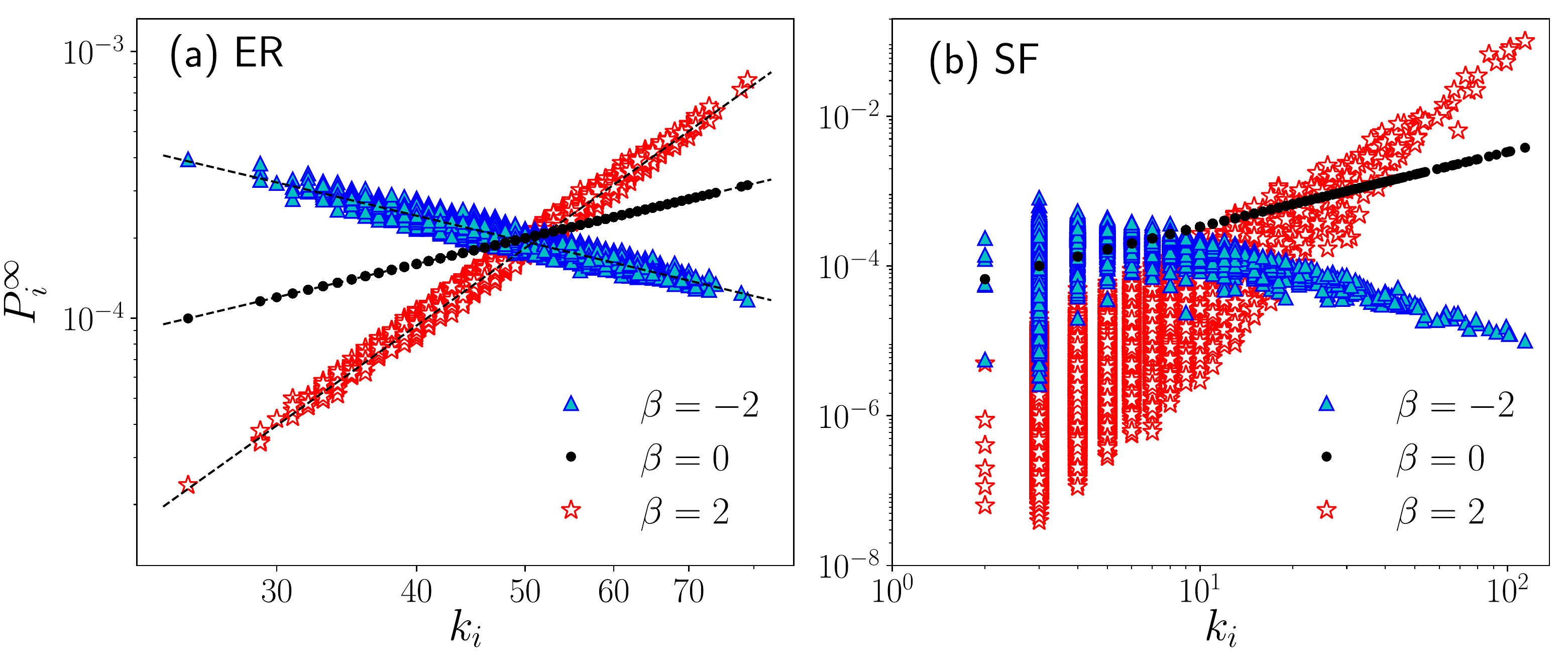}\,
\end{center}
\vspace{-5mm}
\caption{\label{FigureRW2} (Color online) Stationary distribution $P_i^{\infty}$ as a function of $k_i$ for degree biased random walks. The values are obtained by direct evaluation of Eq. (\ref{PinfDBRW}). We use three values of the parameter $\beta$ and we study two types of networks with $N=5000$ nodes.  (a) An Erd\"{o}s-R\'enyi network (ER) with an average degree $\langle k \rangle=50$; the dashed lines represent the results obtained by the mean field approximation. (b) A scale-free network (SF) with $\langle k \rangle=6$.}
\end{figure*}
This type of random walk is a particular case of the preferential navigation with $q_i=k_i$ in Eq. (\ref{wijNBRW}). The resulting strategy is known as degree biased random walks \cite{WangPRE2006,FronczakPRE2009}. For this particular case, the stationary distribution $P_i^{\infty}$ takes the form
\begin{equation}\label{PinfDBRW}
	P_i^{\infty}=\frac{\sum_{l=1}^N (k_i k_l)^\beta A_{il}}{\sum_{l,m=1}^N (k_l k_m)^\beta A_{lm}} \, .
\end{equation}
Degree biased random walks have been studied extensively in the literature in different contexts as varied as routing processes \cite{WangPRE2006},  chemical reactions \cite{KwonPRE2010}, extreme events \cite{KishorePRE2012,LingEPJB2013}, among others \cite{FronczakPRE2009,LambiottePRE2011,Battiston2016}. Additionally, mean field approximations have been explored for diverse cases \cite{FronczakPRE2009,KwonPRE2010,ZhangJSMTE2011}. For example, in networks with no degree correlations is valid the approximation $P_i^{\infty}\approx \frac{k_i^{\beta+1}} {\sum_{l=1}^N k_l^{\beta+1}}$. In Fig. \ref{FigureRW2} we present the values of the stationary distribution $P_i^\infty$  for degree biased random walks on an Erd\"{o}s-R\'enyi network (ER) and, a scale-free  network (SF) of the Barab\'asi-Albert type, in which each node has a degree that follows asymptotically a power-law distribution $p(k)\approx k^\gamma$ \cite{NewmanBook,VespiBook}. We calculate the stationary distribution  by direct evaluation of the Eq.  (\ref{PinfDBRW}) and we depict $P_i^\infty$ as a function of the degree $k_i$. The results reveal that in the ER network is valid the mean-field approximation whereas in a SF network, this is only valid for nodes with high degrees \cite{FronczakPRE2009}.

\subsubsection{Maximal entropy random walks}
Maximum entropy random walks (MERW) are a particular strategy derived from Eq. (\ref{wijNBRW}) for which the random walker uses information of the neighbor nodes. In this case, the transition probability is defined in terms of the components of the eigenvector centrality $\xi_i$ of the node $i$. The value $\xi_i$ is determined by the $i$-th component of the normalized eigenvector $\vec{\xi}$ of the adjacency matrix $\mathbf{A}$ that satisfies $\mathbf{A}\vec{\xi}=\chi\vec{\xi}$, where $\chi$ is the maximum eigenvalue of $\mathbf{A}$. In the study of topological features of networks, the components $\xi_i$ of the eigenvector centrality  quantify the global influence of the node $i$ in the whole structure \cite{NewmanBook}. 
\\[1mm]
In this way, MERW are defined in the formalism of weighted networks with the election of weights $\Omega_{ij}=\xi_i \xi_j A_{ij}$. Then, the value of the strength $S_i$ is
\begin{equation}
S_i=\sum_{l=1}^N \Omega_{il}=
\sum_{l=1}^N \xi_i \xi_l A_{il}=
\xi_i 
\sum_{l=1}^N A_{il}\xi_l =\chi \xi_i^2, 
\end{equation}
where the last result is a consequence of the relation $\sum_{l=1}^N A_{il}\xi_l=\chi \xi_i$ that satisfy the components of  the eigenvector centrality. In this way, by using Eq. (\ref{wijomega}), the  transition rule  $\pi_{i\to j}$ is given by
\begin{equation}\label{wijmaxentropy}
	\pi_{i\to j}=A_{ij}\frac{\xi_i \xi_j}{\chi \xi_i^2}=A_{ij}\frac{\xi_j}{\chi \xi_i},
\end{equation}
relation that defines a maximal entropy random walk \cite{BurdaPRL2009}. Additionally, by using the Eq. (\ref{Pinf}), the stationary distribution of the maximal entropy random walk is
\begin{equation}\label{pinfmaxentropy}
P_i^\infty=\frac{\chi \xi_i^2}{\sum_{l=1}^N \chi \xi_l^2}= \xi_i^2. 
\end{equation}
It is worth to mention that, the MERW defined by the transition probabilities in Eq. (\ref{wijmaxentropy}) maximizes the entropy rate production $h$ of the process given by \cite{BurdaPRL2009} 
\begin{equation}
h=-\sum_{i=1}^N P_i^\infty \sum_{j=1}^N \pi_{i\to j} \log \pi_{i\to j}.
\end{equation}
Combining this expression with Eqs. (\ref{wijmaxentropy}) and (\ref{pinfmaxentropy}), $h=\log \chi$ \cite{BurdaPRL2009}. In this case, the trajectories that follow the random walker are maximally random \cite{BurdaPRL2009,SinatraPRE2011}. Diverse variations of the MERW and applications of this process have been explored in \cite{SinatraPRE2011,OchabPRE2012,LawrencePRE2014,YuanSciRep2014}.
\subsubsection{Random walks for image segmentation}
\begin{figure}[!t]
\begin{center}
\includegraphics*[width=0.48\textwidth]{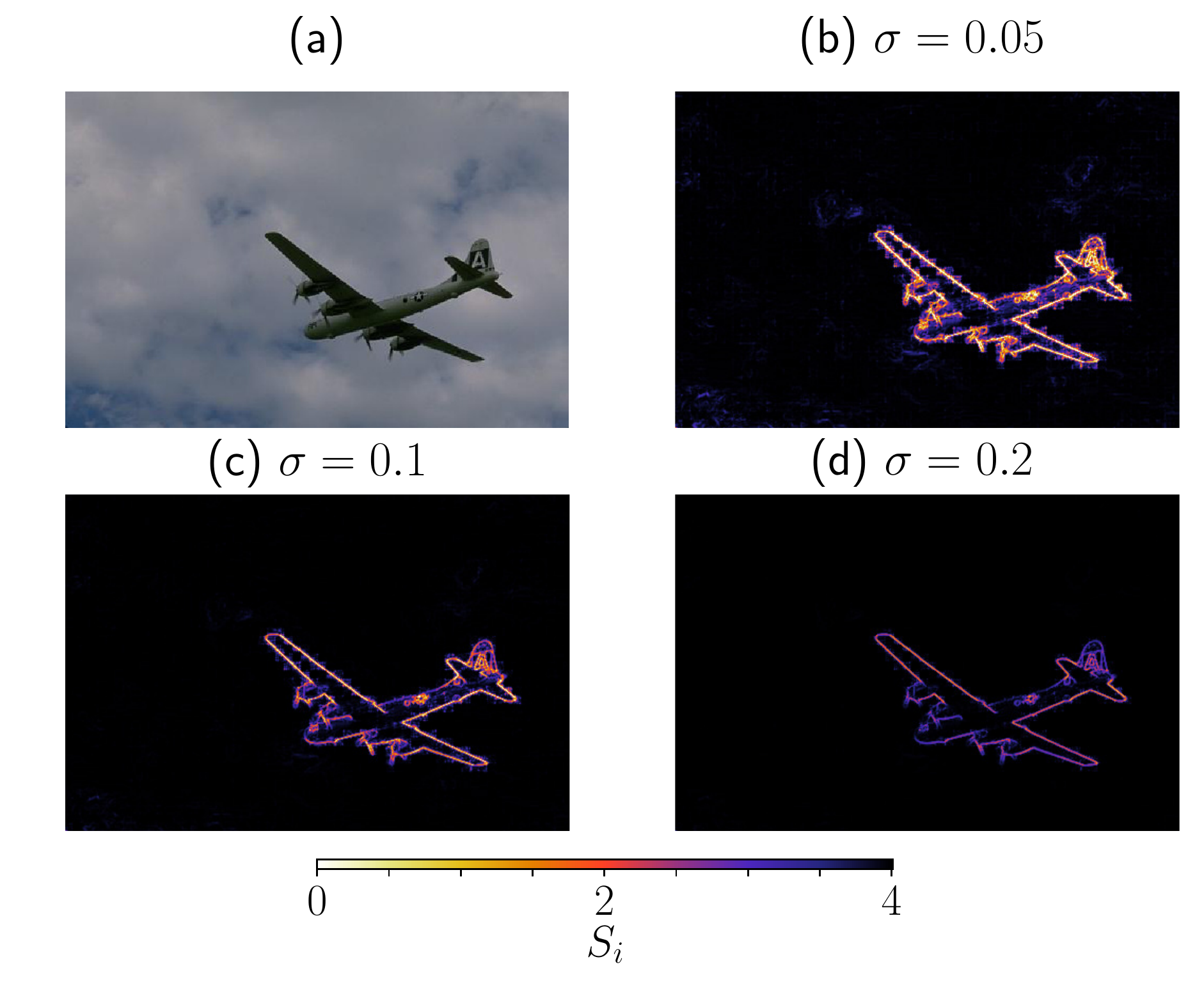}
\end{center}
\vspace{-5mm}
\caption{\label{FigureRW3} (Color online) Strength $S_i$  for random walks in the context of image segmentation. The values are obtained evaluating 
the sum $S_i=\sum_{l=1}^N \Omega_{il}$ with the weights given by Eq. (\ref{GradyStrategy}). In (a) we present the original image  (\#3096 from the Berkeley segmentation database BSD300 \cite{DataBase2001}). In (b)-(d) we present the obtained results for different values of the parameter $\beta$, the colorbar denotes the scale of values for $S_{i}$. Regions with $S_{i}=4$ present little variations in the intensity of a pixel in relation with its nearest neighbors, in these regions the random walker behaves as a normal random walker.}
\end{figure}
An important application of random walks on networks emerges in the context of the processing  and segmentation of digital images  \cite{GradyIEEE2006}. In this case, the statistical description of the diffusive transport from seed regions to specific pixels allows to detect and differentiate objects and structures in a digital image \cite{GradyIEEE2006}. The network is a square lattice where each node represents a pixel and the normalized intensity $\mathcal{I}_i$ of $i$ is a quantity  associated to the norm of the vector $\vec{p}_i$ that contains the values RGB (red, green, blue) of the respective pixel, $0\leq \mathcal{I}_i \leq 1$. In terms of a matrix of weights  $\mathbf{\Omega}$, a local random walker is defined by \cite{GradyIEEE2006}
\begin{equation}\label{GradyStrategy}
\Omega_{ij}=\text{exp}\left[-(\mathcal{I}_i-\mathcal{I}_j)^2/\sigma^2\right] A_{ij}.
\end{equation}
Here, the real parameter $\sigma$ satisfies $\sigma>0$ and the values $A_{ij}$ give the elements of the adjacency matrix of a square lattice associated to the pixels positions and interactions between nearest neighbors. The resulting random walker follows a strategy given by Eq. (\ref{wijomega}) to visit the pixels; this transition probability gives high probability to the pass to pixels with the same intensity and $\sigma$ determines the interaction between the pixels establishing a characteristic scale for the differences of intensity in the model controlling the capacity to hop to sites with a different color. In Fig. \ref{FigureRW3} we plot the strength $S_i$ for each pixel, this quantity is proportional to the stationary probability distribution for a random walker in a digital image that follows a strategy determined by the weights given by Eq. (\ref{GradyStrategy}). It is observed how with this strategy, $S_i$ takes high values in regions with uniform color and low values in the boundaries of the object. In this way, the random walker propagates uniformly in regions with the same color and with low probability passes through the boundary of the object. This property makes this type of weights good candidates for image segmentation algorithms.
\\[2mm]
In addition to the local strategy mentioned before, it is worth mentioning that exists different variations of these models; this is the case of the topological biased random walks for which $\Omega_{ij}=e^{\beta y_{ij}} A_{ij}$ \cite{ZlaticPRE2010}, where the quantity $y_{ij}$ describes the properties of the edge that connects $i$ with $j$. A similar idea is explored for image segmentation in \cite{GradyIEEE2007}, showing the vast applicability of random walks in different scenarios.
\subsection{Non-local random walks}
Non-local random walks on networks are motivated by the possibility of hopping from one node to sites on the network beyond the neighbor nodes in cases where the total structure of the network is available. Random walk strategies with long-range displacements have shown an unprecedented applicability in the context of web searching. The  PageRank introduced to classify pages on the Web \cite{Brin1998} and variations of this non-local strategy have been explored to rank the importance of nodes in a broad range of systems. In the following part, we present diverse non-local strategies on undirected networks that can be expressed in terms of a matrix of weights that includes information about the whole structure of the network to define the dynamical process. As particular examples of this case we have the L\'evy flights on networks \cite{RiascosMateos2012}, the fractional diffusion on networks \cite{RiascosMateosFD2014,RiascosMateosFD2015}, the dynamics of agents moving visiting specific locations in  a city \cite{RiascosMateosPlos2017} and different strategies in the context of the random multi-hopper model  \cite{Estrada2017Multihopper}. The study and possible applications of non-local dynamical processes on networks are relatively new and open questions related with the exploration of the effects that non-locality introduces as well as the search of global quantities that allow us to compare the performance of non-local against local dynamics.
\subsubsection{L\'evy flights on networks}
\label{refSecLevy}
\begin{figure}[!b]
\begin{center}
\includegraphics*[width=0.47\textwidth]{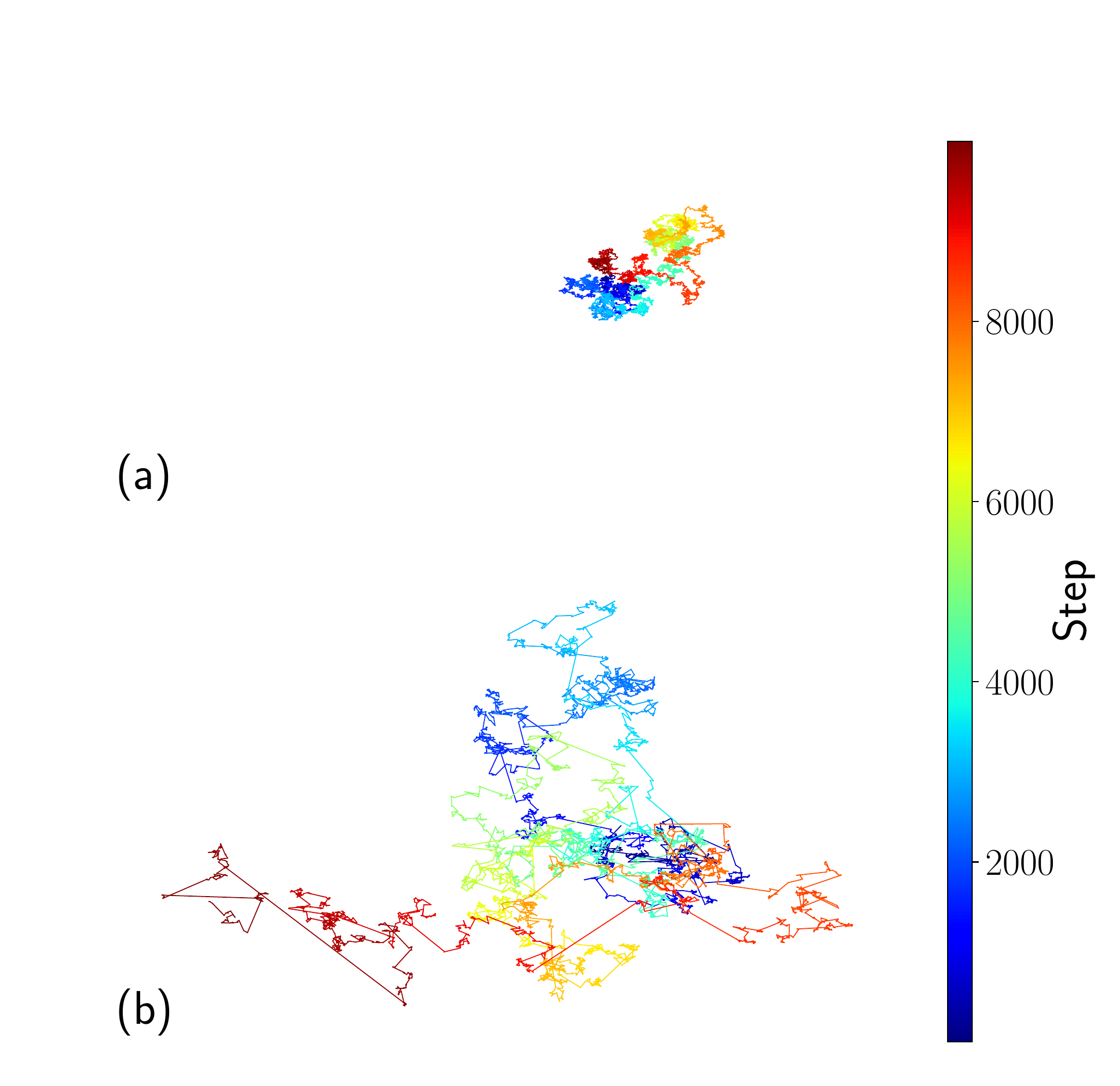}
\end{center}
\vspace{-5mm}
\caption{\label{FigureRW4} (Color online) Monte Carlo simulations of two different types of random walks on a plane. (a) Brownian motion. (b) L\'evy flights. We depict $10^4$ steps for each realization of the dynamics. The colorbar codifies the number of each step that in this case is a measure of a discrete time. }
\end{figure}
The term L\'evy flights makes reference to a random walk with displacements of length $l$ that appear with a probability distribution  $\mathcal{K}(l)$ that asymptotically is described by an inverse power-law relation \cite{Ralf2004,RevModPhysZaburdaev2015}. For L\'evy flights in the $n$-dimensional space $\mathbb{R}^{n}$, $\mathcal{K}(0)=0$ and $\mathcal{K}(l) \sim\frac{1}{l^{n+2\gamma}}$ if   $l\neq 0$ for $0<\gamma<1$. With this definition, the variance of the displacements diverges; this characteristic differentiates L\'evy flights from the Brownian motion for which the variance is finite \cite{Weiss}. In Fig. \ref{FigureRW4} we present Monte Carlo simulations for Brownian motions and L\'evy flights in a plane. L\'evy flights have a fractal behavior consisting of trajectories that alternate between groups described by local movements (similar to the observed in the Brownian motion) interrupted by long-range jumps; this structure is repeated at all levels. In this way, L\'evy flights combine local movements, that appear with high probability, with long-range displacements that emerge with low but non-null probability. These characteristics are illustrated in Fig. \ref{FigureRW4}(b). L\'evy flights constitute an active area of research in different complex systems. For example, L\'evy flights are encountered in the modelling of animal dynamics and foraging \cite{BES2004,Proc2006,BoyerJRSI,ForeignBook,Viswanathan_Plos2017}, human mobility \cite{Brock2006,Brown,Rhee}, among many others \cite{Ralf2004,AnomalousDiff,RevModPhysZaburdaev2015}.

\begin{figure}[!t]
\begin{center}
\includegraphics*[width=0.48\textwidth]{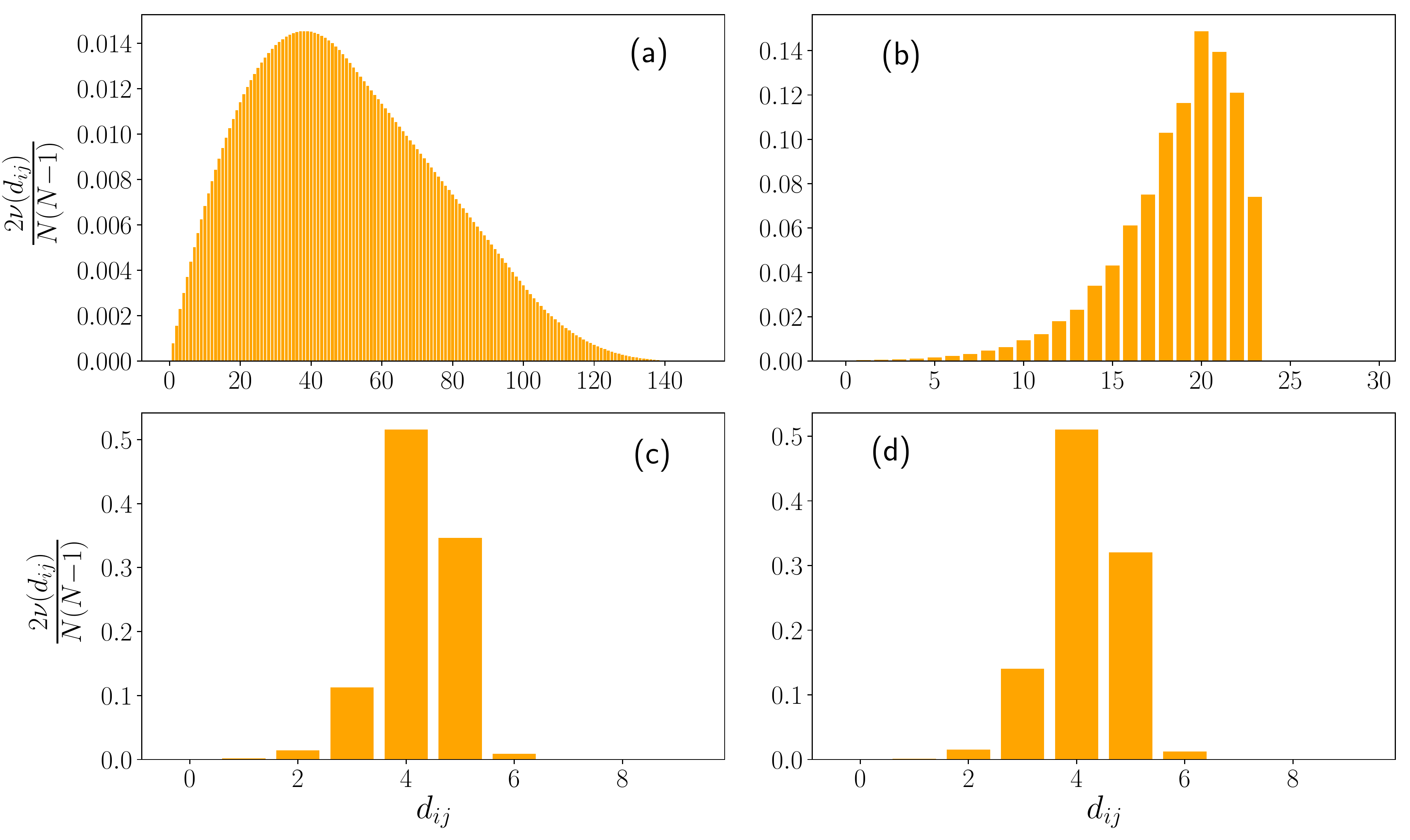}
\end{center}
\vspace{-5mm}
\caption{\label{FigureRW5} (Color online) Frequencies $\nu(d_{ij})$ of the non-null distances $d_{ij}$ in the entries of the distance matrix $\mathbf{D}$. We analyze networks with  $N=5000$. (a) Square lattice with dimensions $50\times 100$. (b) Tree. (c) Erd\H{o}s--R\'enyi network with probability of connection $p=\frac{\log N}{N}$. (d) Scale-free network of the Barab\'asi--Albert type. The results are expressed  as a fraction of the value $N(N-1)/2$  that gives the total number of different non-null entries  in the distance matrix  $\mathbf{D}$.}
\end{figure}

In the context of networks, L\'evy flights are introduced in reference \cite{RiascosMateos2012}. In this case, the transitions are defined in terms of the distance $ d_ {ij} $ that gives the number of lines in the shortest path connecting the nodes $ i $ and $ j $. All the information about the distances between nodes is contained in the distance matrix $\mathbf{D}$ with elements $d_{ij}$ for $i,j=1,2,\ldots,N$. The distance matrix $\mathbf{D}$ contains more information about the structure of the network than the adjacency matrix  $\mathbf{A}$,   but   $\mathbf{D}$ can be calculated efficiently from $\mathbf{A}$  using different algorithms \cite{NewmanBook}. In Fig. \ref{FigureRW5} we depict the relative frequency of distances in the entries of the matrix $\mathbf{D}$ for  large-world networks (square lattice and tree) and small-world networks (Erd\H{o}s--R\'enyi network and scale-free network of the Barab\'asi-Albert type). The histograms reveal the marked difference between the distances in these two types of structures.
\\[3mm]
\begin{figure*}[!t]
\begin{center}
\vspace{2mm}
\includegraphics*[width=0.8\textwidth]{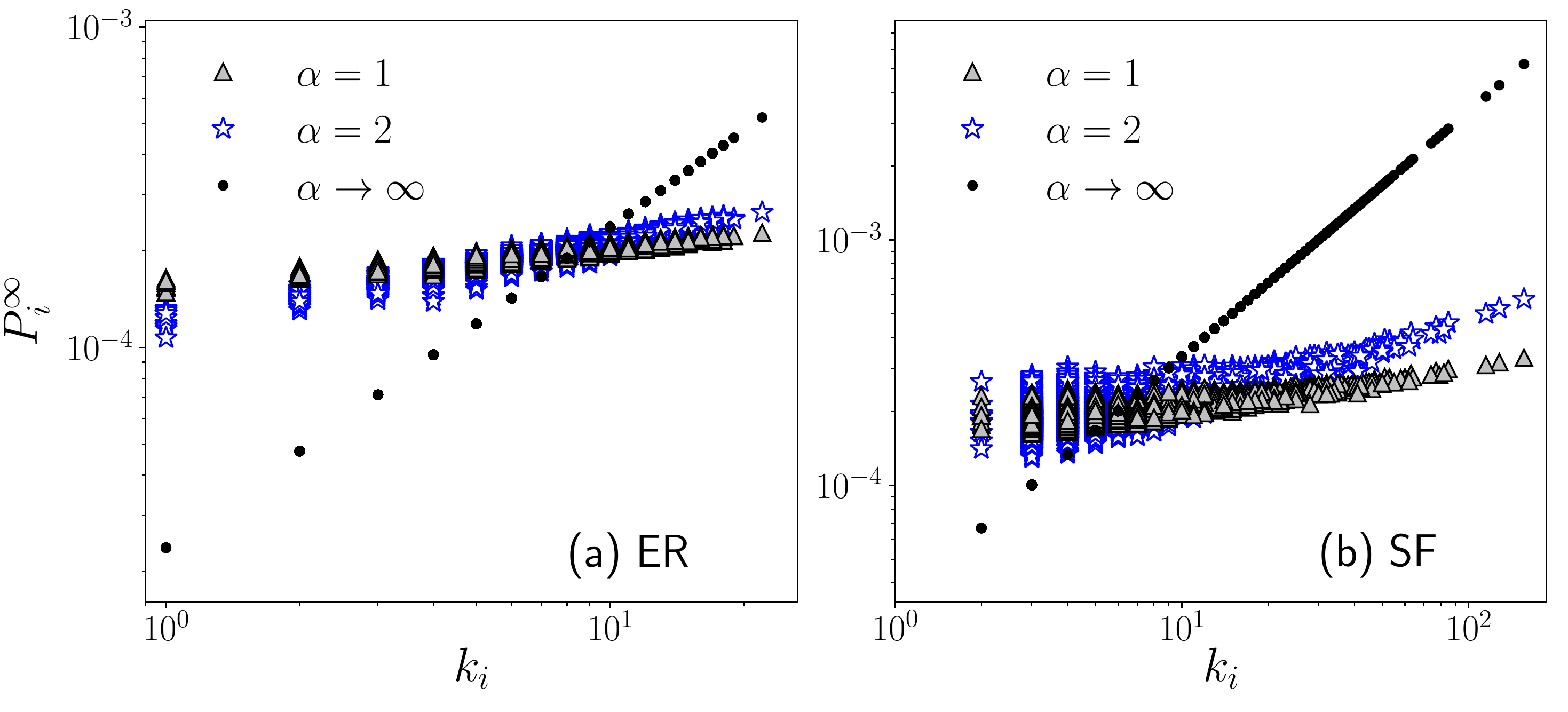}
\end{center}
\vspace{-5mm}
\caption{\label{FigureRW6} (Color online) Stationary distribution $P_i^\infty$  in terms of the degree $k_i$ for L\'evy flights on networks with $N=5000$. The strategies with  $\alpha=1$ and $\alpha=2$ use long-range displacements in the network. The result for the normal walker (limit $ \alpha \to \infty) $ is also depicted.}
\end{figure*}
L\'evy flights on networks can be described in terms of the weights $\Omega_{ij}=d_{ij}^{-\alpha}$ for $i\neq j$ and $\Omega_{ii}=0$. Here $\alpha$ is a real parameter in the interval $0\leq\alpha<\infty$. For the elements of the transition matrix, we have $\pi_{i\to i}=0$ and by using Eq. (\ref{wijomega}) for $i\neq j$ is obtained \cite{RiascosMateos2012}
\begin{equation}\label{wijLevy}
\pi_{i\to j}=\frac{d_{ij}^{-\alpha}}{\sum_{l \neq i}d_{il}^{-\alpha}}.
\end{equation}
The dynamics inspired in L\'evy flights allows long-range transitions on the network. For a finite non-null value of $\alpha$, the transitions to the nearest neighbors appear with high probability, but hops beyond these nodes are allowed generalizing the dynamics observed in the normal random walker in Eq.  (\ref{wijNRW}). In the limit  $\alpha\to\infty$ we have $\lim_{\alpha\to\infty} d_{ij}^{-\alpha}=A_{ij}$, then $\pi_{i\to j}=\frac{A_{ij}}{k_i}$ and the L\'evy strategy recovers  the normal random walk. Another interesting limit case is obtained when $\alpha\to 0$, in this case $\lim_{\alpha\to 0} d_{ij}^{-\alpha}=1$ if $i\neq j$ and the dynamics induces the possibility to reach with equal probability any node of the network \cite{RiascosMateos2012}.
\\[1mm]
Once defined L\'evy flights in terms of the elements $\Omega_{ij}=d_{ij}^{-\alpha}$ for  $i\neq j$; for this particular model we denote the strength $S_i=\sum_{l=1}^N \Omega_{il}$ as $ D_i^{(\alpha)}=\sum_{l\neq i}d_{il}^{-\alpha}$ and by using the Eq. (\ref{Pinf})  we obtain for the stationary distribution
\begin{equation}\label{PinfLevy}
	P_i^{\infty}=\frac{D_{i}^{(\alpha)}}{\sum_{l=1}^N D_{l}^{(\alpha)}}=\frac{ \sum_{l\neq i}d_{il}^{-\alpha}}{\sum_{l\neq m}\sum_{m}d_{lm}^{-\alpha}} \, .
\end{equation}
This result establishes that $P_i^{\infty}$ is proportional to the quantity $ D_i^{(\alpha)}$. In addition, the value $ D_i^{(\alpha)}$,  can be expressed as \cite{RiascosMateos2012}
\begin{equation}\label{Dsum}
  D_{i}^{(\alpha)} =  \sum_{l=1}^{N-1} \frac{1}{l^{\alpha}} n_i^{(l)}=k_i+ \frac{n_i^{(2)}}{2^{\alpha}}+ \frac{n_i^{(3)}}{3^{\alpha}}+\ldots   ,
\end{equation}
where $ n_i^{(l)}$ is the number of nodes at a distance $l$ of the node $i$; in particular, $n_i^{(1)}=k_i$. In this way, by means of the expression in Eq. (\ref{Dsum}) we observe that $D_{i}^{(\alpha)}$ is a generalization of the degree $k_i$ that combines all the information about the structure of the network. 
This \textsl{long-range degree} emerges from the study of L\'evy flights on networks and was introduced in \cite{RiascosMateos2012}.
\\[2mm]
In Fig. \ref{FigureRW6} we depict the stationary distribution obtained from the analytical result in Eq. (\ref{PinfLevy})  for an Erd\H{o}s R\'enyi network and a scale-free network. Also calls the attention that, compared to the normal strategy, L\'evy flights represent a democratic strategy in the sense that the probability of visiting sites with many connections decreases and for sites with a lower degree, this probability increases. Being able to easily reach any node on the network can offer advantages if the goal is the exploration of the entire structure. This aspect is discussed in detail later when the efficiency of the random walker is analyzed.
\\[2mm]
Different aspects of L\'evy flights and their capacity to explore networks have been studied in  \cite{ZhaoPhysA2014,Huang2014132,Weng2015,Weng2016}, as well as in the context of multiplex networks \cite{Guo2016}. A general approach to study the random walker in Eq. (\ref{wijLevy}) and other strategies defined in terms of a function of the distances in a network are analyzed in detail by Estrada et.al. in \cite{Estrada2017Multihopper}. In this context is introduced the exponential strategy that in terms of our matrix of weights formalism is defined by $\Omega_{ij}=e^{-sd_{ij}}$ for $i\neq j$ and using  $s>0$.  By following a similar approach to the presented in Eqs. (\ref{PinfLevy})-(\ref{Dsum}), can be deduced analytical expressions for the stationary probability distribution of the exponential strategy. 
\subsubsection{Gravity law, spatial networks, and human mobility}

In diverse situations networks are embedded in a metric space, then, spatial locations are assigned to each node. This is the case of spatial networks that describe several real systems like social networks, airports and transportation networks, among others \cite{Barthelemy2011,Huang2014132,Barbosa2018}.
\\[1mm]
On the other hand, it has been suggested that migration and human movements are well described in terms of a ``Gravity Law" that models the number of trips from a location $i$ to the location $j$ as $g_{ij}=C\,p_i p_j/l_{ij}^\alpha$. Here $p_i$ and $p_j$ denote the population of the respective locations, $l_{ij}$ is the geometric distance between the nodes $i$, $j$$, C$ is a constant and $\alpha>0$ is a free parameter. This type of model suggests a similar algorithm for a random walker on networks described by the weights
\begin{equation}\label{OmegaijGravity}
\Omega_{ij}=\frac{q_i q_j}{d_{ij}^\alpha} 
\end{equation}
for $i\neq j$. Here the value $q_i$ is a quantity associated to the node $i$ in the network and $d_{ij}$ is the topological distance in the network. The general formalism in terms of weighted networks also applies to the model presented in Eq. (\ref{OmegaijGravity}), but with geometric distances $l_{ij}$. In this model, the structure of the network is absent and it is assumed as a complete graph. 
\\[1mm]
In the gravity law model, the resulting random walker contains characteristics of the biased random walks determined by Eq. (\ref{wijNBRW}) and the L\'evy flights on networks with transition probabilities given by Eq. (\ref{wijLevy}). The random walk defined in Eq. (\ref{OmegaijGravity}) has been explored in order to characterize co-occurrences of words on web pages \cite{YuGIS2014}. In addition, there are different variations of the gravity law in spatial networks (see  \cite{Barthelemy2011} and references therein). Some of these models are described in the weighted network approach by weights $\Omega_{ij}$ proportional to a positive function of the distance $f(d_{ij})$ \cite{Barthelemy2011}.
\\[2mm]
As an example of random walks that take place in a continuous space but can be modeled with the formalism of random walks defined in terms of a matrix of weights, in reference \cite{RiascosMateosPlos2017} is introduced a strategy to visit randomly specific locations in a spatial region modeling characteristics of human mobility in urban settlements. In this case, $N$ points are located in a $2D$ plane and integer numbers 
$i=1,2, \ldots ,N$ label each location. In addition, the coordinates of the locations are known and we denote as $l_{ij}$ the distance between the places $i$ and $j$. The distance $l_{ij}=l_{ji}\geq 0$ can be calculated by different metrics; for example, in some applications could be appropriated the use a Euclidean metric, whereas, in other contexts, a Manhattan distance could be more useful. In order to define a discrete time random walker that at each step visits one of the locations, the transition probability $\pi_{i\to j}^{(\alpha)}(R)$ to hop  from site $i$ to site $j$ is given by \cite{RiascosMateosPlos2017}
\begin{equation}\label{wijRalpha}
\pi_{i\to j}^{(\alpha)}(R)=\frac{\Omega_{ij}^{(\alpha)}(R)}{\sum_{m=1}^{N} \Omega_{im}^{(\alpha)}(R)},
\end{equation}
where the weights $\Omega_{ij}^{(\alpha)}(R)$ are defined by the relation \cite{RiascosMateosPlos2017}
\begin{eqnarray}\label{Omega_ij}
\Omega_{ij}^{(\alpha)}(R)&=
\left\{
\begin{array}{ll}
1 & \rm{for} \quad  0\leq l_{ij}\leq R,\\
\left(R/l_{ij}\right)^\alpha & \rm{for} \quad  R<l_{ij}.\\
\end{array}\right.
\end{eqnarray}
\begin{figure}[!b] 
\begin{center}
\includegraphics*[width=0.48\textwidth]{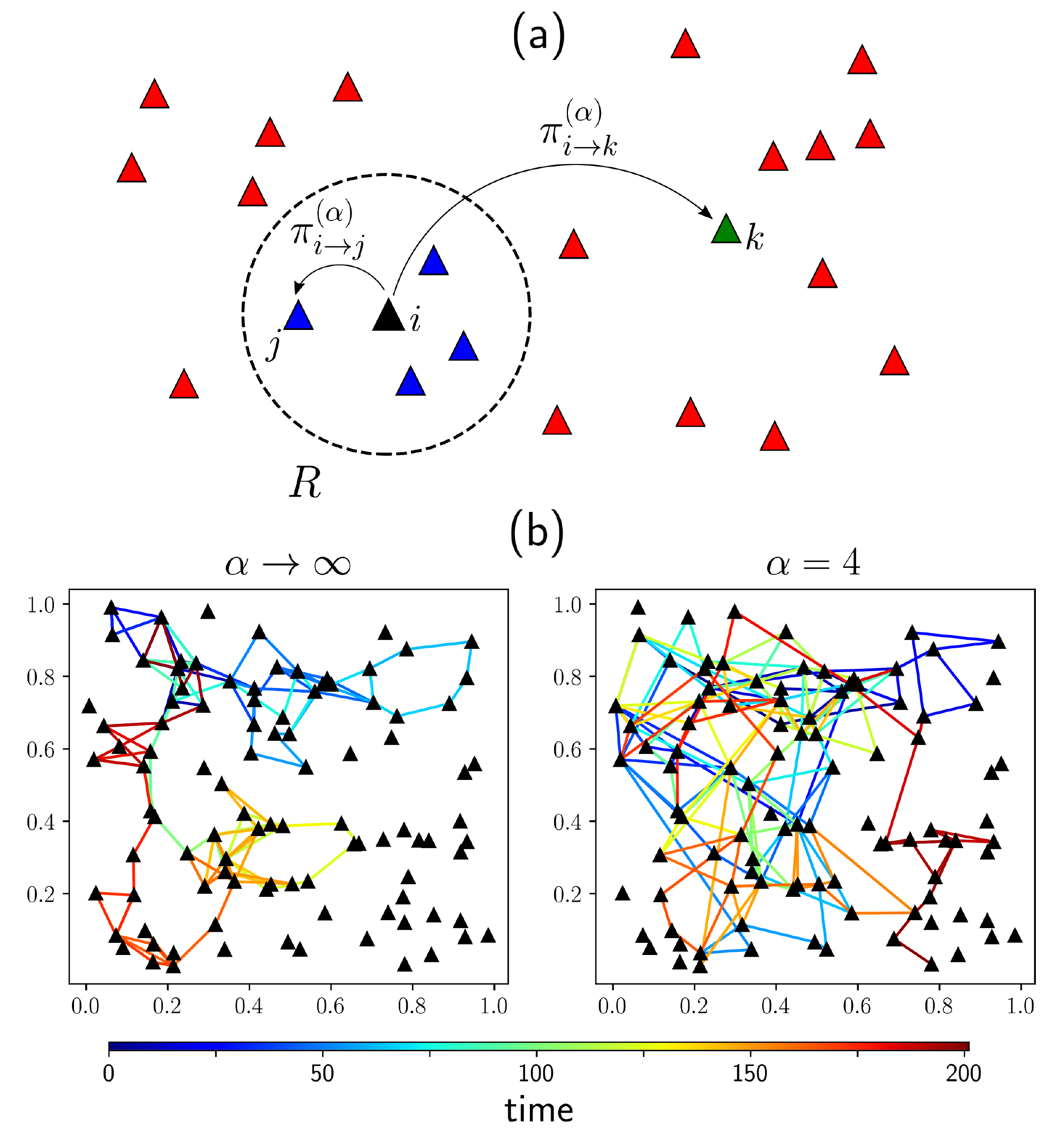}
\end{center}
\caption{\label{FigureRW7} (Color online) A schematic illustration of the random walk strategy as defined in Eq (\ref{wijRalpha}). In (a) we depict random locations on the plane (represented by triangles); the probability  to go from location $i$ to a different site is determined by two types of transition probabilities: First, to a site $j$ inside a circular region of radius $R$ centered in the location $i$, $\pi_{i\to j}^{(\alpha)}(R)$, which is a constant; and second, a transition to a site $k$ outside the circle of radius $R$, $\pi_{i\to j}^{(\alpha)}(R)$ that considers long-range transitions with a power-law decay proportional to $l_{ik}^{-\alpha}$, where $l_{ik}$ is the distance between sites $i$ and $k$. In (b) we show Monte Carlo simulations of a discrete-time random walker that visits $N=100$ specific locations in the region $[0,1]\times[0,1]$ in $\mathbb{R}^2$ following the random strategy defined by the transition probabilities in Eq (\ref{wijRalpha}), with $R=0.17$. We depict the results for $\alpha\to \infty$ and $\alpha=4$. The total number of steps is $t=200$ and the scale in the color bar represents the time at each step.}
\end{figure}
Here $\alpha$ and $R$ are positive real parameters. The radius $R$ determines a neighborhood around which 
the random walker can go from the initial site to any of the locations in this region with equal probability; this 
transition is independent of the distance between the respective sites. That is, if there are $S$ sites inside a circle of radius $R$,
the probability of going to any of these sites is  a constant. Additionally, for places beyond the local 
neighborhood, for distances greater than $R$, the transition probability decays as an inverse power law of the distance and is proportional 
to $l_{ij}^{-\alpha}$ \cite{RiascosMateosPlos2017}. In this way, the parameter $R$ defines a characteristic length of the local neighborhood and 
$\alpha$ controls the capacity of the walker to hop with long-range displacements. In particular, in the limit $\alpha\to \infty$ the dynamics becomes 
local, whereas the case $\alpha\to 0$ gives the possibility to go from one location to any 
different one with the same probability. In this limit, we have $\pi_{i\to j}^{(0)}(R)=N^{-1}$.
This model is then a combination of a rank model \cite{Liben05,Noulas2012,Pan2013} for shorter distances and a 
gravity-like model for larger ones \cite{Simini2012,Barbosa2018}. It is important to mention that in the strategy defined by the weights in Eq. (\ref{Omega_ij}), we choose $\Omega_{ii}^{(\alpha)}(R)\neq 0$, in this way the walker also can stay in the node $i$ with non-null probability. All the results presented are also valid for this case whenever the value of $R$ is such that the random walker can reach any of the $N$ sites used in the definition of the transition matrix.
\\[2mm]
In Fig. \ref{FigureRW7}(a) we illustrate the model for the random strategy introduced in Eq (\ref{wijRalpha}). 
In Fig. \ref{FigureRW7}(b), we present Monte Carlo simulations of the random walker described 
by Eqs (\ref{wijRalpha})-(\ref{Omega_ij}). We generate $N$ random locations (points) on a 2D plane on the 
region $[0,1]\times[0,1]$ in $\mathbb{R}^2$ and, for different values of the exponent $\alpha$, we depict the trajectories 
described by the walkers. In the case of $\alpha\to \infty$, it is observed how the dynamics is local and only 
allows transitions to sites in a neighborhood determined by a radius $R=0.17$ around each location. In this 
case, all the possible trajectories in the limit $t \to \infty$ form a random geometric graph \cite{Dall2002,Estrada2015}; 
we can identify features of this structure in our simulation. On the other hand, finite values 
of $\alpha$ model spatial long-range displacements such as the dynamics illustrated in 
Fig. \ref{FigureRW7}(b) for the case $\alpha=4$. We observe how the introduction of the long-range strategy 
improves the capacity of the random walker to visit and explore more locations in comparison with the local 
dynamics defined by the limit $\alpha\to \infty$ \cite{RiascosMateosPlos2017}.

\subsubsection{Fractional transport}
\label{subs_fractional}
\begin{figure*}[!t] 
\begin{center}
\includegraphics*[width=0.77\textwidth]{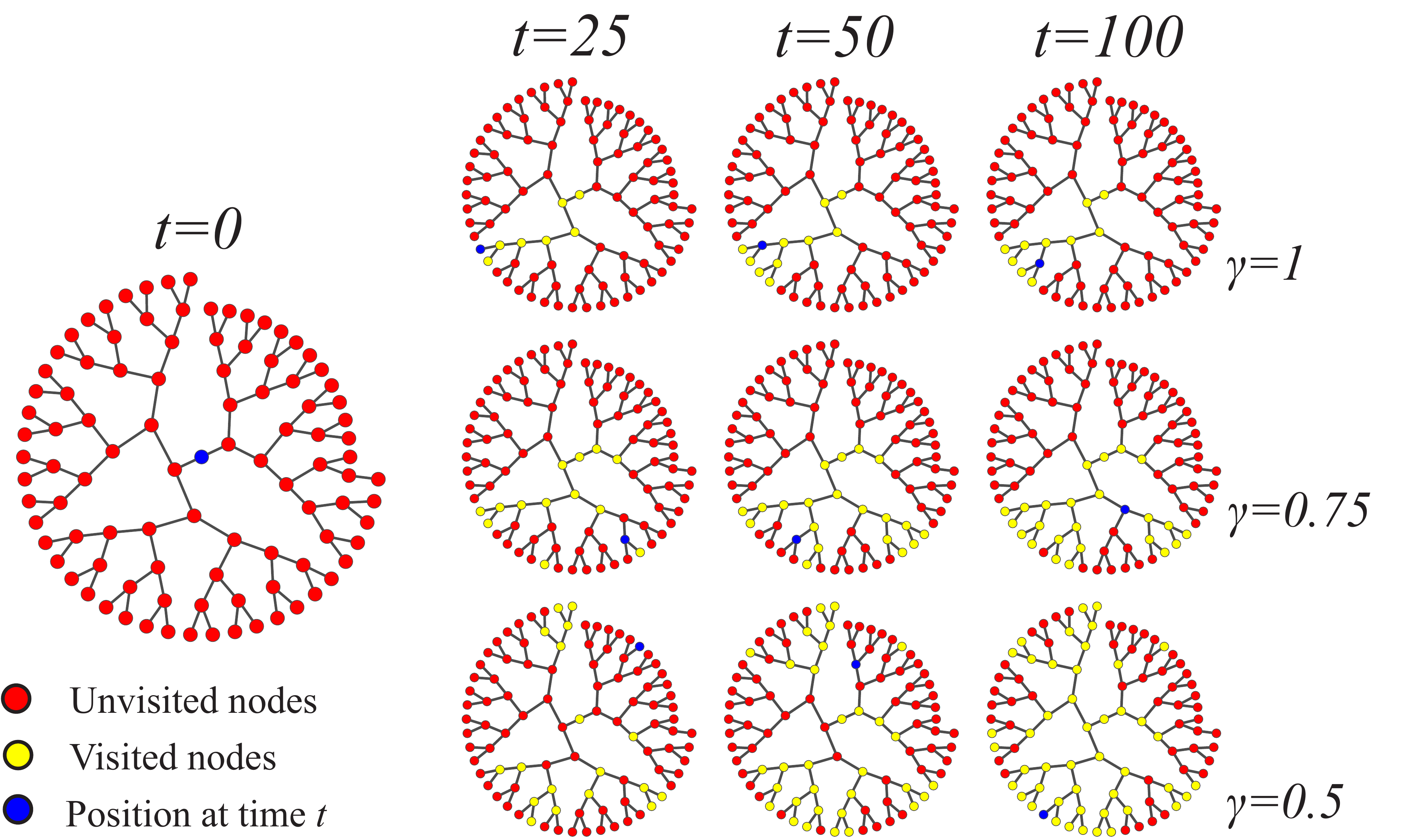}
\end{center}
\vspace{-4mm}
\caption{\label{FigureRW8}(Color online) Monte Carlo simulation of a discrete-time fractional random walker on a 
tree with transition probabilities given by Eq. (\ref{wijfrac}). The dynamics starts at $t = 0$ from an arbitrary node.  We show three discrete times $t=25,50,100$ for three values of the parameter $\gamma=1,0.75,0.5$. The case $\gamma=1$ corresponds to a normal random walker whereas the cases with $\gamma=0.75,0.5$ correspond to a fractional random walk leading to anomalous transport. We represent with different colors the unvisited nodes, visited nodes and the position of the random walker at time $t$. } 
\end{figure*}
The fractional transport on networks is defined in terms of a power of the Laplacian matrix $\mathbf{L}$ with elements given by $L_{ij}=\delta_{ij} k_i-A_{ij}$, where $\delta_{ij}$ denotes the Kronecker delta; in particular, $L_{ii}=k_i$ . The Laplacian matrix is introduced in graph theory and in the modeling of dynamical processes on networks  \cite{VespiBook,ArenasPhysRep2008,Estrada2015icn,Mohar1991lsg,Mohar1997sal,MulkenPR502,LapSpectra,Estrada2012,FoussBook2016}. In addition, the matrix $\mathbf{L}$ is interpreted as a discrete form of the Laplacian operator $(-\nabla^2)$ \cite{NewmanBook,Mohar1991lsg,Mohar1997sal}. In the context of the fractional diffusion on networks is introduced the fractional Laplacian  matrix $\mathbf{L}^{\gamma}$, where $\gamma$ is a real number ($0<\gamma<1$). The resulting process models the fractional dynamics on general networks \cite{RiascosMateosFD2014,RiascosMateosFD2015}.
\\[3mm]
Since the Laplacian matrix $\mathbf{L}$ is a symmetric matrix, by using the Gram-Schmidt orthonormalization of the eigenvectors of $\mathbf{L}$, we obtain a set of eigenvectors $\{\left|\Psi_j\right\rangle\}_{j=1}^N$  that satisfy the eigenvalue equation 
$\mathbf{L}\left|\Psi_j\right\rangle=\mu_j\left|\Psi_j\right\rangle$ for $j=1,\ldots,N$ 
and $\left\langle\Psi_i|\Psi_j\right\rangle=\delta_{ij}$, where $\mu_j$ are the eigenvalues, which are real and nonnegative. For connected networks, the smallest eigenvalue is $\mu_1=0$ and $\mu_m>0$ for $m=2,\ldots,N$ \cite{VanMieghem}. We define the matrix $\mathbf{Q}$  with elements $Q_{ij}=\left\langle i|\Psi_j\right\rangle$  and the diagonal matrix $\mathbf{\Lambda}=\textrm{diag}(0,\mu_2,\ldots,\mu_N)$. These matrices satisfy $\mathbf{L}\,\mathbf{Q}=\mathbf{Q}\,\mathbf{\Lambda}$, therefore $\mathbf{L}=\mathbf{Q}\mathbf{\Lambda}\mathbf{Q}^{\dag}$, where $\mathbf{Q}^{\dag}$ denotes the conjugate transpose of $\mathbf{Q}$. Therefore  \cite{bellman1960}
\begin{equation}\label{LfracDef}
	\mathbf{L}^{\gamma}=\mathbf{Q} \mathbf{\Lambda}^{\gamma} \mathbf{Q}^{\dag}
	=\sum_{m=2}^N \mu_m^{\gamma}\left|\Psi_m\right\rangle\left\langle \Psi_m\right| ,
\end{equation}
where $\mathbf{\Lambda}^{\gamma}=\textrm{diag}(0,\mu_2^{\gamma},\ldots,\mu_N^{\gamma})$. It is worth noticing that the diagonal elements of the fractional Laplacian matrix defined in Eq. (\ref{LfracDef}) introduce a generalization of the degree $k_i=(\mathbf{L})_{ii}$ to the fractional case. In this way, the fractional degree $k_i^{(\gamma)}$ of the node $i$ is given by \cite{RiascosMateosFD2014}
\begin{equation}\label{FracDegreeGeneral}
 k_i^{(\gamma)}\equiv(\mathbf{L}^{\gamma})_{ii}=\sum_{m=2}^N \mu_m^{\gamma}\langle i\left |\Psi_m\right\rangle\left\langle \Psi_m\right| i\rangle.
\end{equation}
The fractional random walk is the random walk associated to the fractional diffusion in networks \cite{RiascosMateosFD2014}. In the formalism of weighted networks is defined by the elements $\Omega_{ii}=0$ and, for $i\neq j$
\begin{equation}
\Omega_{ij}=-(\mathbf{L}^\gamma)_{ij}
\end{equation}
with $0<\gamma\leq 1$. On the other hand, the elements of the Laplacian matrix satisfy $k_i=-\sum_{l\neq i} L_{il}$ and, in the fractional case we have $k_i^{(\gamma)}=-\sum_{l\neq i} (\mathbf{L}^\gamma)_{il}$. As  result the strength of the node $i$ is given by
\begin{equation}\label{FracDegree}
S_i=\sum_{l=1}^N \Omega_{il}=
-\sum_{l\neq i}(\mathbf{L}^\gamma)_{il}=k_i^{(\gamma)},
\end{equation}
then, by using Eq. (\ref{wijomega}), the transition probability $\pi_{i\to j}$ is given by
\begin{equation}\label{wijfrac}
\pi_{i\to j}=\delta_{ij}-\frac{(\mathbf{L}^\gamma)_{ij}}{k_i^{(\gamma)}}\, .
\end{equation}
In the limit $\gamma\to 1$, the normal random walk strategy is recovered. In addition, by using the Eq. (\ref{Pinf}), the stationary distribution is
\begin{equation}\label{PinfFrac}
P_i^\infty=\frac{k_i^{(\gamma)}}{\sum_{l=1}^N k_l^{(\gamma)}}\, . 
\end{equation}
This is a generalization of the result $P_i^{\infty}\propto k_i$ for normal random walks discussed before and recovered from Eq. (\ref{PinfFrac}) when $\gamma=1$.
\\[2mm]
The fractional random walk is the process associated to the fractional diffusion on networks and the transition probabilities in Eq. (\ref{wijfrac}) define a navigation strategy with long-range displacements on the network \cite{RiascosMateosFD2014}. The case of infinite $n$-dimensional lattices with periodic boundary conditions has been addressed in different in contexts \cite{Michelitsch2016,Michelitsch2016Chaos,Michelitsch2017PhysA,Michelitsch2017PhysARecurrence,Michelitsch2018}.  For this type of periodic structures, it is obtained the analytical relation \cite{Michelitsch2017PhysA}
\begin{equation}\label{LevyLattices}
\pi_{i\to j}\sim d_{ij}^{-n-2\gamma}\qquad \text{for} \quad d_{ij}\gg 1.
\end{equation}
The result in Eq. (\ref{LevyLattices}) establishes a connection between L\'evy flights on networks \cite{RiascosMateos2012} and the fractional strategy defined by Eq. (\ref{wijfrac}). On the other hand, in networks with constant degree $k$, the fractional Laplacian can be expressed as \cite{RiascosMateosFD2015}
\begin{equation}\label{LfracAserie}
(\mathbf{L}^{\gamma})_{ij}
=\sum_{m=0}^\infty {\gamma \choose m}(-1)^m k^{\gamma-m}(\mathbf{A}^m)_{ij}
\end{equation}
where ${x \choose y}\equiv\frac{\Gamma(x+1)}{\Gamma(y+1)\Gamma(x-y+1)}$ and $\Gamma(x)$ denotes the Gamma function \cite{HandbookAbramowitz}.  The result in Eq.  (\ref{LfracAserie}) relates the fractional Laplacian matrix with the integer powers of the adjacency matrix $\mathbf{A}^m$ for $m=1, 2,\ldots$ for which the element $(\mathbf{A}^m)_{ij}$  is the number of all the possible trajectories connecting the nodes $i$, $j$ with $m$ links \cite{GodsilBook}.  In this way, the fractional strategy defined by the transition matrix with elements $\pi_{i\to j}$ in Eq. (\ref{wijfrac}) incorporates global information about all the possible trajectories connecting the nodes $i$ and $j$ \cite{RiascosMateosFD2015}. 
\\[2mm]
In order to illustrate the effect of the fractional dynamics of a random walker on a network,  
in Fig. \ref{FigureRW8} we present Monte Carlo simulations of discrete-time random walks on a 
tree. The discrete time $t$ denotes the number of steps of the random walker as it moves from one node to the next node
on the network. Given the topology of the network, we calculate the adjacency matrix 
and the corresponding Laplacian matrix $\mathbf{L}$ of the network. Then we obtain its eigenvalues and eigenvectors 
that allow us in turn to get the fractional Laplacian matrix $\mathbf{L}^{\gamma}$. Finally, using Eq. (\ref{wijfrac}), 
we determine the transition probabilities for different values of the parameter $\gamma$.
The dynamics starts at $t = 0$ from an arbitrary node. We show three discrete times $t=25,50,100$ for three values of the
parameter $\gamma=1, 0.75, 0.5$. Here, we depict one representative realization of a random walker as it navigates 
from one node to another randomly. The case $\gamma=1$ corresponds to normal random walk leading to normal diffusion. In 
this case, the random walker can move only locally to nearest neighbors and, as can be seen in the figure, the walker
revisits very frequently the same nodes and therefore the exploration of the network is redundant and not very 
efficient. The cases with $\gamma = 0.75, 0.5$ correspond to a fractional random walk leading to anomalous diffusion.
In this case, the random walker can navigate in a long-range fashion from one 
node to another arbitrarily distant node. This allows us to explore more efficiently the network since the walker does not
tend to revisit the same nodes; on the contrary, it tends to explore and navigates distant new regions each time. 
All this can be seen in the figure for different times, and allow us to make a comparison between a random walker 
using regular dynamics and a fractional dynamics \cite{RiascosMateosFD2015}. A detailed analysis of the fractional Laplacian of graphs and its relation with long-range navigation on networks and applications is presented in references \cite{RiascosMateosFD2014,RiascosMateosFD2015,Michelitsch2016,Michelitsch2016Chaos,Michelitsch2017PhysA,deNigris2016,DeNigris2017,Michelitsch2017PhysARecurrence}.
\\[2mm]
The introduction of the fractional random walks is motivated by the search of an equivalent on networks of the fractional diffusion and its relation with L\'evy flights. Recently, other types of functions of matrices with local information have shown interesting properties associated with long-range dynamics and the global structure of networks; this is the case of the concept communicability \cite{Estrada2012} and the accessibility random walk introduced in \cite{ArrudaPRE2014}. Particular functions of matrices can be used to define  different types of long-range strategies and characterized with the formalism reviewed in this work.
\\[2mm]
As a generalization of Eq. (\ref{wijfrac}), other functions of the Laplacian $g(\mathbf{L})$ can be applied to define random walk strategies on networks. The functions $g(x)$ to define random walk strategies should satisfy the following conditions \cite{RiascosMichelitsch2017_gL}
\begin{itemize}
\item  {\bf Condition I:} The matrix $g(\mathbf{L})$ must be positive semidefinite, i.e., the eigenvalues of $g(\mathbf{L})$ are restricted to be positive or zero. In this way, the property of the Laplacian eigenvalues $\mu_i\geq 0$ for $i=2,\ldots,N$ is preserved by the function $g(x)$.
\item  {\bf Condition II:} The elements of the matrix  $g(\mathbf{L})$ denoted as $g_{ij}(\mathbf{L})$, for $i,j=1,2,\ldots ,N$, should satisfy
\begin{equation}\label{gsumnull}
\sum_{j=1}^N g_{ij}(\mathbf{L})=0.
\end{equation}
Therefore, the function $g(x)$ maintains the property $\sum_{j=1}^N L_{ij}=0$ associated to the elements of the Laplacian matrix.
\item {\bf Condition III:} All the non-diagonal elements of  $g(\mathbf{L})$ must satisfy
\begin{equation}\label{EqconditionIII}
g_{ij}(\mathbf{L})\leq 0.
\end{equation}
\end{itemize}
For this type of functions, transition probabilities are defined by the relation
\begin{equation}\label{wij_g}
\pi_{i \to j}=\delta_{ij}-\frac{g_{ij}(\mathbf{L})}{\mathcal{K}_i},
\end{equation}
where we use the generalized degrees $\mathcal{K}_i$ defined by the diagonal elements of $g(\mathbf{L})$ that satisfy \cite{RiascosMichelitsch2017_gL}
\begin{equation}
\mathcal{K}_i=g_{ii}(\mathbf{L})=-\sum_{l\neq i}g_{il}(\mathbf{L}).
\end{equation}
Examples of functions that satisfy the conditions I-III are the fractional Laplacian of a graph $g(\mathbf{L})=\mathbf{L}^{\gamma}$ with $0<\gamma<1$, the logarithmic function  $g(\mathbf{L})=\log\left(\mathbb{I}+\alpha \mathbf{L}\right)$ for $\alpha>0$ and the function $g(\mathbf{L})=\mathbb{I}-e^{-a\mathbf{L}}$ with $a>0$. In all these cases is observed that the random walker hops with  long-range displacements on the network \cite{RiascosMichelitsch2017_gL}. 
\\[2mm]
In terms of the formalism of the matrix of weights, the generalized random walk strategy in Eq. 
(\ref{wij_g}) can be analyzed by using the weights $\Omega_{ij}=-g_{ij}(\mathbf{L})$ for $i\neq j$ and $\Omega_{ii}=0$. In this way, as a consequence of the condition in Eq. (\ref{EqconditionIII}), the weights satisfy $\Omega_{ij}\geq 0$;  also, the strength of each node is given by the generalized degree $\mathcal{K}_i$ allowing us to write the stationary probability distribution of the process as
\begin{equation}
P_i^\infty=\frac{\mathcal{K}_i}{\sum_{l=1}^N \mathcal{K}_l}.
\end{equation}
In the general case described in Eq. (\ref{wij_g}), the values of $g_{ij}(\mathbf{L})$ can be obtained by using the spectral methods described before for the fractional Laplacian (see \cite{RiascosMichelitsch2017_gL} for details).
\\[2mm]
\begin{table*}[t]
\centering
\footnotesize
\begin{tabular}{clccl}
\multicolumn{5}{c}{\bf LOCAL STRATEGIES} \\
\hline
\multicolumn{2}{c}{Strategy} & \multicolumn{1}{c}{Weights $\Omega_{ij}$, $i\neq j$.} & \multicolumn{1}{c}{Parameters} & \multicolumn{1}{c}{References} \\
\hline
1.  & Normal random walk            & $A_{ij}$ & 									 & \cite{Lovasz1996,Hughes,NohRieger}\\ 
2.  & Biased random walk             & $(q_i q_j)^\beta A_{ij}$ & $\beta \in \mathbb{R}$, $q_i>0$                              & \\
3.  & Degree biased random walk      & $(k_i k_j)^\beta A_{ij}$ & $\beta \in \mathbb{R}$ 				 & \cite{FronczakPRE2009}\\
4.  & Maximal entropy random walk    & $\xi_i \xi_j A_{ij}$ &  							 & \cite{BurdaPRL2009,SinatraPRE2011,YuanSciRep2014} \\
5.  & Random walks for image segmentation   & $e^{-\frac{1}{\sigma^2}(\mathcal{I}_i-\mathcal{I}_j)^2}A_{ij}$ & $\sigma>0$   		 & \cite{GradyIEEE2006}\\
6.  & Topologically biased random walk           & $e^{\beta y_{ij}} A_{ij}$ &  $\beta \in \mathbb{R}$, $y_{ij}=y_{ji}$ & \cite{ZlaticPRE2010,GradyIEEE2007} \\
\hline
\multicolumn{5}{c}{\bf NON-LOCAL STRATEGIES} \\
\hline
1.  & L\'evy Flights                 & $d_{ij}^{-\alpha}$        & $0\leq \alpha<\infty$         	      &\cite{RiascosMateos2012,YuanPRE2013,ZhaoPhysA2014,Estrada2017Multihopper}\\ 
2.  & Exponential strategy                  & $e^{-s d_{ij}}$            & $s>0$         	             &\cite{Estrada2017Multihopper}\\    
3.  & Gravity Law                    & $q_i q_j/l_{ij}^\alpha$   & $0\leq \alpha<\infty$       	 & \cite{Gonzalez2008,Barthelemy2011,YuGIS2014}      \\                           
4.  & Fractional Diffusion           & $-(\mathbf{L}^\gamma)_{ij}$ & $0 < \gamma<1$ 			         & \cite{RiascosMateosFD2014,RiascosMateosFD2015}\\
5.  & General functions of the Laplacian          & $-g_{ij}(\mathbf{L})$ &                               & \cite{RiascosMichelitsch2017_gL}\\
6. & Communicability                & $(e^{-\beta \mathbf{A}})_{ij}$ & $\beta>0$              			 & \cite{Estrada2009,Estrada2012}\\
\hline
\end{tabular}
\caption{\label{Table1} Diverse types of random walks described in terms of a symmetric matrix of weights $\mathbf{\Omega}$ with transition probabilities $\pi_{i\to j}$ defined by Eq. (\ref{wijomega}). For all the random walk strategies $\Omega_{ii}=0$ and, the non-diagonal elements are presented in the table with a short description of the quantities and parameters involved in the definition. The detailed description of each random walk strategy is presented in Section \ref{SectRWS}.}
\end{table*}
We conclude this section with a compilation of the types of random walk strategies represented by specific types of weighted networks. In Table \ref{Table1} we summarize the matrices of weights that define the local and non-local strategies analyzed in this section. Each model is presented with the respective parameters that define the random walker and key references to works analyzing these strategies.

\section{Mean first passage time and global characterization}
\label{MFPTtheory}
Once described a general formalism that allows us to define different types of local and non-local random walks strategies on networks and analytical results for their respective stationary distributions; in this section, we explore the mean first passage time (MFPT) \cite{SRedner}, that gives the average number of steps needed by the random walker to reach a specific node for the first time. We also study global times to quantify and compare the capacity of local and non-local random walks to explore different types of networks.  
\subsection{MFPT}
In order to calculate the MFPT for strategies defined in terms of weighted networks, we use a similar approach to the formalism presented in \cite{Hughes,NohRieger} where normal random walks are studied. We start representing the probability $P_{ij}(t)$ in the master equation in Eq. (\ref{master}) as
\begin{equation}\label{EquF}
P_{ij}(t) = \delta_{t0} \delta_{ij} + \sum_{t'=0}^t   P_{jj}(t-t')  F_{ij}(t') \ .
\end{equation}
The first term in Eq. (\ref {EquF}) represents the initial condition and $F_{ij}(t)$ is the probability to start in the node $i$ and reach the node $j$ for the first time after $t$ steps, by definition $F_{ij}(0)=0$. Now, by using the discrete Laplace transform $\tilde{f}(s) \equiv\sum_{t=0}^\infty e^{-st} f(t)$, the relation in Eq. (\ref{EquF}) takes the form 
\begin{equation}\label{LaplTransF}
\widetilde{F}_{ij} (s) = (\widetilde{P}_{ij}(s) - \delta_{ij}) / 
\widetilde{P}_{jj} (s) \ .
\end{equation}
By definition, using the quantity $F_{ij}(t)$, the MFPT $\langle T_{ij}\rangle$  for a random walker that starts in the node $i$ and reach for the first time the node $j$ is given by \cite{Hughes}
\begin{equation}
\langle T_{ij}\rangle \equiv \sum_{t=0}^{\infty} t F_{ij} (t) = -\widetilde{F}'_{ij}(0).
\end{equation}
Now, by means of the moments $R^{(n)}_{ij}$ of the probability $P_{ij}(t)$ defined as
\begin{equation}
R^{(n)}_{ij}\equiv \sum_{t=0}^{\infty} t^n ~ \{P_{ij}(t)-P_j^\infty\},
\end{equation}
the expansion in series of $\widetilde{P}_{ij}(s)$ is
\begin{equation}
\widetilde{P}_{ij}(s) =P_j^\infty\frac{1}{(1-e^{-s})}
+ \sum_{n=0}^\infty (-1)^n R^{(n)}_{ij} \frac{s^n}{n!} \ .
\end{equation}
Introducing this result in Eq. (\ref{LaplTransF}), the MFPT is obtained
\begin{equation}\label{Tij}
\langle T_{ij} \rangle =\frac{1}{P_j^\infty}\left[R^{(0)}_{jj}-R^{(0)}_{ij}+\delta_{ij}\right] .
\end{equation}
In Eq. (\ref{Tij}) there are three different terms: the mean first return time $\langle T_{ii} \rangle=1/P_i^\infty$, the quantity  
\begin{equation}
\tau_j\equiv R_{jj}^{(0)}/ P_j^\infty, 
\end{equation}
which is a time independent of the initial node and the time $R_{ij}^{(0)}/ P_j^\infty$ that depends on $i$ and $j$. Furthermore, from the detailed balance condition is obtained $\frac{R^{(n)}_{ij}}{P_j^\infty}=\frac{R^{(n)}_{ji}}{P_i^\infty}$,  as consequence
\begin{equation}
	\langle T_{ij} \rangle-\langle T_{ji}\rangle=\tau_j-\tau_i ,
\end{equation}
relation that describes the asymmetry of navigation \cite{NohRieger}. The time $\tau_i$ is interpreted as the average time needed to reach the node $i$ from a randomly chosen initial node of the network; on the other hand, the quantity $C_i\equiv\tau_i^{-1}$ is the random walk centrality introduced for the analysis of random walks with local information \cite{NohRieger}. The centrality $C_i$  combines information of the network and the random walk strategy implemented to visit nodes and gives a high value to nodes easy to reach and small values to nodes for which the random walker takes, in average, many steps to hit the node for the first time starting from any node of the network \cite{NohRieger,RiascosMateos2012}.
\\[2mm]
Additional to the times $\langle T_{ij} \rangle$ and $\tau_i$, from Eq. (\ref{Tij}) we have
\begin{equation}\label{TijPinf_Kemeny}
	\sum_{j=1}^N \langle T_{ij} \rangle P_j^{\infty}=\sum_{j=1}^N R_{jj}^{(0)}-\sum_{j=1}^N R_{ij}^{(0)}+1=\sum_{j=1}^N R_{jj}^{(0)}+1\, .
\end{equation}
The quantity $K\equiv \sum_{m=1}^N R_{mm}^{(0)}$ in the context of stochastic processes is denominated Kemeny's constant \cite{Kemeny,MFPTanalytic}. As result of the relation in Eq. (\ref{TijPinf_Kemeny})
\begin{equation}\label{Kconst}
	K= \sum_{m=1}^N R_{mm}^{(0)}=\sum_{j\neq i}\langle T_{ij}\rangle P_j^\infty \, ,
\end{equation}
equation that establishes a connection between the Kemeny's constant of Markovian processes and the global time obtained by averaging the mean first passage times $\langle T_{ij} \rangle$ weighted with the stationary distribution $P_j^{\infty}$.
\subsection{Linear algebra approach}
\label{LAsect}
Once defined general quantities that characterize the performance of a random walk strategy to explore a network, it is important to have an algorithm that allows us to calculate these values by using the information consigned in the transition probability matrix  $\mathbf{\Pi}$ in Eq. (\ref{wijomega}), defined in terms of the matrix of weights $\mathbf{\Omega}$ and that essentially contains all the information about the random walker. Therefore, in the following part we deduce expressions for the MFPT $\langle T_{ij} \rangle$, the time $\tau_i$ and the Kemeny's constant $K$ in terms of the eigenvalues and eigenvectors of the transition matrix $\mathbf{\Pi}$. 
\\[2mm]
In order to calculate $\tau_i$ and $\langle T_{ij} \rangle$ is necessary to find $P_{ij}(t)$. We start with the matrical form of Eq. (\ref{master})
\begin{equation}	
\vec{P}(t)=\vec{P}(0)\mathbf{\Pi}^t  \, .
\end{equation}
Here $\vec{P}(t)$ is the probability vector at time $t$. Using Dirac's notation
\begin{equation}\label{ProbVector}
P_{ij}(t)=\left\langle i\right|\mathbf{\Pi}^t \left|j\right\rangle,
\end{equation}
where $\{\left|m\right\rangle \}_{m=1}^N$ represents the canonical base of $\mathbb{R}^N$. 
\\[1mm]
Due to the existence of a detailed balance condition, the matrix $\mathbf{\Pi}$ can be diagonalized and its spectrum has real values \cite{VKampen}. For right eigenvectors of $\mathbf{\Pi}$ we have $\mathbf{\Pi}\left|\phi_i\right\rangle=\lambda_i\left|\phi_i\right\rangle $ for $i=1,..,N$, where the set of eigenvalues is ordered in the form $\lambda_1=1$ and $1>\lambda_2\geq..\geq\lambda_N\geq -1 $. On the other hand, from right eigenvectors we define a matrix $\mathbf{Z}$ with elements $Z_{ij}=\left\langle i|\phi_j\right\rangle$. The matrix $\mathbf{Z}$ is invertible, and a new set of vectors $\left\langle \bar{\phi}_i\right|$ is obtained by means of $(\mathbf{Z}^{-1})_{ij}=\left\langle \bar{\phi}_i |j\right\rangle $, then
\begin{equation}\label{cond1}
\delta_{ij}=(\mathbf{Z}^{-1}\mathbf{Z})_{ij}=\sum_{l=1}^N \left\langle\bar{\phi}_i|l\right\rangle \left\langle l|\phi_j\right\rangle=\langle\bar{\phi}_i|\phi_j\rangle \, 
\end{equation}
and 
\begin{equation}\label{cond2}
\mathbb{I}=\mathbf{Z}\mathbf{Z}^{-1}=\sum_{l=1}^N \left|\phi_l\right\rangle \left\langle \bar{\phi}_l \right| \, ,
\end{equation}
where $\mathbb{I}$ is the $N\times N$ identity matrix. 
\\[2mm]
In different cases, especially when it is necessary to calculate numerically the eigenvalues and eigenvectors of the transition matrix, it is convenient to use the symmetry of the matrix of weights $\mathbf{\Omega}$. In this way, the eigenvectors $\left|\phi_l\right\rangle$ and $\left\langle \bar{\phi}_l \right|$ can alternatively be deduced from the analysis the  symmetric matrix $\mathbf{M}$ with elements
\begin{equation}
M_{ij}=\Omega_{ij}/\sqrt{S_i S_j}.
\end{equation}
From an orthonormal set of eigenvectors $\left|\varphi_l\right\rangle$ that satisfy $\mathbf{M} \left|\varphi_l\right\rangle=\lambda_l \left|\varphi_l\right\rangle$ for $l=1,\ldots , N$, it is obtained $	\left|\phi_l\right\rangle=\mathbf{S}^{-1/2} \left|\varphi_l\right\rangle$ and $\left\langle \bar{\phi}_l \right |=\left\langle \varphi_l \right |\mathbf{S}^{1/2}$ where $\mathbf{S}$ is the $N\times N$ diagonal matrix  $\mathbf{S}=\textrm{diag}(S_1,\ldots,S_N)$.
\\[2mm]
Once obtained the spectrum and the left and right eigenvectors of the transition matrix, we can deduce different analytical expressions for quantities that characterize the random walker. By using the diagonal matrix $\mathbf{\Delta} \equiv \textrm{diag}(\lambda_1,\ldots,\lambda_N)$ is obtained $\mathbf{\Pi}=\mathbf{Z}\mathbf{\Delta}\mathbf{Z}^{-1}$, therefore Eq. (\ref{ProbVector}) takes the form
\begin{equation}\label{PtSpect}
	P_{ij}(t)=\left\langle i\right|\mathbf{Z}\mathbf{\Delta}^t\mathbf{Z}^{-1}\left|j\right\rangle
	= \sum_{l=1}^N\lambda_{l}^t\left\langle i|\phi_l\right\rangle \left\langle \bar{\phi}_l|j\right\rangle  \, .
\end{equation}
From Eq. (\ref{PtSpect}), the stationary probability distribution $P_j^{\infty}=\left\langle i|\phi_1\right\rangle \left\langle \bar{\phi}_1|j\right\rangle$, where the result $\left\langle i|\phi_1\right\rangle=\rm{constant}$ makes $P_j^{\infty}$ independent of the initial condition. Now, by means of the definition of $R_{ij}^{(0)}$, we have
\begin{equation}\label{RijSpect}
R_{ij}^{(0)}=\sum_{l=2}^N\frac{1}{1-\lambda_l}\left\langle i|\phi_l\right\rangle \left\langle\bar{\phi}_l|j\right\rangle \, .
\end{equation}
Therefore, the time $\tau_i$ is given by
\begin{equation}\label{TauiSpect}
	\tau_i=\sum_{l=2}^N\frac{1}{1-\lambda_l}\frac{\left\langle i|\phi_l\right\rangle \left\langle\bar{\phi}_l|i\right\rangle}{\left\langle i|\phi_1\right\rangle \left\langle\bar{\phi}_1|i\right\rangle}\, ,
\end{equation}
and, for $i \neq j$ in  Eq. (\ref{Tij}), the MFPT $\left\langle T_{ij}\right\rangle$ is
\begin{equation}\label{TijSpect}
\left\langle T_{ij}\right\rangle
=\sum_{l=2}^N\frac{1}{1-\lambda_l}\frac{\left\langle j|\phi_l\right\rangle \left\langle\bar{\phi}_l|j\right\rangle-\left\langle i|\phi_l\right\rangle \left\langle\bar{\phi}_l|j\right\rangle}{\left\langle j|\phi_1\right\rangle \left\langle\bar{\phi}_1|j\right\rangle}\, ,
\end{equation}
whereas $\left\langle T_{ii}\right\rangle=(\left\langle i|\phi_1\right\rangle \left\langle\bar{\phi}_1|i\right\rangle)^{-1}$. Finally, from Eqs. (\ref{Kconst}) and (\ref{RijSpect}) is obtained the Kemeny's constant
\begin{equation}\label{KconstSpect}
K=\sum_{m=1}^N\sum_{l=2}^N \frac{1}{1-\lambda_l} \left\langle\bar{\phi}_l|m\right\rangle \left\langle m|\phi_l\right\rangle =\sum_{l=2}^N \frac{1}{1-\lambda_l}
\end{equation}
result that only depends on the eigenvalues of the transition matrix $\mathbf{\Pi}$. 
\subsection{Global characterization}
\begin{figure*}[!t]
\begin{center}
\includegraphics*[width=0.8\textwidth]{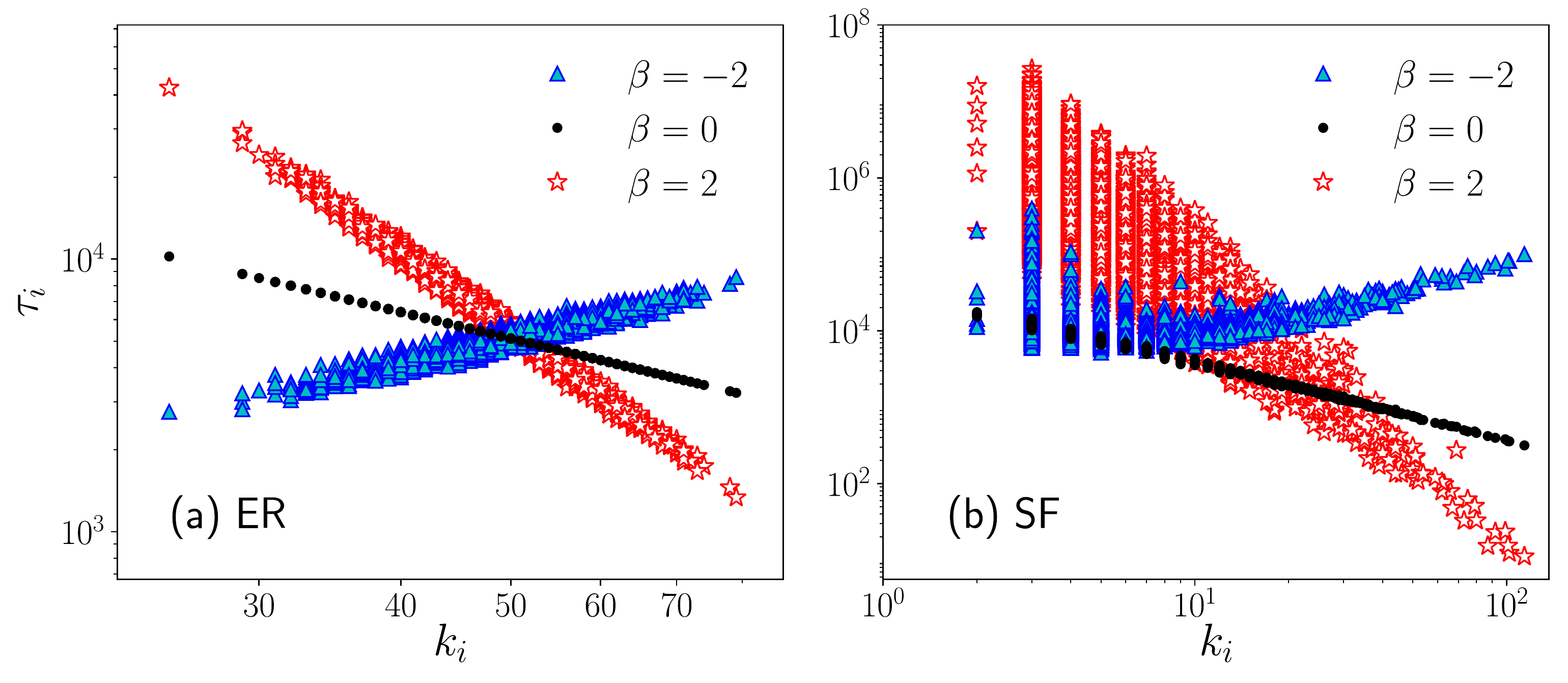}
\end{center}
\vspace{-5mm}
\caption{\label{FigureRW9} (Color online) Time $\tau_i$ vs. $k_i$ for degree biased random walks on networks obtained using Eq. (\ref{TauiSpect}). (a) Erd\"{o}s-R\'enyi (ER) network with $N=5000$ and $\left\langle k\right\rangle=50$. (b) Scale-free (SF) network with $N=5000$ $\langle k \rangle=6$. We use three values of the parameter $\beta$.}
\end{figure*} 
In this part we define global times that quantify the capacity of a random walk to reach any site of the network; by using these global times is possible to compare the efficiency of the different strategies defined through Eq. (\ref{wijomega}). Global quantities like entropy rates \cite{LatoraPRE2008,BurdaPRL2009}, the global mean first passage time \cite{TejedorPRE2009}, and the cover time \cite{VespiBook,Hughes} have been used to study random walks on networks. We use the global quantity \cite{RiascosMateos2012}
\begin{equation}\label{tauglobal}
\tau\equiv\frac{1}{N}\sum_{i=1}^{N}\tau_i \,  ,
\end{equation}
that gives an estimate of the  average time to reach any site of the network. The values $\tau_i$  can present a huge dispersion due to the fact that in some irregular networks there are nodes easily accessible to the random walker and other sites that are hardly reached; despite this fact, the mean value of the times $\tau_i$ is an important quantity that characterize the capacity of a random walker to visit the nodes of a network. In the following section we explore the time $\tau$ for different random walk strategies. 
\\[2mm]
On the other hand, in the particular case of random walks on weighted networks for which the value $S_i=\sum_{l=1}^N\Omega_{il}$ is constant, the stationary distribution given by Eq. (\ref{Pinf}) is $P_{i}^{\infty}=1/N$. In this type of regular cases, using Eqs. (\ref{TauiSpect}) and (\ref{tauglobal}) we have for the global time $\tau$
\begin{equation}\label{EffectRegular}
\tau_{\rm{reg}}=\frac{1}{N}\sum_{i=1}^{N}\frac{R_{ii}^{(0)}}{ P_i^{\infty}}=\sum_{i=1}^{N}R_{ii}^{(0)}=\sum_{l=2}^N \frac{1}{1-\lambda_l} \, .
\end{equation}
Then, in regular cases $\tau_{\rm{reg}}$ is equal to the Kemeny's constant. Examples of this simplification are the normal random walks on a complete graph. This case illustrates the best scenario for the exploration of a network by means of normal random walks since all the nodes are connected. For a complete graph $A_{ij}=1-\delta_{ij}$ and $\pi_{i\to j}=\frac{1-\delta_{ij}}{N-1}$ \cite{VanMieghem}. The eigenvalues of the matrix $\mathbf{\Pi}$ are $\lambda_1=1$ and $\lambda_2=\ldots=\lambda_N=-(N-1)^{-1}$, then the Kemeny's constant given by Eq. (\ref{EffectRegular}) for unbiased random walks on a complete network is
\begin{equation}
\tau_{0}=\frac{(N-1)^2}{N} \, ,
\end{equation}
this is the lowest value that the time $\tau$ can take.
\section{Efficiencies of particular strategies}
In this section, we apply the results in Eqs. (\ref{TauiSpect})-(\ref{KconstSpect}) that allow to calculate exact values of $\langle T_{ij}\rangle$, $\tau_j$ and $K$  for random walks on weighted networks. In particular, we analyze the global time $\tau$ and the Kemeny’s constant for the preferential navigation, the L\'evy flights on networks, the fractional transport and the model in Eq. (\ref{wijRalpha}). All these are defined in terms of a matrix of weights through the approach presented in Section 
\ref{SectRWS}. Similar methods can be implemented to study different types of random walks described in Table \ref{Table1}.

\subsection{Preferential navigation}
The preferential navigation defined in Eq. (\ref{wijNBRW}) can represent different types  of local random walkers; in particular, degree biased random walkers with transitions described in Section \ref{SecDBRW}. In order to characterize this process, in Fig. \ref{FigureRW9} we depict the values of the time $\tau_i$ for two different networks, the respective stationary distribution was presented in Fig. \ref{FigureRW2}.  The obtained  values of $\tau_i$ give the average number of steps needed by a degree biased random walker to reach the node $i$ from a random site in the network for different values of the parameter $\beta$. In the ER network is observed the validity of the result $\tau_i\approx 1/P_i^{\infty}$ obtained by a mean field approximation \cite{FronczakPRE2009}. On the other hand, in the SF network is not valid this approximation and it is observed that, compared with the case $\beta=0$ that recovers the normal random walk, any degree biased random walk is a bad strategy to reach efficiently nodes with a lower degree in the SF network. Our findings also reveal that, in comparison with the result for $\beta=0$, in the ER network the value $\beta=-2$ reduces the number of steps needed to reach nodes with few connections, whereas the parameter $\beta=2$ reduces the value of $\tau_i$ in nodes with large degree $k_i$.
\\[2mm]
\begin{figure*}[!t]
\begin{center}
\includegraphics*[width=0.90\textwidth]{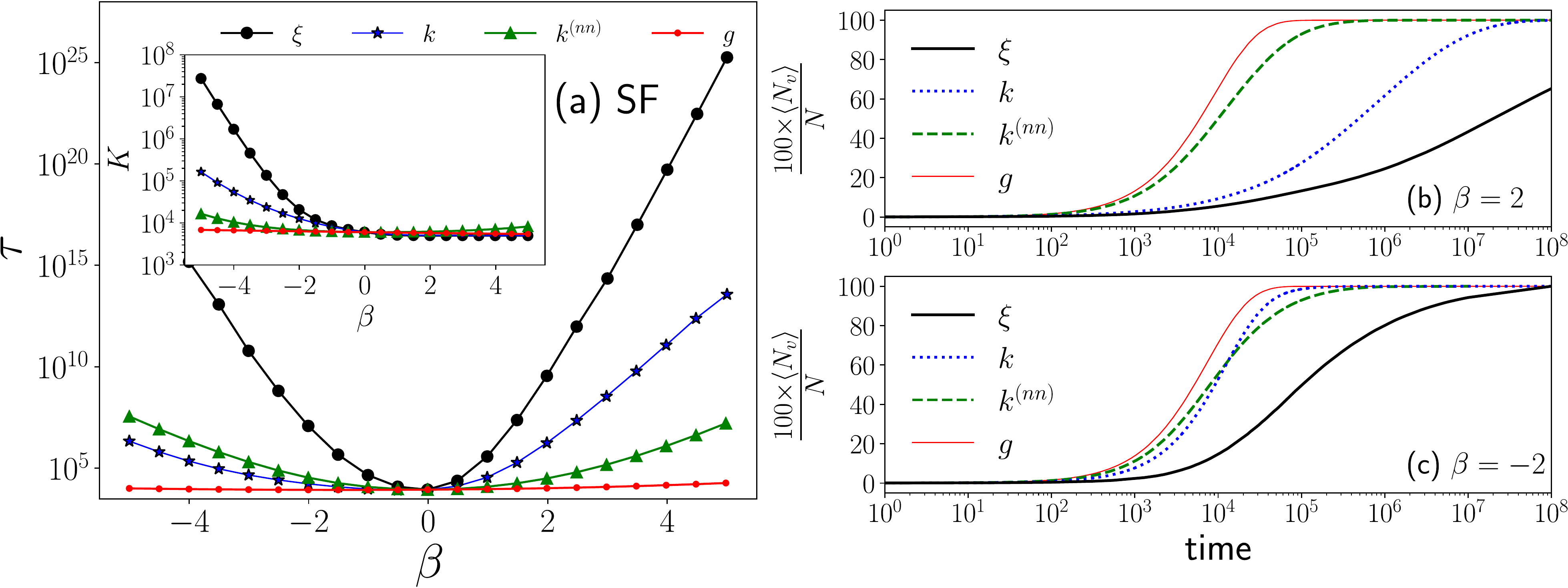}
\end{center}
\vspace{-3mm}
\caption{\label{FigureRW10} (Color online) Preferential random walks on a scale-free network (SF) of the Barab\'asi-Albert type with $N=5000$ nodes. (a) Global time $\tau$ vs. $\beta$ for different types of random walks defined by  Eq. (\ref{wijNBRW}) with $q_i$ given by the node degree $k_i$, the average degree of neighbors $k_i^{(nn)}$, the closeness centrality $g_i$ and the eigenvector centrality $\xi_i$; in the inset we  plot the values of the Kemeny's constant $K$ vs. $\beta$. The continuous lines are used as a guide. In (b) and (c) we present Monte Carlo simulations of the number of  visited sites $N_v$ as a function of time for node biased random walks with $\beta=-2$ and $\beta=2$. The average of the number of visited nodes $\langle N_v\rangle$ is obtained from $1000$ different realizations of the random walker, the results are expressed as a fraction of the total number of nodes multiplied by 100.
}
\end{figure*} 	
Now, we calculate the global time $\tau$ for different cases of the preferential navigation defined by Eq. (\ref{wijNBRW}). We start generating the network and obtaining the transition matrix $\mathbf{\Pi}$ for specific values of the parameter $\beta$ using the definition in Eq. (\ref{wijNBRW}). Once obtained $\mathbf{\Pi}$ we calculate the respective left and right eigenvectors $ \left\langle \bar{\phi}_l \right |$ and $ \left|\phi_l\right\rangle$  by the method described in Section \ref{LAsect}. Then, we use the Eq. (\ref{TauiSpect}) to calculate the values of $\tau_i$ and finally, the mean value of  the times $\tau_i$ gives $\tau$. In a similar way, the Kemeny's constant is obtained from the spectrum of $\mathbf{\Pi}$ by means of Eq. (\ref{KconstSpect}).  This process is repeated for different types of local strategies. We study cases for which the value $q_i$ in Eq. (\ref{wijNBRW}) is determined by common quantities used to describe the role of the node $i$ in each network, we explore the effect of the following $q_i$ choices given by:
\begin{itemize}
\item The node degree $k_i$, therefore, the dynamics is the degree biased random walk explored in Figs. \ref{FigureRW2} and \ref{FigureRW9}.
\item The average degree of the neighbors of $i$ given by $k_i^{(nn)}=(\sum_{l=1}^N A_{il}k_l)/k_i$. In this case, the transition probabilities $\pi_{i \to j}$ depend on the average degree of the first neighbors of the node $j$.
\item The closeness centrality $g_i$ of the node $i$ given by $g_i=(\sum_{j\neq i} d_{ij})^{-1}$ where $d_{ij}$ is the distance between $i$ and $j$ \cite{NewmanBook}. 
\item The eigenvector centrality $\xi_i$ of the node $i$. For the particular case $\beta=1$, the probabilities $\pi_{i\to j}$ 
are determined by Eq. (\ref{wijmaxentropy}) that defines a maximal entropy random walk.
\end{itemize}

From the quantities $k_i$, $k_i^{(nn)}$, $g_i$, $\xi_i$, we analyze the global time $\tau$ for the preferential strategy with different values of the parameter $\beta$ that modules the biased random walk. From the numerical value of $\tau$ we can compare the efficiency of the strategies to visit the nodes on the network.  In Fig. \ref{FigureRW10}  we analyze each of these strategies in a scale-free network. In Fig. \ref{FigureRW10}(a) we depict the time $\tau$ as a function of $\beta$. In this case, we observe that unbiased random walks  ($\beta=0$) have the lowest values of $\tau$. In addition, for the random walk with $q_i=g_i$ the values of $\tau$ do not change significantly with variations of the parameter $\beta$. In addition, it is observed that, for a given value of $\beta$, the strategy defined by $q_i=\xi_i$ has the largest values of $\tau$ making this method to visit the nodes of the network inefficient to reach easily any node of the structure. In addition, the results reveal that, in comparison to the unbiased case, some node biased random walks need much more time to explore the network.  
\\[2mm]
Furthermore, we are interested in the values of $K$ and $\tau$ as a measure of the efficiency of the random walker to explore sites on the network. In this way, we use Monte Carlo simulations of each  preferential random walk with the values $\beta=-2$ and $\beta=2$. We depict the obtained results in Figs. \ref{FigureRW10}(b)-(c) for the average number or visited nodes $\langle N_v \rangle$ as a function of time. Based on our findings for $\tau$, we know that in some cases, for example when $q_i=k_i$ or $q_i=\xi_i$, the average time $\tau$ is much bigger for $\beta=2$ than the result obtained for $\beta=-2$. These results and the behavior observed in the Monte Carlo simulations are in agreement  with the predictions that $\tau$ gives for the different orders of magnitude of the time needed to visit any node on the network. In contrast, the Kemeny's constant (presented in the inset in Fig. \ref{FigureRW10}(a)) does not describe the results obtained with Monte Carlo simulations.  In this way,  we can infer that for the preferential navigation, the eigenvectors of the transition matrix $\mathbf{\Pi}$ contain relevant information about the efficiency of the processes, by considering only the spectrum of $\mathbf{\Pi}$ this information is lost. This does not apply in regular cases for which $P_i^\infty=1/N$, where $\tau_{\rm{reg}}=K$, and as a consequence important information to describe the efficiency of the process is contained in the spectrum of $\mathbf{\Pi}$. 
\\[2mm]
All the analysis discussed for the preferential random walk in Fig. \ref{FigureRW10} reveals that the time $\tau$ is a measure that describes appropriately the global efficiency of random walk strategies. On the other hand, the Kemeny's constant is a useful simplification to analyze only regular cases when the strength $S_i=\sum_{l=1}^N  \Omega_{il}$ is constant.
\subsection{L\'evy flights}
\begin{figure*}[!t]
\begin{center}
\includegraphics*[width=0.8\textwidth]{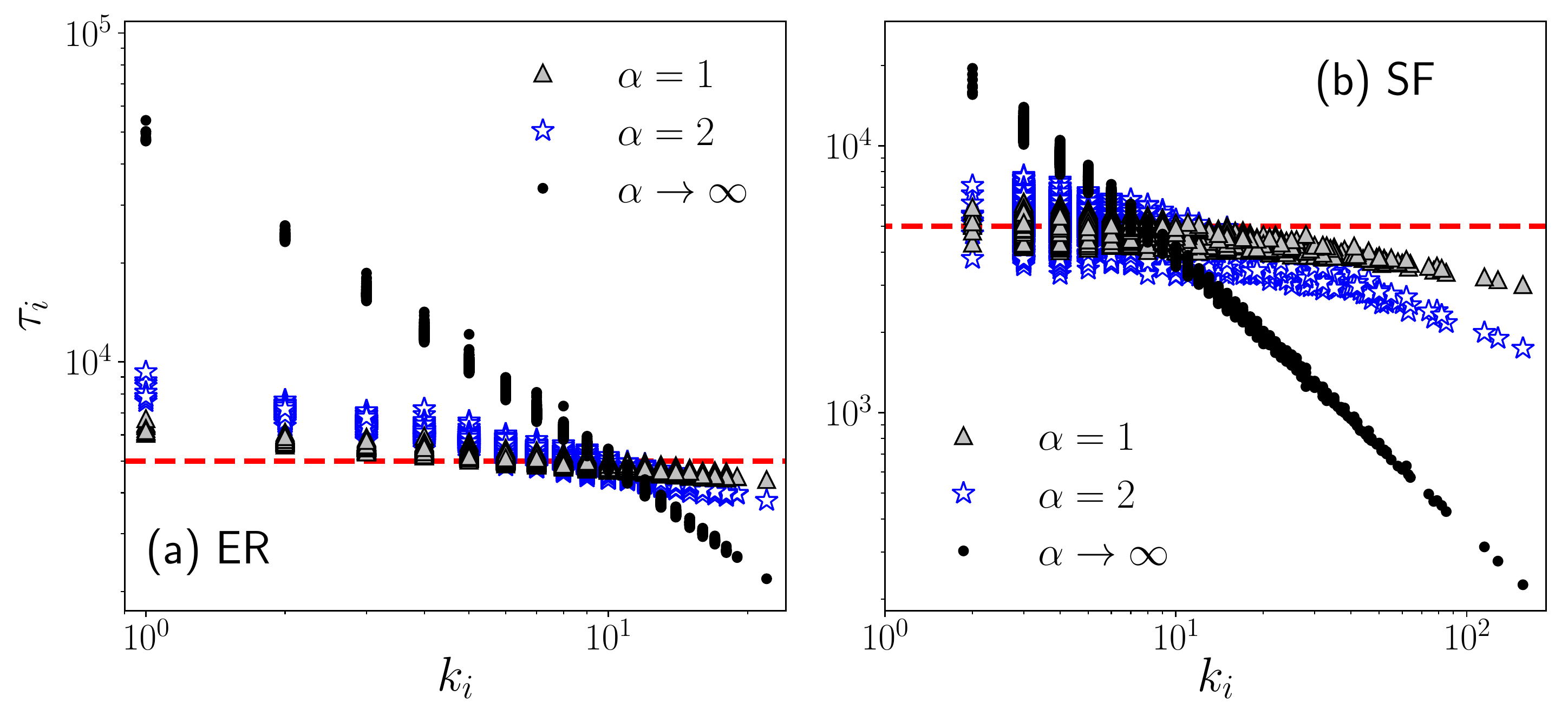}
\end{center}
\vspace{-5mm}
\caption{\label{FigureRW11} (Color online) Time $\tau_i$ vs. $k_i$ for L\'evy flights on networks; the quantities are obtained evaluating numerically Eq. (\ref{TauiSpect}). We use three values of the parameter $\alpha$ and we study two types of networks with $N=5000$.  (a)  Erd\"{o}s-R\'enyi network (ER) at the percolation limit $p=(\log N)/N$. (b) Scale-free (SF) network  with $\langle k \rangle=6$. Dashed lines denote the value $\tau_0$ obtained for $\alpha=0$.}
\end{figure*}
In this part, we analyze the efficiency of L\'evy flights to explore a network. We use the formalism introduced in Section \ref{LAsect} valid for all the strategies represented in terms of the matrix of weights $\mathbf{\Omega}$. In this way we have exact analytical values that allow us to evaluate the mean first passage time $\langle T_{ij}\rangle$, the time $\tau_i$ and the global quantity $\tau$ by using Eqs. (\ref{TauiSpect})-(\ref{tauglobal}). In Fig. \ref{FigureRW11} we present the results obtained for  $\tau_i$ as a function of $k_i$ for an Erd\H{o}s--R\'enyi network and a scale-free network. In a similar way to the results observed for the preferential random walk in Fig. \ref{FigureRW9}; we obtain that for L\'evy flights on these small-world networks $R^{(0)}_{ii}\approx 1$ and, as a consequence  $\tau_i\approx 1/P_i^\infty$.
\\[2mm]
On the other hand, the efficiency of L\'evy flights can be explored with the quantity $\tau$. In networks for which the long-range degree $D_i^{(\alpha)}$ is a constant for all the nodes on the network, the stationary distribution is $P_i^\infty=\frac{1}{N}$, i.e. each node has the same probability to be visited in the limit $t\to \infty$. This is the case of some regular networks like rings, square lattices with periodic boundary conditions, complete graphs, among others. For this type of networks $\tau$ is the Kemeny's constant given by Eq. (\ref{KconstSpect}), result that only depends on the spectrum of the transition matrix $\mathbf{\Pi}$ with elements given by Eq. (\ref{wijomega}). In the following part we calculate analytically the value of $\tau$ for L\'evy flights on rings and we explore numerically the efficiency in other structures.
\\[2mm]
First, we study L\'evy flights on a ring. In this particular case, the matrix of weights $\mathbf{\Omega}$ and the transition matrix $\mathbf{\Pi}$  are circulant matrices \cite{VanMieghem}. The eigenvalues and eigenvectors of circulant matrices are well known \cite{VanMieghem,CirculantReview2006} and, in this way, we can obtain exact analytical expressions for the different quantities presented in Section \ref{MFPTtheory}. For example, for a ring   with an even number of nodes $N$, the spectrum of the transition matrix $\mathbf{\Pi}$ is
\begin{equation}\label{LFSpectCycleEven}
	\lambda_l=\frac{2\sum_{n=1}^{N/2-1}n^{-\alpha} \cos\left[n \theta_l\right] +\left(\frac{N}{2}\right)^{-\alpha}\cos\left[\frac{N}{2}\theta_l\right]}{2\sum_{n=1}^{N/2-1}n^{-\alpha}+\left(\frac{N}{2}\right)^{-\alpha}},  
\end{equation}
with $\theta_l=2\pi(l-1)/N$. In a similar way, for an odd value of $N$, we have
\begin{equation}\label{LFSpectCycleOdd}
	\lambda_l=\frac{\sum_{n=1}^{(N-1)/2}n^{-\alpha} \cos\left[n\theta_l\right]}{\sum_{n=1}^{(N-1)/2}n^{-\alpha}}.  
\end{equation}
Introducing the eigenvalues in Eqs. (\ref{LFSpectCycleEven}) and (\ref{LFSpectCycleOdd}) in the relation in Eq. (\ref{EffectRegular}), we obtain analytically the value of $\tau$ for L\'evy flights on a ring. In particular, when $\alpha\to\infty$ we deduce $\tau$ for normal random walks on a ring
\begin{equation}\label{TauRing_NRW}
	\tau=\sum_{l=2}^{N}\frac{1}{1-\cos\left[\frac{2\pi}{N}(l-1)\right]}.
\end{equation}
\begin{figure}[!t]
\begin{center}
\includegraphics*[width=0.47\textwidth]{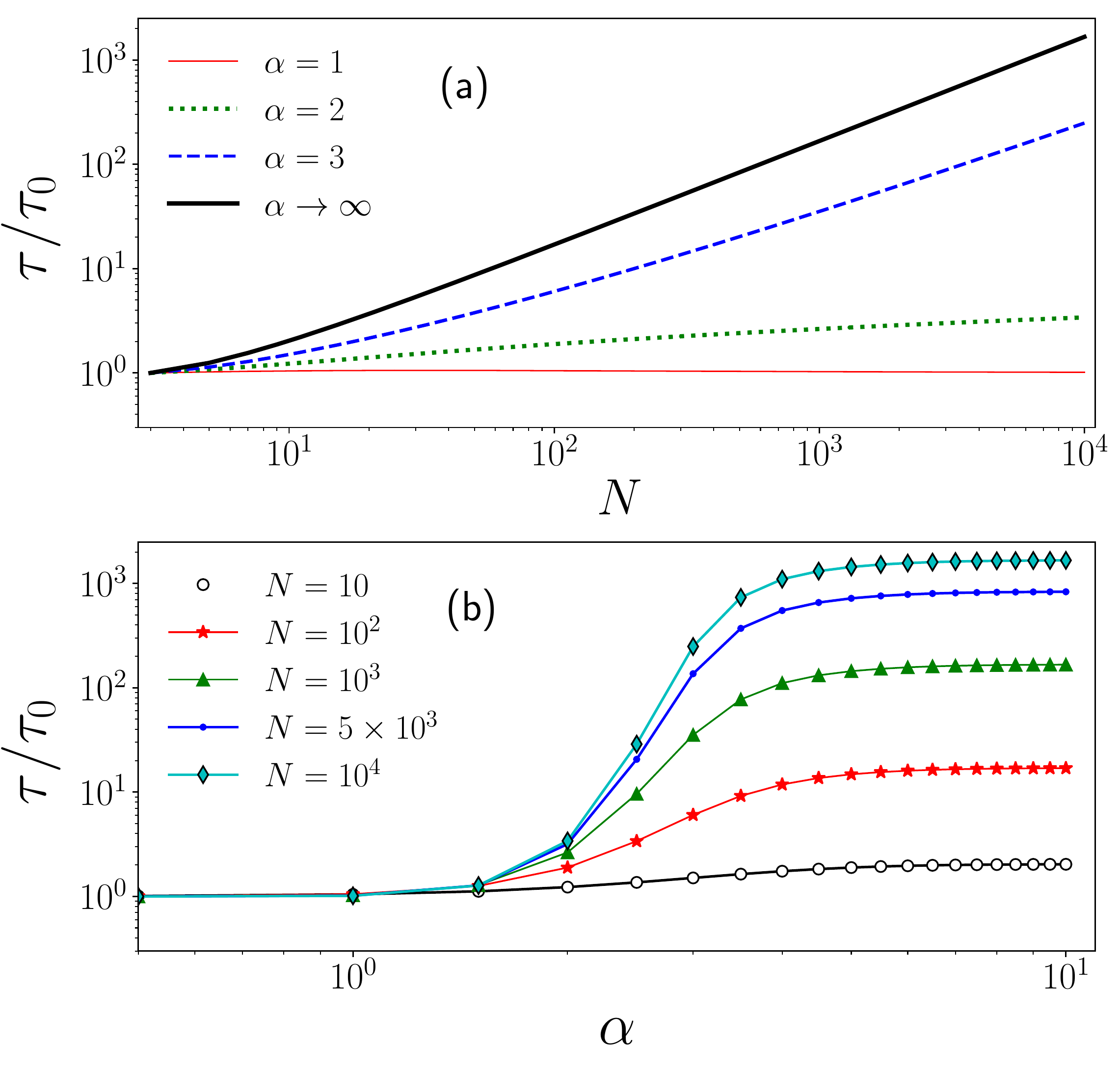}
\end{center}
\vspace{-8mm}
\caption{\label{FigureRW12} (Color online) Global time $\tau$ for L\'evy flights to reach any site of a finite ring. The results are expressed in terms of the value $\tau_0=(N-1)^2/N$ obtained for a complete graph. The values are calculated using the analytical expressions for the spectra in Eqs. (\ref{LFSpectCycleEven})-(\ref{LFSpectCycleOdd}) and the Eq. (\ref{EffectRegular}). In (a) the time $\tau$ is presented as a function of $N$ for  $\alpha=1, 2,3$ and the limit case $\alpha\to \infty$; these results are complemented in (b) where the time $\tau$ is plotted as a function of $\alpha$ for L\'evy flights with different values of $N$.}
\end{figure}
\begin{figure}[!t]
\begin{center}
\includegraphics*[width=0.47\textwidth]{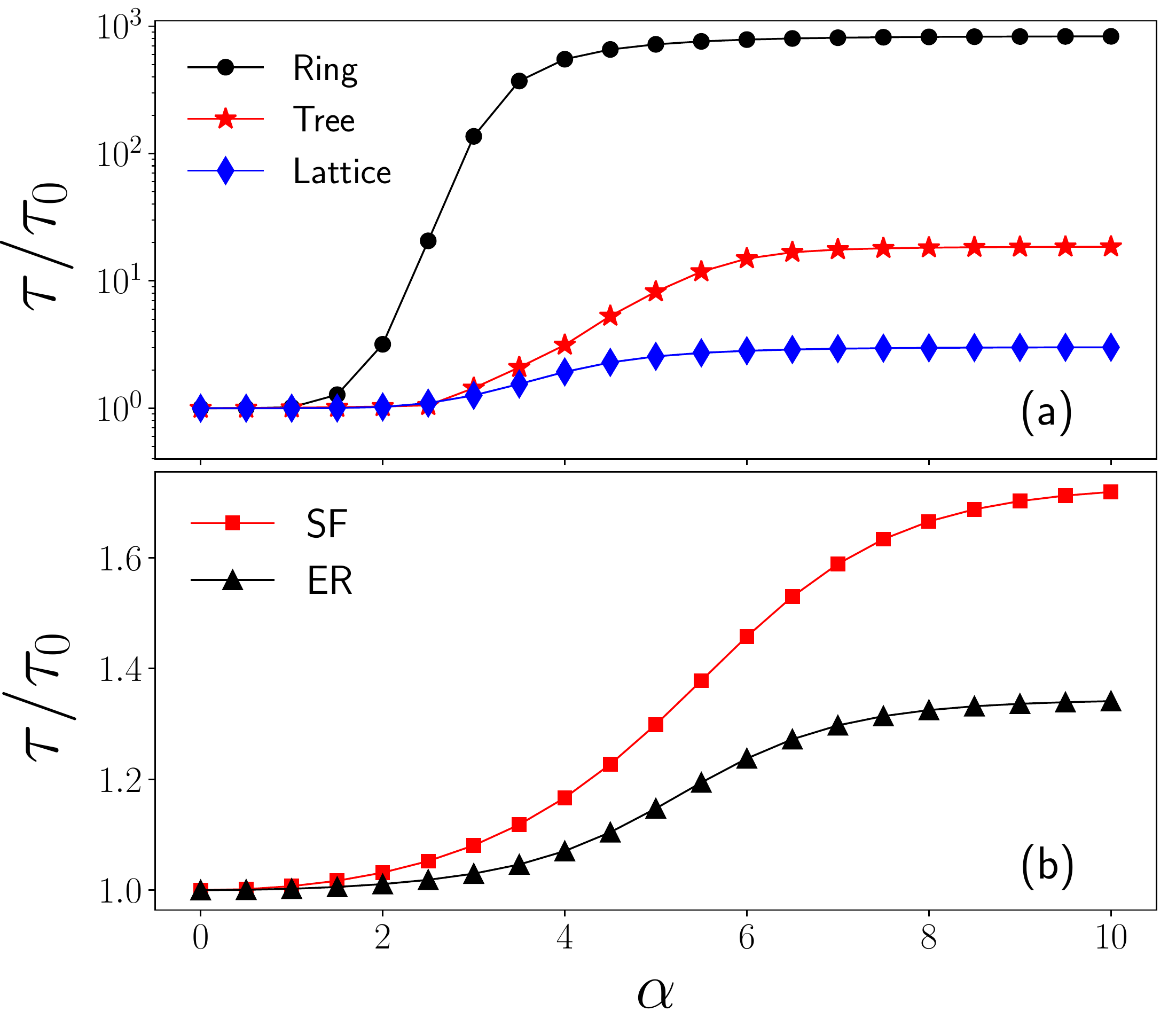}
\end{center}
\vspace{-8mm}
\caption{\label{FigureRW13} (Color online) Global time $\tau$ that gives the average number of steps needed for L\'evy flights to reach any site of the network in different structures; the results are expressed in terms of the value $\tau_0=(N-1)^2/N$. In figures (a) and (b) we depict the time $\tau$ for different values of $\alpha$ in networks with $N=5000$. We calculated the values numerically using the Eqs. (\ref{TauiSpect}) and (\ref{tauglobal}) for large-world networks (ring, tree, and a square lattice of size $50\times 100$ and periodic boundary conditions) and small-world networks 
(Erd\H{o}s--R\'enyi network (ER) in the percolation limit and a scale-free network (SF) of the  Barab\'asi--Albert type).}
\end{figure}
In Fig. \ref{FigureRW12} we analyze the values of $\tau$ for L\'evy flights on finite rings. In Fig. \ref{FigureRW12}(a) we present the result for $\tau/\tau_0$ as a function of the ring size $N$ for different values of $\alpha$. From the obtained results, we observe that in the limit $\alpha\to \infty$, the time $\tau$ behaves as $\tau/\tau_0 \sim N$, therefore $\tau \sim N^2$. In a similar way, the reduction of the parameter $\alpha$ gradually changes $\tau$ to $\tau\sim N^{1+\delta}$, where $\delta$ takes values in the interval $0<\delta<1$ for $0<\alpha<\infty$. Finally, in the limit  $\alpha\to 0$, we have $\tau\sim N$. In addition, in Fig. \ref{FigureRW12}(b) we depict the results for  $\tau/\tau_0$ as a function of $\alpha$ for different values of the ring size $N$. In this case, it is observed how the relation $\tau/\tau_0$  maintains a similar behavior for different values of the number of nodes  $ N$. In the interval $0<\alpha<2$  is observed that $\tau/\tau_0$ are close to the value $\tau_0$ obtained in the limit $ \alpha \to  0$ or in a complete graph. Moreover, in the range $2<\alpha\leq 5$, $\tau/\tau_0$ presents an increase that can be of several orders. For $ \alpha> 6$, the results are again constant and close to the limit $ \alpha \to \infty$ given by Eq. (\ref{TauRing_NRW}). 
\\[2mm]
Once analyzed the L\'evy flights on rings, for which the eigenvalues of the transition have the analytic form presented in Eqs. (\ref{LFSpectCycleEven}) and (\ref{LFSpectCycleOdd}), it is important to explore the value of $\tau$ for other structures. In Fig. \ref{FigureRW13}, we present the results obtained for the average number of steps $\tau$  as a function of the parameter  $\alpha$ for L\'evy flights on different types of networks. In regular networks (ring and square lattice), the value of $\tau$ is obtained from Eq. (\ref{EffectRegular}). For the tree, the ER and the SF networks, the results are calculated using Eqs. (\ref{TauiSpect}) and (\ref{tauglobal}). The obtained values for $\tau$  suggest that, compared with the normal random walker, in large-world networks the average number of steps required to reach any node in the network is lower for the L\'evy flight strategy. In small-world networks, the differences are smaller, but even in this case, the L\'evy strategy improves the results obtained for the normal random walker. This result is reasonable due to the fact that in large-world networks,  L\'evy flights define a dynamics that induces small-world property. In the case of small-world networks, the nodes in the network are separated by short distances and, in this way, L\'evy flights and the normal random walk explore the network with efficiencies of the same order of magnitude \cite{RiascosMateos2012}.
\subsection{Fractional transport}
\begin{figure}[t!]
\begin{center}
\includegraphics[width=0.47\textwidth]{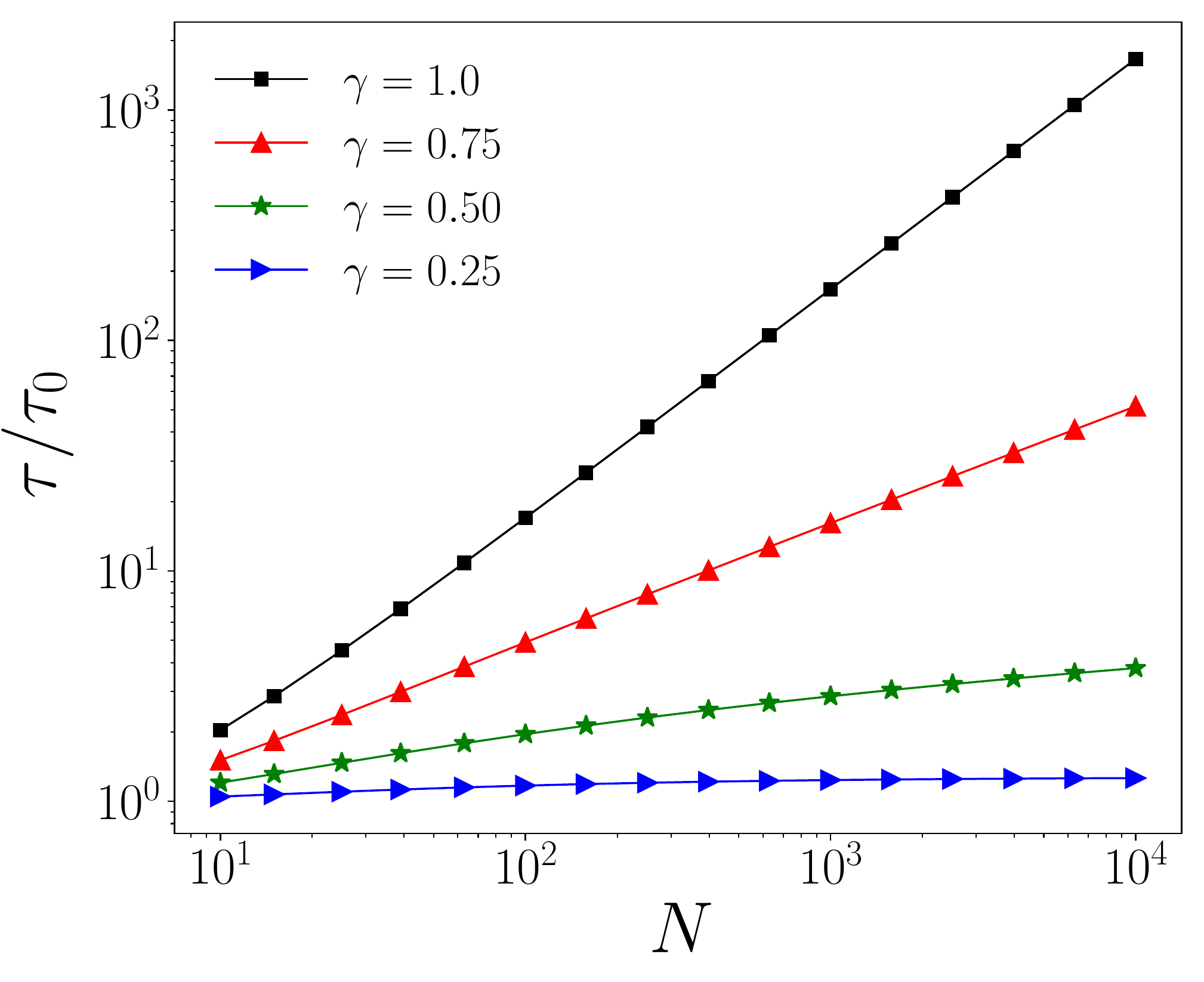}
\end{center}
\vspace{-6mm}
\caption{\label{FigureRW14} Global time $\tau$ as a function of the number of nodes $N$ for the fractional transport on rings. We obtain the results  for the time $\tau$ by direct evaluation of the Eq. (\ref{TauRingEq}). We express the time $\tau$ in relation to $\tau_0=(N-1)^2/N$ for different values of the parameter $\gamma$ that defines  each strategy. Solid lines are used as a guide.}
\end{figure}

In this part we apply the approach described before for preferential random walks and L\'evy flights to the case of the fractional transport on networks with transition probabilities between nodes given by the Eq. (\ref{wijfrac}). In the case of rings with $N$ nodes, the eigenvalues of the transition matrices can be obtained analytically  \cite{RiascosMateosFD2015,RiascosMichelitsch2017_gL}. For this regular structure, the fractional degree in Eq. (\ref{FracDegree}) is a constant $k^{(\gamma)}$ given by \cite{RiascosMichelitsch2017_gL}
\begin{equation}\label{degreeGLring}
k^{(\gamma)}=(\mathbf{L}^\gamma)_{ii}=\frac{1}{N}\sum_{l=1}^N \left(2-2\cos\left[\frac{2\pi}{N}(l-1)\right]\right)^\gamma
\end{equation}
and the eigenvalues $\{\lambda_i\}_{i=1} ^N$ for the  transition matrix $\mathbf{\Pi}$, with elements  in Eq. (\ref{wijfrac}), are
\begin{equation}\label{lambdaGLring}
\lambda_i=1-\frac{1}{k^{(\gamma)}}\, \left(2-2\cos\left[\frac{2\pi}{N}(i-1)\right]\right)^\gamma.
\end{equation}
Now, as a consequence of the results in Eqs.  (\ref{degreeGLring}) and (\ref{lambdaGLring}), the time $\tau$ that characterizes the  global performance of the random strategy in Eq. (\ref{wijfrac}) to explore a ring coincides with the Kemeny's constant and is given by \cite{RiascosMichelitsch2017_gL}
\begin{equation}\label{TauRingEq}
\tau=k^{(\gamma)} \sum_{m=2}^N\frac{1}{\left(2-2\cos\phi_m\right)^\gamma},
\end{equation}
where $\phi_i\equiv\frac{2\pi}{N}(i-1)$.
\\[2mm]
In Fig. \ref{FigureRW14} we represent the numerical values of the global time $\tau/\tau_0$ obtained for the fractional transport on rings. The results are calculated by direct evaluation of the result in Eq. (\ref{TauRingEq}).  We explore the parameter $\gamma$ that defines the fractional strategy for different values of the size of the ring $N$. In Fig. \ref{FigureRW14} we observe that the dynamics with $0<\gamma<1$ always improves the capacity to explore the ring in comparison with a normal random walk recovered in the case $\gamma=1$. This effect is observed in the reduction of the quantity $\tau/\tau_0$ for $\gamma=0.25,\, 0.5$ and $\gamma=0.75$. On the other hand, in the limit $\gamma\to 0$ the dynamics is equivalent to a normal random walker on a fully connected network  allowing, with the same probability, transitions from one node to any site of the ring \cite{RiascosMateosFD2015}. A similar behavior to this limit is also observed for the case $\gamma=0.25$ for all the values of $N$ analyzed.

\begin{figure}[t!]
\begin{center}
\includegraphics[width=0.47\textwidth]{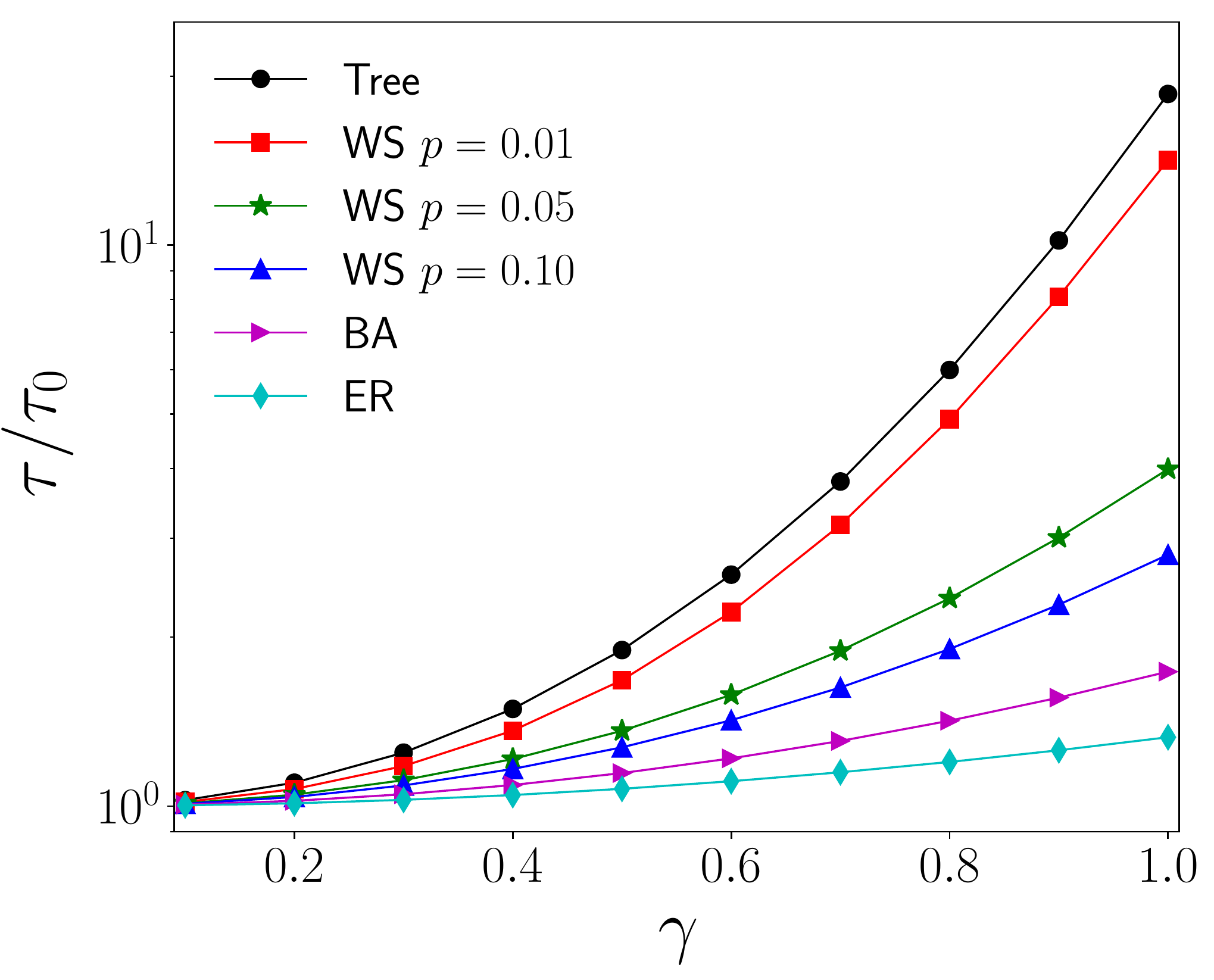}
\end{center}
\vspace{-6mm}
\caption{\label{FigureRW15} Global time $\tau$ for the fractional transport on connected networks with $N=5000$ nodes: a tree, random networks generated from the Watts-Strogatz (WS) model with rewiring probabilities $p=0.01$,  $p=0.05$, $p=0.1$, a scale-free (SF) network of the Barab\'asi-Albert type and a random network of the Erd\H{o}s-R\'enyi (ER) type at the percolation limit $p=\log{N}/N$.  We obtain the results  for the time $\tau$ by numerical evaluation of the Eqs. (\ref{TauiSpect}) and (\ref{tauglobal}). We express $\tau$ in relation to the value $\tau_0=(N-1)^2/N$. Solid lines are used as a guide. }
\end{figure}

Through the evaluation of the global time $\tau$ we can analyze the fractional dynamics in different types of large-world and small-world networks. Unlike the previous cases explored for rings, other types of networks have not the same fractional degree $k^{(\gamma)}_i$ for all the nodes $i=1,2,\ldots,N$. In this way, the efficiency or global performance of the random walker is quantified by the time $\tau$ given by the average of the times Eq. (\ref{TauiSpect}) that depends on the eigenvectors and eigenvalues of the transition matrix $\mathbf{\Pi}$ with elements given by Eq. (\ref{wijfrac}). 
\\[2mm]
In Fig. \ref{FigureRW15} we show the global time $\tau$ for networks with $N=5000$ nodes. We analyze a deterministic tree created by an iterative method for which an initial node ramifies with two leaves that also repeat this process until the size $N$, the final structure is a large-world network with average distances $\langle d\rangle$ between nodes that scale as the size of the network. On the other hand, we analyze random networks generated with the Watts-Strogatz model for which an initially regular network is generated and then rewired uniformly randomly with probability $p$; for values of $p\to 0$ this random network has the large world property of the original lattice; however, the rewiring introduces shortcuts  that reduce the average path lengths with the increasing of $p$ \cite{WattsStrogatz}. In addition, small-world networks generated with the Erd\H{o}s-R\'enyi model and  a scale-free (SF) network of the Barab\'asi-Albert type are explored \cite{ErdosRenyi,BarabasiAlbert}. We observe that the generalized dynamics defined in terms of the fractional Laplacian $\mathbf{L}^\gamma$ with $0<\gamma<1$ always improves the efficiency to explore the networks, the effects are marked in large-world networks with a significant change in the  value $\tau/\tau_0$, but the dynamics also improves the results for small-world networks.
\\[2mm]
As we discussed in Section \ref{subs_fractional}, the random walk defined in terms of the fractional Laplacian $\mathbf{L}^\gamma$ is a particular case of non-local strategies expressed using functions of the Laplacian $g(\mathbf{L})$. The same approach presented here applies to other types of strategies \cite{RiascosMichelitsch2017_gL};  for example, when we use weights defined in terms of the logarithmic function  $\log\left(\mathbb{I}+\alpha \mathbf{L}\right)$ for $\alpha>0$ and the function $\mathbb{I}-e^{-a\mathbf{L}}$ with $a>0$.
\subsection{Random walks to visit specific locations}
\begin{figure}[!t]
\begin{center}
\includegraphics*[width=0.47\textwidth]{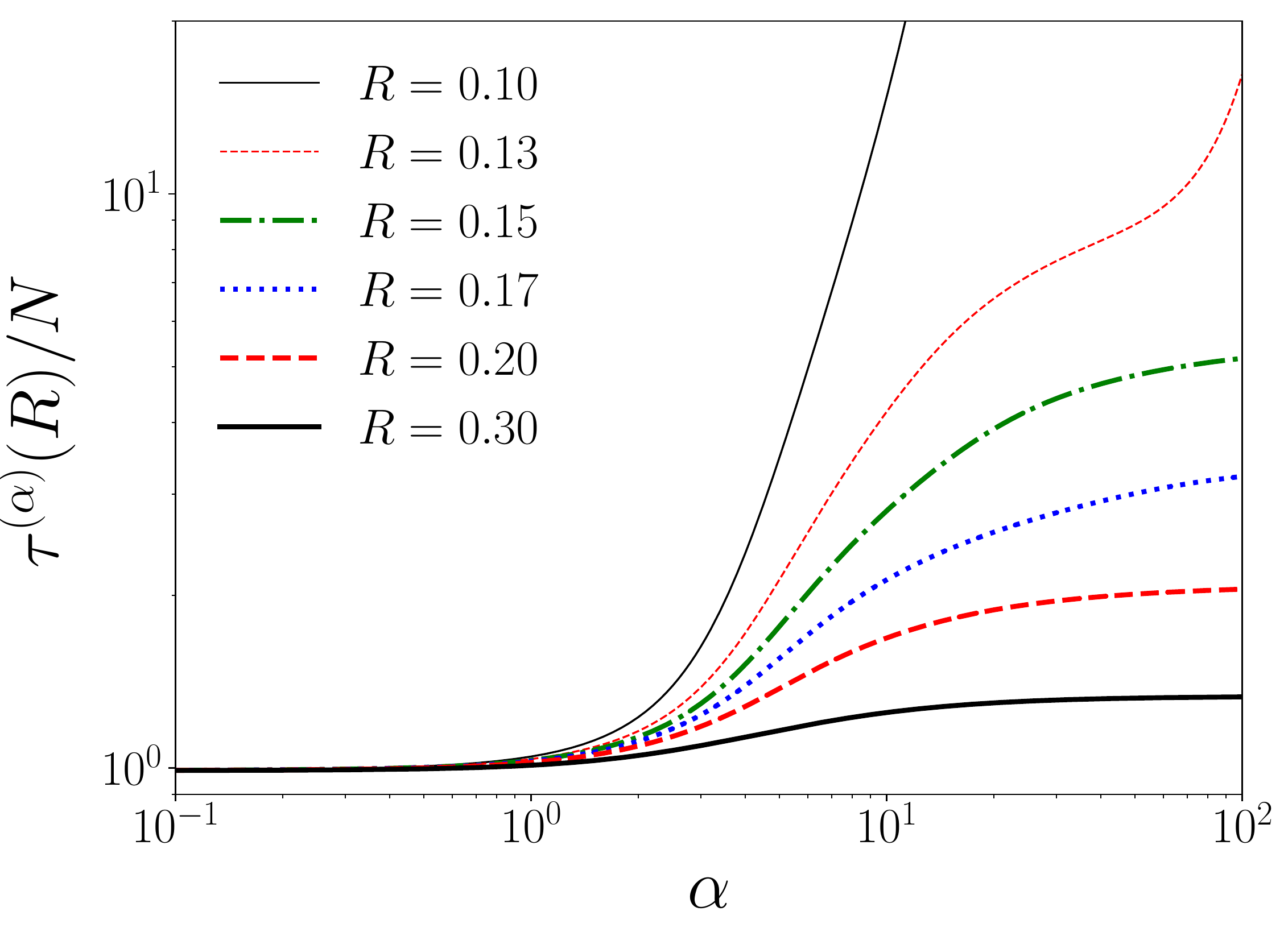}\,
\end{center}
\caption{\label{FigureRW16} (Color online) Global time to visit $N$ locations. The value $\tau^{(\alpha)}(R)$ gives the average number of steps needed to reach 
any of the $N$ sites, independently of the initial condition; we use $N=100$ random sites in the region $[0,1]\times [0,1]$ in $\mathbb{R}^2$. The results are obtained from the analytical expressions in Eqs (\ref{TauiSpect})-(\ref{tauglobal}) and the numerical evaluation of the eigenvectors and eigenvalues of the transition matrix with elements $w_{i\to j}^{(\alpha)}(R)$.}
\end{figure}

We end this section with an application of the formalism developed in terms of a matrix of weights but now for a model in the context of human mobility that not necessarily requires the definition of a network. We analyze the random walk strategy with transition probabilities defined by Eqs. (\ref{wijRalpha})-(\ref{Omega_ij}) to visit specific locations in a region. In this case, the matrix of weights with elements $\Omega_{ij}^{(\alpha)}(R)$ depends on the characteristic length $R$, that defines a local neighborhood, and the parameter $\alpha$ that controls a dynamics with similar characteristics to the L\'evy flights on networks but now to visit the locations. In order to quantify the capacity of the random walker to visit  the $N$ locations in space, we use the 
time $\tau^{(\alpha)}(R)$, that gives the average number of steps needed to reach any of the $N$ sites independently of the initial condition by numerical evaluation of the Eqs. (\ref{TauiSpect}) and (\ref{tauglobal}). In Fig. \ref{FigureRW16} we present the time $\tau^{(\alpha)}(R)$  for different values of the  parameters $\alpha$ and $R$ to visit $N=100$ random locations on the plane. The values are obtained  using  the exact analytical results in terms of  the eigenvectors and eigenvalues of the transition matrix defined by Eq (\ref{wijRalpha}). It is observed how, for $\alpha>>1$, different values of $R$ define diverse ways to visit the $N$ sites in the plane; in particular, $R <<1$ characterizes a local strategy that requires many steps to reach the locations. Strategies with $\alpha\leq 1$ are optimal and in this interval the results are independent of the parameter $R$. The results observed with the help of the global time $\tau^{\alpha}(R)$ suggest that long-range strategies always improve the capacity of the random walker to reach any of the $N$ locations \cite{RiascosMateosPlos2017}.
\section{Conclusions}
In this work, we presented a general approach to examine different random walk strategies on undirected networks described as a discrete time Markovian process with transition probabilities defined in terms of a symmetric matrix of weights. This formalism is explored for different types of random walks with transitions restricted to nearest neighbors in the case of local strategies and hops from one node to any site of the network in the case non-local strategies. We obtained the stationary probability distribution and other quantities that characterize the capacity of each random walk strategy to reach the nodes of the network like the mean first passage time and average times to reach a specific node. All these quantities are obtained in terms of the eigenvectors and eigenvalues of the transition matrix $\mathbf{\Pi}$; in a similar way, we calculate global quantities like the Kemeny's constant and the average time to reach any site of the network $\tau$. We explore in detail these quantities for the preferential random walk, the L\'evy flights and the fractional transport on networks. For these cases, we conclude that the value of $\tau$ is a good measure that allows comparing the performance of random walk strategies since it quantifies the efficiency of the random walker to explore new sites on the network. On the other hand, the Kemeny's constant is a useful simplification that depends only on the spectrum of the transition matrix and describes correctly the efficiency of random walk strategies with a constant stationary probability distribution. 
\\[2mm]
Finally, it is worth mentioning that the formalism introduced here can be implemented to the study of other processes described in terms of a matrix of weights. An application in a context different from the field of network science is presented for the analysis of a random walk introduced in the context of human mobility and implemented to visit locations in the plane. In this case, the global time $\tau$ help us to understand how strategies that combine local and non-local displacements improve the capacity of the exploration of specific locations in a region in the plane.

\onecolumngrid


\section*{References}



\begin{thebibliography}{108}%
\makeatletter
\providecommand \@ifxundefined [1]{%
 \@ifx{#1\undefined}
}%
\providecommand \@ifnum [1]{%
 \ifnum #1\expandafter \@firstoftwo
 \else \expandafter \@secondoftwo
 \fi
}%
\providecommand \@ifx [1]{%
 \ifx #1\expandafter \@firstoftwo
 \else \expandafter \@secondoftwo
 \fi
}%
\providecommand \natexlab [1]{#1}%
\providecommand \enquote  [1]{``#1''}%
\providecommand \bibnamefont  [1]{#1}%
\providecommand \bibfnamefont [1]{#1}%
\providecommand \citenamefont [1]{#1}%
\providecommand \href@noop [0]{\@secondoftwo}%
\providecommand \href [0]{\begingroup \@sanitize@url \@href}%
\providecommand \@href[1]{\@@startlink{#1}\@@href}%
\providecommand \@@href[1]{\endgroup#1\@@endlink}%
\providecommand \@sanitize@url [0]{\catcode `\\12\catcode `\$12\catcode
  `\&12\catcode `\#12\catcode `\^12\catcode `\_12\catcode `\%12\relax}%
\providecommand \@@startlink[1]{}%
\providecommand \@@endlink[0]{}%
\providecommand \url  [0]{\begingroup\@sanitize@url \@url }%
\providecommand \@url [1]{\endgroup\@href {#1}{\urlprefix }}%
\providecommand \urlprefix  [0]{URL }%
\providecommand \Eprint [0]{\href }%
\providecommand \doibase [0]{http://dx.doi.org/}%
\providecommand \selectlanguage [0]{\@gobble}%
\providecommand \bibinfo  [0]{\@secondoftwo}%
\providecommand \bibfield  [0]{\@secondoftwo}%
\providecommand \translation [1]{[#1]}%
\providecommand \BibitemOpen [0]{}%
\providecommand \bibitemStop [0]{}%
\providecommand \bibitemNoStop [0]{.\EOS\space}%
\providecommand \EOS [0]{\spacefactor3000\relax}%
\providecommand \BibitemShut  [1]{\csname bibitem#1\endcsname}%
\let\auto@bib@innerbib\@empty
%</preamble>
\bibitem [{\citenamefont {Klafter}\ and\ \citenamefont
  {Sokolov}(2011)}]{KlafterSokolov}%
  \BibitemOpen
  \bibfield  {author} {\bibinfo {author} {\bibfnamefont {J.}~\bibnamefont
  {Klafter}}\ and\ \bibinfo {author} {\bibfnamefont {I.}~\bibnamefont
  {Sokolov}},\ }\href@noop {} {\emph {\bibinfo {title} {First Steps in Random
  Walks: From Tools to Applications}}}\ (\bibinfo  {publisher} {Oxford
  University Press, Oxford},\ \bibinfo {year} {2011})\BibitemShut {NoStop}%
\bibitem [{\citenamefont {Masuda}\ \emph {et~al.}(2017)\citenamefont {Masuda},
  \citenamefont {Porter},\ and\ \citenamefont {Lambiotte}}]{MasudaPhysRep2017}%
  \BibitemOpen
  \bibfield  {author} {\bibinfo {author} {\bibfnamefont {N.}~\bibnamefont
  {Masuda}}, \bibinfo {author} {\bibfnamefont {M.~A.}\ \bibnamefont {Porter}},
  \ and\ \bibinfo {author} {\bibfnamefont {R.}~\bibnamefont {Lambiotte}},\
  }\href {https://doi.org/10.1016/j.physrep.2017.07.007} {\bibfield  {journal}
  {\bibinfo  {journal} {Phys. Rep.}\ }\textbf {\bibinfo {volume} {716--717}},\
  \bibinfo {pages} {1} (\bibinfo {year} {2017})}\BibitemShut {NoStop}%
\bibitem [{\citenamefont {Redner}(2001)}]{SRedner}%
  \BibitemOpen
  \bibfield  {author} {\bibinfo {author} {\bibfnamefont {S.}~\bibnamefont
  {Redner}},\ }\href@noop {} {\emph {\bibinfo {title} {A Guide to First-Passage
  Processes}}}\ (\bibinfo  {publisher} {Cambridge University Press},\ \bibinfo
  {address} {New York},\ \bibinfo {year} {2001})\BibitemShut {NoStop}%
\bibitem [{\citenamefont {van Kampen}(1992)}]{VKampen}%
  \BibitemOpen
  \bibfield  {author} {\bibinfo {author} {\bibfnamefont {N.~G.}\ \bibnamefont
  {van Kampen}},\ }\href@noop {} {\emph {\bibinfo {title} {Stochastic Processes
  in Physics and Chemistry}}}\ (\bibinfo  {publisher} {North Holland},\
  \bibinfo {address} {The Netherlands},\ \bibinfo {year} {1992})\BibitemShut
  {NoStop}%
\bibitem [{\citenamefont {Viswanathan}\ \emph {et~al.}(2011)\citenamefont
  {Viswanathan}, \citenamefont {da~Luz}, \citenamefont {Raposo},\ and\
  \citenamefont {Stanley}}]{ForeignBook}%
  \BibitemOpen
  \bibfield  {author} {\bibinfo {author} {\bibfnamefont {G.~M.}\ \bibnamefont
  {Viswanathan}}, \bibinfo {author} {\bibfnamefont {M.~G.~E.}\ \bibnamefont
  {da~Luz}}, \bibinfo {author} {\bibfnamefont {E.~P.}\ \bibnamefont {Raposo}},
  \ and\ \bibinfo {author} {\bibfnamefont {H.~E.}\ \bibnamefont {Stanley}},\
  }\href@noop {} {\emph {\bibinfo {title} {The Physics of Foraging}}}\
  (\bibinfo  {publisher} {Cambridge University Press},\ \bibinfo {address} {New
  York},\ \bibinfo {year} {2011})\BibitemShut {NoStop}%
\bibitem [{\citenamefont {Wosniack}\ \emph {et~al.}(2017)\citenamefont
  {Wosniack}, \citenamefont {Santos}, \citenamefont {Raposo}, \citenamefont
  {Viswanathan},\ and\ \citenamefont {da~Luz}}]{Viswanathan_Plos2017}%
  \BibitemOpen
  \bibfield  {author} {\bibinfo {author} {\bibfnamefont {M.~E.}\ \bibnamefont
  {Wosniack}}, \bibinfo {author} {\bibfnamefont {M.~C.}\ \bibnamefont
  {Santos}}, \bibinfo {author} {\bibfnamefont {E.~P.}\ \bibnamefont {Raposo}},
  \bibinfo {author} {\bibfnamefont {G.~M.}\ \bibnamefont {Viswanathan}}, \ and\
  \bibinfo {author} {\bibfnamefont {M.~G.~E.}\ \bibnamefont {da~Luz}},\ }\href
  {\doibase 10.1371/journal.pcbi.1005774} {\bibfield  {journal} {\bibinfo
  {journal} {PLoS Comput. Biol.}\ }\textbf {\bibinfo {volume} {13}},\ \bibinfo
  {pages} {e1005774} (\bibinfo {year} {2017})}\BibitemShut {NoStop}%
\bibitem [{\citenamefont {Newman}(2010)}]{NewmanBook}%
  \BibitemOpen
  \bibfield  {author} {\bibinfo {author} {\bibfnamefont {M.~E.~J.}\
  \bibnamefont {Newman}},\ }\href@noop {} {\emph {\bibinfo {title} {Networks:
  An Introduction}}}\ (\bibinfo  {publisher} {Oxford University Press},\
  \bibinfo {address} {Oxford},\ \bibinfo {year} {2010})\BibitemShut {NoStop}%
\bibitem [{\citenamefont {Barab\'asi}\ and\ \citenamefont
  {P\'osfai}(2016)}]{barabasi2016book}%
  \BibitemOpen
  \bibfield  {author} {\bibinfo {author} {\bibfnamefont {A.-L.}\ \bibnamefont
  {Barab\'asi}}\ and\ \bibinfo {author} {\bibfnamefont {M.}~\bibnamefont
  {P\'osfai}},\ }\href@noop {} {\emph {\bibinfo {title} {Network science}}}\
  (\bibinfo  {publisher} {Cambridge University Press},\ \bibinfo {address}
  {Cambridge},\ \bibinfo {year} {2016})\BibitemShut {NoStop}%
\bibitem [{\citenamefont {Latora}\ \emph {et~al.}(2017)\citenamefont {Latora},
  \citenamefont {Nicosia},\ and\ \citenamefont {Russo}}]{Latora2017}%
  \BibitemOpen
  \bibfield  {author} {\bibinfo {author} {\bibfnamefont {V.}~\bibnamefont
  {Latora}}, \bibinfo {author} {\bibfnamefont {V.}~\bibnamefont {Nicosia}}, \
  and\ \bibinfo {author} {\bibfnamefont {G.}~\bibnamefont {Russo}},\
  }\href@noop {} {\emph {\bibinfo {title} {Complex Networks: Principles,
  Methods and Applications}}}\ (\bibinfo  {publisher} {Cambridge University
  Press},\ \bibinfo {year} {2017})\BibitemShut {NoStop}%
\bibitem [{\citenamefont {Barrat}\ \emph {et~al.}(2008)\citenamefont {Barrat},
  \citenamefont {Barth\'elemy},\ and\ \citenamefont {Vespignani}}]{VespiBook}%
  \BibitemOpen
  \bibfield  {author} {\bibinfo {author} {\bibfnamefont {A.}~\bibnamefont
  {Barrat}}, \bibinfo {author} {\bibfnamefont {M.}~\bibnamefont
  {Barth\'elemy}}, \ and\ \bibinfo {author} {\bibfnamefont {A.}~\bibnamefont
  {Vespignani}},\ }\href@noop {} {\emph {\bibinfo {title} {Dynamical Processes
  on Complex Networks}}}\ (\bibinfo  {publisher} {Cambridge University Press},\
  \bibinfo {address} {Cambridge},\ \bibinfo {year} {2008})\BibitemShut
  {NoStop}%
\bibitem [{\citenamefont {Van~Mieghem}(2011)}]{VanMieghem}%
  \BibitemOpen
  \bibfield  {author} {\bibinfo {author} {\bibfnamefont {P.}~\bibnamefont
  {Van~Mieghem}},\ }\href@noop {} {\emph {\bibinfo {title} {Graph Spectra for
  Complex Networks}}}\ (\bibinfo  {publisher} {Cambridge University Press},\
  \bibinfo {address} {New York},\ \bibinfo {year} {2011})\BibitemShut {NoStop}%
\bibitem [{\citenamefont {Hughes}(1996)}]{Hughes}%
  \BibitemOpen
  \bibfield  {author} {\bibinfo {author} {\bibfnamefont {B.~D.}\ \bibnamefont
  {Hughes}},\ }\href@noop {} {\emph {\bibinfo {title} {Random Walks and Random
  Environments: Vol. 1: Random Walks}}}\ (\bibinfo  {publisher} {Oxford
  University Press, New York},\ \bibinfo {year} {1996})\BibitemShut {NoStop}%
\bibitem [{\citenamefont {Lov\'asz}(1996)}]{Lovasz1996}%
  \BibitemOpen
  \bibfield  {author} {\bibinfo {author} {\bibfnamefont {L.}~\bibnamefont
  {Lov\'asz}},\ }in\ \href@noop {} {\emph {\bibinfo {booktitle} {Combinatorics,
  Paul Erd\H{o}s is Eighty}}},\ Vol.~\bibinfo {volume} {2},\ \bibinfo {editor}
  {edited by\ \bibinfo {editor} {\bibfnamefont {D.}~\bibnamefont {{Mikl\'os}}},
  \bibinfo {editor} {\bibfnamefont {V.~T.}\ \bibnamefont {{S\'os}}}, \ and\
  \bibinfo {editor} {\bibfnamefont {T.}~\bibnamefont {{Sz\H{o}nyi}}}}\
  (\bibinfo  {publisher} {J\'anos Bolyai Mathematical Society},\ \bibinfo
  {address} {Budapest},\ \bibinfo {year} {1996})\ pp.\ \bibinfo {pages}
  {353--398}\BibitemShut {NoStop}%
\bibitem [{\citenamefont {M\"ulken}\ and\ \citenamefont
  {Blumen}(2011)}]{MulkenPR502}%
  \BibitemOpen
  \bibfield  {author} {\bibinfo {author} {\bibfnamefont {O.}~\bibnamefont
  {M\"ulken}}\ and\ \bibinfo {author} {\bibfnamefont {A.}~\bibnamefont
  {Blumen}},\ }\href {\doibase 10.1016/j.physrep.2011.01.002} {\bibfield
  {journal} {\bibinfo  {journal} {Phys. Rep.}\ }\textbf {\bibinfo {volume}
  {502}},\ \bibinfo {pages} {37} (\bibinfo {year} {2011})}\BibitemShut
  {NoStop}%
\bibitem [{\citenamefont {Noh}\ and\ \citenamefont {Rieger}(2004)}]{NohRieger}%
  \BibitemOpen
  \bibfield  {author} {\bibinfo {author} {\bibfnamefont {J.~D.}\ \bibnamefont
  {Noh}}\ and\ \bibinfo {author} {\bibfnamefont {H.}~\bibnamefont {Rieger}},\
  }\href {\doibase 10.1103/PhysRevLett.92.118701} {\bibfield  {journal}
  {\bibinfo  {journal} {Phys. Rev. Lett.}\ }\textbf {\bibinfo {volume} {92}},\
  \bibinfo {pages} {118701} (\bibinfo {year} {2004})}\BibitemShut {NoStop}%
\bibitem [{\citenamefont {Fronczak}\ and\ \citenamefont
  {Fronczak}(2009)}]{FronczakPRE2009}%
  \BibitemOpen
  \bibfield  {author} {\bibinfo {author} {\bibfnamefont {A.}~\bibnamefont
  {Fronczak}}\ and\ \bibinfo {author} {\bibfnamefont {P.}~\bibnamefont
  {Fronczak}},\ }\href {\doibase 10.1103/PhysRevE.80.016107} {\bibfield
  {journal} {\bibinfo  {journal} {Phys. Rev. E}\ }\textbf {\bibinfo {volume}
  {80}},\ \bibinfo {pages} {016107} (\bibinfo {year} {2009})}\BibitemShut
  {NoStop}%
\bibitem [{\citenamefont {Tejedor}\ \emph {et~al.}(2009)\citenamefont
  {Tejedor}, \citenamefont {B\'enichou},\ and\ \citenamefont
  {Voituriez}}]{TejedorPRE2009}%
  \BibitemOpen
  \bibfield  {author} {\bibinfo {author} {\bibfnamefont {V.}~\bibnamefont
  {Tejedor}}, \bibinfo {author} {\bibfnamefont {O.}~\bibnamefont {B\'enichou}},
  \ and\ \bibinfo {author} {\bibfnamefont {R.}~\bibnamefont {Voituriez}},\
  }\href {\doibase 10.1103/PhysRevE.80.065104} {\bibfield  {journal} {\bibinfo
  {journal} {Phys. Rev. E}\ }\textbf {\bibinfo {volume} {80}},\ \bibinfo
  {pages} {065104} (\bibinfo {year} {2009})}\BibitemShut {NoStop}%
\bibitem [{\citenamefont {Alessandretti}\ \emph {et~al.}(2017)\citenamefont
  {Alessandretti}, \citenamefont {Sun}, \citenamefont {Baronchelli},\ and\
  \citenamefont {Perra}}]{Baronchelli2017}%
  \BibitemOpen
  \bibfield  {author} {\bibinfo {author} {\bibfnamefont {L.}~\bibnamefont
  {Alessandretti}}, \bibinfo {author} {\bibfnamefont {K.}~\bibnamefont {Sun}},
  \bibinfo {author} {\bibfnamefont {A.}~\bibnamefont {Baronchelli}}, \ and\
  \bibinfo {author} {\bibfnamefont {N.}~\bibnamefont {Perra}},\ }\href
  {\doibase 10.1103/PhysRevE.95.052318} {\bibfield  {journal} {\bibinfo
  {journal} {Phys. Rev. E}\ }\textbf {\bibinfo {volume} {95}},\ \bibinfo
  {pages} {052318} (\bibinfo {year} {2017})}\BibitemShut {NoStop}%
\bibitem [{\citenamefont {{De Domenico}}\ \emph {et~al.}(2016)\citenamefont
  {{De Domenico}}, \citenamefont {Granell}, \citenamefont {Porter},\ and\
  \citenamefont {Arenas}}]{Manlio2016}%
  \BibitemOpen
  \bibfield  {author} {\bibinfo {author} {\bibfnamefont {M.}~\bibnamefont {{De
  Domenico}}}, \bibinfo {author} {\bibfnamefont {C.}~\bibnamefont {Granell}},
  \bibinfo {author} {\bibfnamefont {M.~A.}\ \bibnamefont {Porter}}, \ and\
  \bibinfo {author} {\bibfnamefont {A.}~\bibnamefont {Arenas}},\ }\href
  {\doibase 10.1038/nphys3865} {\bibfield  {journal} {\bibinfo  {journal}
  {Nature Phys.}\ }\textbf {\bibinfo {volume} {12}},\ \bibinfo {pages} {901}
  (\bibinfo {year} {2016})}\BibitemShut {NoStop}%
\bibitem [{\citenamefont {Durrett}(2010)}]{Durrett2010}%
  \BibitemOpen
  \bibfield  {author} {\bibinfo {author} {\bibfnamefont {R.}~\bibnamefont
  {Durrett}},\ }\href {\doibase 10.1073/pnas.0914402107} {\bibfield  {journal}
  {\bibinfo  {journal} {Proc. Natl. Acad. Sci. USA}\ }\textbf {\bibinfo
  {volume} {107}},\ \bibinfo {pages} {4491} (\bibinfo {year}
  {2010})}\BibitemShut {NoStop}%
\bibitem [{\citenamefont {Pastor-Satorras}\ \emph {et~al.}(2015)\citenamefont
  {Pastor-Satorras}, \citenamefont {Castellano}, \citenamefont {Van~Mieghem},\
  and\ \citenamefont {Vespignani}}]{RomualdoRMP2015}%
  \BibitemOpen
  \bibfield  {author} {\bibinfo {author} {\bibfnamefont {R.}~\bibnamefont
  {Pastor-Satorras}}, \bibinfo {author} {\bibfnamefont {C.}~\bibnamefont
  {Castellano}}, \bibinfo {author} {\bibfnamefont {P.}~\bibnamefont
  {Van~Mieghem}}, \ and\ \bibinfo {author} {\bibfnamefont {A.}~\bibnamefont
  {Vespignani}},\ }\href {\doibase 10.1103/RevModPhys.87.925} {\bibfield
  {journal} {\bibinfo  {journal} {Rev. Mod. Phys.}\ }\textbf {\bibinfo {volume}
  {87}},\ \bibinfo {pages} {925} (\bibinfo {year} {2015})}\BibitemShut
  {NoStop}%
\bibitem [{\citenamefont {Sarkar}\ and\ \citenamefont
  {Moore}(2011)}]{Sarkar2011}%
  \BibitemOpen
  \bibfield  {author} {\bibinfo {author} {\bibfnamefont {P.}~\bibnamefont
  {Sarkar}}\ and\ \bibinfo {author} {\bibfnamefont {A.~W.}\ \bibnamefont
  {Moore}},\ }\enquote {\bibinfo {title} {Random walks in social networks and
  their applications: A survey},}\ in\ \href@noop {} {\emph {\bibinfo
  {booktitle} {Social Network Data Analytics}}},\ \bibinfo {editor} {edited by\
  \bibinfo {editor} {\bibfnamefont {C.~C.}\ \bibnamefont {Aggarwal}}}\
  (\bibinfo  {publisher} {Springer},\ \bibinfo {address} {Boston},\ \bibinfo
  {year} {2011})\ pp.\ \bibinfo {pages} {43--77}\BibitemShut {NoStop}%
\bibitem [{\citenamefont {Blanchard}\ and\ \citenamefont
  {Volchenkov}(2011)}]{BlanchardBook2011}%
  \BibitemOpen
  \bibfield  {author} {\bibinfo {author} {\bibfnamefont {P.}~\bibnamefont
  {Blanchard}}\ and\ \bibinfo {author} {\bibfnamefont {D.}~\bibnamefont
  {Volchenkov}},\ }\href@noop {} {\emph {\bibinfo {title} {Random walks and
  diffusions on graphs and databases. An introduction}}},\ Springer {S}eries in
  Synergetics\ (\bibinfo  {publisher} {Springer},\ \bibinfo {address}
  {Berlin},\ \bibinfo {year} {2011})\BibitemShut {NoStop}%
\bibitem [{\citenamefont {Riascos}\ and\ \citenamefont
  {Mateos}(2017)}]{RiascosMateosPlos2017}%
  \BibitemOpen
  \bibfield  {author} {\bibinfo {author} {\bibfnamefont {A.~P.}\ \bibnamefont
  {Riascos}}\ and\ \bibinfo {author} {\bibfnamefont {J.~L.}\ \bibnamefont
  {Mateos}},\ }\href {\doibase 10.1371/journal.pone.0184532} {\bibfield
  {journal} {\bibinfo  {journal} {PLoS One}\ }\textbf {\bibinfo {volume}
  {12}},\ \bibinfo {pages} {e0184532} (\bibinfo {year} {2017})}\BibitemShut
  {NoStop}%
\bibitem [{\citenamefont {Riascos}\ and\ \citenamefont
  {Mateos}(2012)}]{RiascosMateos2012}%
  \BibitemOpen
  \bibfield  {author} {\bibinfo {author} {\bibfnamefont {A.~P.}\ \bibnamefont
  {Riascos}}\ and\ \bibinfo {author} {\bibfnamefont {J.~L.}\ \bibnamefont
  {Mateos}},\ }\href {\doibase 10.1103/PhysRevE.86.056110} {\bibfield
  {journal} {\bibinfo  {journal} {Phys. Rev. E}\ }\textbf {\bibinfo {volume}
  {86}},\ \bibinfo {pages} {056110} (\bibinfo {year} {2012})}\BibitemShut
  {NoStop}%
\bibitem [{\citenamefont {Zhao}\ \emph {et~al.}(2014)\citenamefont {Zhao},
  \citenamefont {Weng},\ and\ \citenamefont {Huang}}]{ZhaoPhysA2014}%
  \BibitemOpen
  \bibfield  {author} {\bibinfo {author} {\bibfnamefont {Y.}~\bibnamefont
  {Zhao}}, \bibinfo {author} {\bibfnamefont {T.}~\bibnamefont {Weng}}, \ and\
  \bibinfo {author} {\bibfnamefont {D.}~\bibnamefont {Huang}},\ }\href
  {\doibase 10.1016/j.physa.2013.11.004} {\bibfield  {journal} {\bibinfo
  {journal} {Physica A: Stat. Mech. Appl.}\ }\textbf {\bibinfo {volume}
  {396}},\ \bibinfo {pages} {212} (\bibinfo {year} {2014})}\BibitemShut
  {NoStop}%
\bibitem [{\citenamefont {Huang}\ \emph {et~al.}(2014)\citenamefont {Huang},
  \citenamefont {Chen},\ and\ \citenamefont {Wang}}]{Huang2014132}%
  \BibitemOpen
  \bibfield  {author} {\bibinfo {author} {\bibfnamefont {W.}~\bibnamefont
  {Huang}}, \bibinfo {author} {\bibfnamefont {S.}~\bibnamefont {Chen}}, \ and\
  \bibinfo {author} {\bibfnamefont {W.}~\bibnamefont {Wang}},\ }\href {\doibase
  10.1016/j.physa.2013.09.014} {\bibfield  {journal} {\bibinfo  {journal}
  {Physica A: Stat. Mech. Appl.}\ }\textbf {\bibinfo {volume} {393}},\ \bibinfo
  {pages} {132} (\bibinfo {year} {2014})}\BibitemShut {NoStop}%
\bibitem [{\citenamefont {Weng}\ \emph {et~al.}(2015)\citenamefont {Weng},
  \citenamefont {Small}, \citenamefont {Zhang},\ and\ \citenamefont
  {Hui}}]{Weng2015}%
  \BibitemOpen
  \bibfield  {author} {\bibinfo {author} {\bibfnamefont {T.}~\bibnamefont
  {Weng}}, \bibinfo {author} {\bibfnamefont {M.}~\bibnamefont {Small}},
  \bibinfo {author} {\bibfnamefont {J.}~\bibnamefont {Zhang}}, \ and\ \bibinfo
  {author} {\bibfnamefont {P.}~\bibnamefont {Hui}},\ }\href {\doibase
  10.1038/srep17309} {\bibfield  {journal} {\bibinfo  {journal} {Sci. Rep.}\
  }\textbf {\bibinfo {volume} {5}},\ \bibinfo {pages} {17309} (\bibinfo {year}
  {2015})}\BibitemShut {NoStop}%
\bibitem [{\citenamefont {Weng}\ \emph {et~al.}(2016)\citenamefont {Weng},
  \citenamefont {Zhang}, \citenamefont {Khajehnejad}, \citenamefont {Small},
  \citenamefont {Zheng},\ and\ \citenamefont {Hui}}]{Weng2016}%
  \BibitemOpen
  \bibfield  {author} {\bibinfo {author} {\bibfnamefont {T.}~\bibnamefont
  {Weng}}, \bibinfo {author} {\bibfnamefont {J.}~\bibnamefont {Zhang}},
  \bibinfo {author} {\bibfnamefont {M.}~\bibnamefont {Khajehnejad}}, \bibinfo
  {author} {\bibfnamefont {M.}~\bibnamefont {Small}}, \bibinfo {author}
  {\bibfnamefont {R.}~\bibnamefont {Zheng}}, \ and\ \bibinfo {author}
  {\bibfnamefont {P.}~\bibnamefont {Hui}},\ }\href {\doibase 10.1038/srep37547}
  {\bibfield  {journal} {\bibinfo  {journal} {Sci. Rep.}\ }\textbf {\bibinfo
  {volume} {6}},\ \bibinfo {pages} {37547} (\bibinfo {year}
  {2016})}\BibitemShut {NoStop}%
\bibitem [{\citenamefont {Guo}\ \emph {et~al.}(2016)\citenamefont {Guo},
  \citenamefont {Cozzo}, \citenamefont {Zheng},\ and\ \citenamefont
  {Moreno}}]{Guo2016}%
  \BibitemOpen
  \bibfield  {author} {\bibinfo {author} {\bibfnamefont {Q.}~\bibnamefont
  {Guo}}, \bibinfo {author} {\bibfnamefont {E.}~\bibnamefont {Cozzo}}, \bibinfo
  {author} {\bibfnamefont {Z.}~\bibnamefont {Zheng}}, \ and\ \bibinfo {author}
  {\bibfnamefont {Y.}~\bibnamefont {Moreno}},\ }\href {\doibase
  10.1038/srep37641} {\bibfield  {journal} {\bibinfo  {journal} {Sci. Rep.}\
  }\textbf {\bibinfo {volume} {6}},\ \bibinfo {pages} {37641} (\bibinfo {year}
  {2016})}\BibitemShut {NoStop}%
\bibitem [{\citenamefont {Zheng}\ \emph {et~al.}(2017)\citenamefont {Zheng},
  \citenamefont {Xiao}, \citenamefont {Wang}, \citenamefont {Zhang},\ and\
  \citenamefont {Jiang}}]{Zheng2017}%
  \BibitemOpen
  \bibfield  {author} {\bibinfo {author} {\bibfnamefont {Z.}~\bibnamefont
  {Zheng}}, \bibinfo {author} {\bibfnamefont {G.}~\bibnamefont {Xiao}},
  \bibinfo {author} {\bibfnamefont {G.}~\bibnamefont {Wang}}, \bibinfo {author}
  {\bibfnamefont {G.}~\bibnamefont {Zhang}}, \ and\ \bibinfo {author}
  {\bibfnamefont {K.}~\bibnamefont {Jiang}},\ }\href {\doibase
  10.1155/2017/8217361} {\bibfield  {journal} {\bibinfo  {journal} {Math.
  Probl. Eng.}\ }\textbf {\bibinfo {volume} {2017}},\ \bibinfo {pages} {Article
  ID 8217361} (\bibinfo {year} {2017})}\BibitemShut {NoStop}%
\bibitem [{\citenamefont {Estrada}\ \emph {et~al.}(2018)\citenamefont
  {Estrada}, \citenamefont {Delvenne}, \citenamefont {Hatano}, \citenamefont
  {Mateos}, \citenamefont {Metzler}, \citenamefont {Riascos},\ and\
  \citenamefont {Schaub}}]{Estrada2017Multihopper}%
  \BibitemOpen
  \bibfield  {author} {\bibinfo {author} {\bibfnamefont {E.}~\bibnamefont
  {Estrada}}, \bibinfo {author} {\bibfnamefont {J.-C.}\ \bibnamefont
  {Delvenne}}, \bibinfo {author} {\bibfnamefont {N.}~\bibnamefont {Hatano}},
  \bibinfo {author} {\bibfnamefont {J.~L.}\ \bibnamefont {Mateos}}, \bibinfo
  {author} {\bibfnamefont {R.}~\bibnamefont {Metzler}}, \bibinfo {author}
  {\bibfnamefont {A.~P.}\ \bibnamefont {Riascos}}, \ and\ \bibinfo {author}
  {\bibfnamefont {M.~T.}\ \bibnamefont {Schaub}},\ }\href {\doibase
  10.1093/comnet/cnx043} {\bibfield  {journal} {\bibinfo  {journal} {J. Compl.
  Net.}\ }\textbf {\bibinfo {volume} {6}},\ \bibinfo {pages} {382} (\bibinfo
  {year} {2018})}\BibitemShut {NoStop}%
\bibitem [{\citenamefont {Riascos}\ and\ \citenamefont
  {Mateos}(2014)}]{RiascosMateosFD2014}%
  \BibitemOpen
  \bibfield  {author} {\bibinfo {author} {\bibfnamefont {A.~P.}\ \bibnamefont
  {Riascos}}\ and\ \bibinfo {author} {\bibfnamefont {J.~L.}\ \bibnamefont
  {Mateos}},\ }\href {\doibase 10.1103/PhysRevE.90.032809} {\bibfield
  {journal} {\bibinfo  {journal} {Phys. Rev. E}\ }\textbf {\bibinfo {volume}
  {90}},\ \bibinfo {pages} {032809} (\bibinfo {year} {2014})}\BibitemShut
  {NoStop}%
\bibitem [{\citenamefont {Riascos}\ and\ \citenamefont
  {Mateos}(2015{\natexlab{a}})}]{RiascosMateosFD2015}%
  \BibitemOpen
  \bibfield  {author} {\bibinfo {author} {\bibfnamefont {A.~P.}\ \bibnamefont
  {Riascos}}\ and\ \bibinfo {author} {\bibfnamefont {J.~L.}\ \bibnamefont
  {Mateos}},\ }\href {http://stacks.iop.org/1742-5468/2015/i=7/a=P07015}
  {\bibfield  {journal} {\bibinfo  {journal} {J. Stat. Mech.}\ }\textbf
  {\bibinfo {volume} {2015}},\ \bibinfo {pages} {P07015} (\bibinfo {year}
  {2015}{\natexlab{a}})}\BibitemShut {NoStop}%
\bibitem [{\citenamefont {Michelitsch}\ \emph
  {et~al.}(2016{\natexlab{a}})\citenamefont {Michelitsch}, \citenamefont
  {Collet}, \citenamefont {Riascos}, \citenamefont {Nowakowski},\ and\
  \citenamefont {Nicolleau}}]{Michelitsch2016Chaos}%
  \BibitemOpen
  \bibfield  {author} {\bibinfo {author} {\bibfnamefont {T.~M.}\ \bibnamefont
  {Michelitsch}}, \bibinfo {author} {\bibfnamefont {B.~A.}\ \bibnamefont
  {Collet}}, \bibinfo {author} {\bibfnamefont {A.~P.}\ \bibnamefont {Riascos}},
  \bibinfo {author} {\bibfnamefont {A.~F.}\ \bibnamefont {Nowakowski}}, \ and\
  \bibinfo {author} {\bibfnamefont {F.~C. G.~A.}\ \bibnamefont {Nicolleau}},\
  }\href {\doibase 10.1016/j.chaos.2016.09.009} {\bibfield  {journal} {\bibinfo
   {journal} {Chaos Solitons \& Fractals}\ }\textbf {\bibinfo {volume} {92}},\
  \bibinfo {pages} {43 } (\bibinfo {year} {2016}{\natexlab{a}})}\BibitemShut
  {NoStop}%
\bibitem [{\citenamefont {Michelitsch}\ \emph
  {et~al.}(2017{\natexlab{a}})\citenamefont {Michelitsch}, \citenamefont
  {Collet}, \citenamefont {Riascos}, \citenamefont {Nowakowski},\ and\
  \citenamefont {Nicolleau}}]{Michelitsch2017PhysA}%
  \BibitemOpen
  \bibfield  {author} {\bibinfo {author} {\bibfnamefont {T.~M.}\ \bibnamefont
  {Michelitsch}}, \bibinfo {author} {\bibfnamefont {B.~A.}\ \bibnamefont
  {Collet}}, \bibinfo {author} {\bibfnamefont {A.~P.}\ \bibnamefont {Riascos}},
  \bibinfo {author} {\bibfnamefont {A.~F.}\ \bibnamefont {Nowakowski}}, \ and\
  \bibinfo {author} {\bibfnamefont {F.~C. G.~A.}\ \bibnamefont {Nicolleau}},\
  }\href {\doibase 10.1088/1751-8121/aa5173} {\bibfield  {journal} {\bibinfo
  {journal} {J. Phys. A: Math. Theor.}\ }\textbf {\bibinfo {volume} {50}},\
  \bibinfo {pages} {055003} (\bibinfo {year} {2017}{\natexlab{a}})}\BibitemShut
  {NoStop}%
\bibitem [{\citenamefont {Michelitsch}\ \emph
  {et~al.}(2017{\natexlab{b}})\citenamefont {Michelitsch}, \citenamefont
  {Collet}, \citenamefont {Riascos}, \citenamefont {Nowakowski},\ and\
  \citenamefont {Nicolleau}}]{Michelitsch2017PhysARecurrence}%
  \BibitemOpen
  \bibfield  {author} {\bibinfo {author} {\bibfnamefont {T.~M.}\ \bibnamefont
  {Michelitsch}}, \bibinfo {author} {\bibfnamefont {B.~A.}\ \bibnamefont
  {Collet}}, \bibinfo {author} {\bibfnamefont {A.~P.}\ \bibnamefont {Riascos}},
  \bibinfo {author} {\bibfnamefont {A.~F.}\ \bibnamefont {Nowakowski}}, \ and\
  \bibinfo {author} {\bibfnamefont {F.~C. G.~A.}\ \bibnamefont {Nicolleau}},\
  }\href {\doibase 10.1088/1751-8121/aa9008} {\bibfield  {journal} {\bibinfo
  {journal} {J. Phys. A: Math. Theor.}\ }\textbf {\bibinfo {volume} {50}},\
  \bibinfo {pages} {505004} (\bibinfo {year} {2017}{\natexlab{b}})}\BibitemShut
  {NoStop}%
\bibitem [{\citenamefont {de~Nigris}\ \emph {et~al.}(2016)\citenamefont
  {de~Nigris}, \citenamefont {Hastir},\ and\ \citenamefont
  {Lambiotte}}]{deNigris2016}%
  \BibitemOpen
  \bibfield  {author} {\bibinfo {author} {\bibfnamefont {S.}~\bibnamefont
  {de~Nigris}}, \bibinfo {author} {\bibfnamefont {A.}~\bibnamefont {Hastir}}, \
  and\ \bibinfo {author} {\bibfnamefont {R.}~\bibnamefont {Lambiotte}},\ }\href
  {\doibase 10.1140/epjb/e2016-60947-3} {\bibfield  {journal} {\bibinfo
  {journal} {Eur. Phys. J. B}\ }\textbf {\bibinfo {volume} {89}},\ \bibinfo
  {pages} {114} (\bibinfo {year} {2016})}\BibitemShut {NoStop}%
\bibitem [{\citenamefont {de~Nigris}\ \emph
  {et~al.}(2017{\natexlab{a}})\citenamefont {de~Nigris}, \citenamefont
  {Carletti},\ and\ \citenamefont {Lambiotte}}]{DeNigris2017}%
  \BibitemOpen
  \bibfield  {author} {\bibinfo {author} {\bibfnamefont {S.}~\bibnamefont
  {de~Nigris}}, \bibinfo {author} {\bibfnamefont {T.}~\bibnamefont {Carletti}},
  \ and\ \bibinfo {author} {\bibfnamefont {R.}~\bibnamefont {Lambiotte}},\
  }\href {\doibase 10.1103/PhysRevE.95.022113} {\bibfield  {journal} {\bibinfo
  {journal} {Phys. Rev. E}\ }\textbf {\bibinfo {volume} {95}},\ \bibinfo
  {pages} {022113} (\bibinfo {year} {2017}{\natexlab{a}})}\BibitemShut
  {NoStop}%
\bibitem [{\citenamefont {de~Nigris}\ \emph
  {et~al.}(2017{\natexlab{b}})\citenamefont {de~Nigris}, \citenamefont
  {Bautista}, \citenamefont {Abry}, \citenamefont {Avrachenkov},\ and\
  \citenamefont {Goncalves}}]{SdeNigris2017}%
  \BibitemOpen
  \bibfield  {author} {\bibinfo {author} {\bibfnamefont {S.}~\bibnamefont
  {de~Nigris}}, \bibinfo {author} {\bibfnamefont {E.}~\bibnamefont {Bautista}},
  \bibinfo {author} {\bibfnamefont {P.}~\bibnamefont {Abry}}, \bibinfo {author}
  {\bibfnamefont {K.}~\bibnamefont {Avrachenkov}}, \ and\ \bibinfo {author}
  {\bibfnamefont {P.}~\bibnamefont {Goncalves}},\ }in\ \href {\doibase
  10.23919/EUSIPCO.2017.8081228} {\emph {\bibinfo {booktitle} {2017 25th
  European Signal Processing Conference (EUSIPCO)}}}\ (\bibinfo {year} {2017})\
  pp.\ \bibinfo {pages} {356--360}\BibitemShut {NoStop}%
\bibitem [{\citenamefont {Riascos}\ and\ \citenamefont
  {Mateos}(2015{\natexlab{b}})}]{RiascosMateos2015}%
  \BibitemOpen
  \bibfield  {author} {\bibinfo {author} {\bibfnamefont {A.~P.}\ \bibnamefont
  {Riascos}}\ and\ \bibinfo {author} {\bibfnamefont {J.~L.}\ \bibnamefont
  {Mateos}},\ }\href {\doibase 10.1103/PhysRevE.92.052814} {\bibfield
  {journal} {\bibinfo  {journal} {Phys. Rev. E}\ }\textbf {\bibinfo {volume}
  {92}},\ \bibinfo {pages} {052814} (\bibinfo {year}
  {2015}{\natexlab{b}})}\BibitemShut {NoStop}%
\bibitem [{\citenamefont {Riascos}\ \emph {et~al.}(2018)\citenamefont
  {Riascos}, \citenamefont {Michelitsch}, \citenamefont {Collet}, \citenamefont
  {Nowakowski},\ and\ \citenamefont {Nicolleau}}]{RiascosMichelitsch2017_gL}%
  \BibitemOpen
  \bibfield  {author} {\bibinfo {author} {\bibfnamefont {A.~P.}\ \bibnamefont
  {Riascos}}, \bibinfo {author} {\bibfnamefont {T.~M.}\ \bibnamefont
  {Michelitsch}}, \bibinfo {author} {\bibfnamefont {B.~A.}\ \bibnamefont
  {Collet}}, \bibinfo {author} {\bibfnamefont {A.~F.}\ \bibnamefont
  {Nowakowski}}, \ and\ \bibinfo {author} {\bibfnamefont {F.~C. G.~A.}\
  \bibnamefont {Nicolleau}},\ }\href
  {http://stacks.iop.org/1742-5468/2018/i=4/a=043404} {\bibfield  {journal}
  {\bibinfo  {journal} {J. Stat. Mech.}\ }\textbf {\bibinfo {volume} {2018}},\
  \bibinfo {pages} {043404} (\bibinfo {year} {2018})}\BibitemShut {NoStop}%
\bibitem [{\citenamefont {Weiss}(1994)}]{Weiss}%
  \BibitemOpen
  \bibfield  {author} {\bibinfo {author} {\bibfnamefont {G.~H.}\ \bibnamefont
  {Weiss}},\ }\href@noop {} {\emph {\bibinfo {title} {Aspects and Applications
  of the Random Walk}}}\ (\bibinfo  {publisher} {North-Holland},\ \bibinfo
  {address} {Amsterdam},\ \bibinfo {year} {1994})\BibitemShut {NoStop}%
\bibitem [{\citenamefont {Zhang}\ \emph {et~al.}(2013)\citenamefont {Zhang},
  \citenamefont {Shan},\ and\ \citenamefont {Chen}}]{ZhangPRE2013}%
  \BibitemOpen
  \bibfield  {author} {\bibinfo {author} {\bibfnamefont {Z.}~\bibnamefont
  {Zhang}}, \bibinfo {author} {\bibfnamefont {T.}~\bibnamefont {Shan}}, \ and\
  \bibinfo {author} {\bibfnamefont {G.}~\bibnamefont {Chen}},\ }\href {\doibase
  10.1103/PhysRevE.87.012112} {\bibfield  {journal} {\bibinfo  {journal} {Phys.
  Rev. E}\ }\textbf {\bibinfo {volume} {87}},\ \bibinfo {pages} {012112}
  (\bibinfo {year} {2013})}\BibitemShut {NoStop}%
\bibitem [{\citenamefont {Condamin}\ \emph {et~al.}(2007)\citenamefont
  {Condamin}, \citenamefont {B\'enichou},\ and\ \citenamefont
  {Moreau}}]{Condamin2007}%
  \BibitemOpen
  \bibfield  {author} {\bibinfo {author} {\bibfnamefont {S.}~\bibnamefont
  {Condamin}}, \bibinfo {author} {\bibfnamefont {O.}~\bibnamefont
  {B\'enichou}}, \ and\ \bibinfo {author} {\bibfnamefont {M.}~\bibnamefont
  {Moreau}},\ }\href {\doibase 10.1103/PhysRevE.75.021111} {\bibfield
  {journal} {\bibinfo  {journal} {Phys. Rev. E}\ }\textbf {\bibinfo {volume}
  {75}},\ \bibinfo {pages} {021111} (\bibinfo {year} {2007})}\BibitemShut
  {NoStop}%
\bibitem [{\citenamefont {Telcs}(1989)}]{Telcs1989}%
  \BibitemOpen
  \bibfield  {author} {\bibinfo {author} {\bibfnamefont {A.}~\bibnamefont
  {Telcs}},\ }\href {\doibase 10.1007/BF00339997} {\bibfield  {journal}
  {\bibinfo  {journal} {Probab. Theory Relat. Fields}\ }\textbf {\bibinfo
  {volume} {82}},\ \bibinfo {pages} {435} (\bibinfo {year} {1989})}\BibitemShut
  {NoStop}%
\bibitem [{\citenamefont {Yang}(2005)}]{YangPRE2005}%
  \BibitemOpen
  \bibfield  {author} {\bibinfo {author} {\bibfnamefont {S.-J.}\ \bibnamefont
  {Yang}},\ }\href {\doibase 10.1103/PhysRevE.71.016107} {\bibfield  {journal}
  {\bibinfo  {journal} {Phys. Rev. E}\ }\textbf {\bibinfo {volume} {71}},\
  \bibinfo {pages} {016107} (\bibinfo {year} {2005})}\BibitemShut {NoStop}%
\bibitem [{\citenamefont {Sanders}(2009)}]{SandersPRE2009}%
  \BibitemOpen
  \bibfield  {author} {\bibinfo {author} {\bibfnamefont {D.~P.}\ \bibnamefont
  {Sanders}},\ }\href {\doibase 10.1103/PhysRevE.80.036119} {\bibfield
  {journal} {\bibinfo  {journal} {Phys. Rev. E}\ }\textbf {\bibinfo {volume}
  {80}},\ \bibinfo {pages} {036119} (\bibinfo {year} {2009})}\BibitemShut
  {NoStop}%
\bibitem [{\citenamefont {Kishore}\ \emph {et~al.}(2011)\citenamefont
  {Kishore}, \citenamefont {Santhanam},\ and\ \citenamefont
  {Amritkar}}]{KishorePRE2011}%
  \BibitemOpen
  \bibfield  {author} {\bibinfo {author} {\bibfnamefont {V.}~\bibnamefont
  {Kishore}}, \bibinfo {author} {\bibfnamefont {M.~S.}\ \bibnamefont
  {Santhanam}}, \ and\ \bibinfo {author} {\bibfnamefont {R.~E.}\ \bibnamefont
  {Amritkar}},\ }\href {\doibase 10.1103/PhysRevLett.106.188701} {\bibfield
  {journal} {\bibinfo  {journal} {Phys. Rev. Lett.}\ }\textbf {\bibinfo
  {volume} {106}},\ \bibinfo {pages} {188701} (\bibinfo {year}
  {2011})}\BibitemShut {NoStop}%
\bibitem [{\citenamefont {Meyer}\ \emph {et~al.}(2012)\citenamefont {Meyer},
  \citenamefont {Agliari}, \citenamefont {B\'enichou},\ and\ \citenamefont
  {Voituriez}}]{AgliariBenichouPRE2012}%
  \BibitemOpen
  \bibfield  {author} {\bibinfo {author} {\bibfnamefont {B.}~\bibnamefont
  {Meyer}}, \bibinfo {author} {\bibfnamefont {E.}~\bibnamefont {Agliari}},
  \bibinfo {author} {\bibfnamefont {O.}~\bibnamefont {B\'enichou}}, \ and\
  \bibinfo {author} {\bibfnamefont {R.}~\bibnamefont {Voituriez}},\ }\href
  {\doibase 10.1103/PhysRevE.85.026113} {\bibfield  {journal} {\bibinfo
  {journal} {Phys. Rev. E}\ }\textbf {\bibinfo {volume} {85}},\ \bibinfo
  {pages} {026113} (\bibinfo {year} {2012})}\BibitemShut {NoStop}%
\bibitem [{\citenamefont {Wang}\ \emph {et~al.}(2006)\citenamefont {Wang},
  \citenamefont {Wang}, \citenamefont {Yin}, \citenamefont {Xie},\ and\
  \citenamefont {Zhou}}]{WangPRE2006}%
  \BibitemOpen
  \bibfield  {author} {\bibinfo {author} {\bibfnamefont {W.-X.}\ \bibnamefont
  {Wang}}, \bibinfo {author} {\bibfnamefont {B.-H.}\ \bibnamefont {Wang}},
  \bibinfo {author} {\bibfnamefont {C.-Y.}\ \bibnamefont {Yin}}, \bibinfo
  {author} {\bibfnamefont {Y.-B.}\ \bibnamefont {Xie}}, \ and\ \bibinfo
  {author} {\bibfnamefont {T.}~\bibnamefont {Zhou}},\ }\href {\doibase
  10.1103/PhysRevE.73.026111} {\bibfield  {journal} {\bibinfo  {journal} {Phys.
  Rev. E}\ }\textbf {\bibinfo {volume} {73}},\ \bibinfo {pages} {026111}
  (\bibinfo {year} {2006})}\BibitemShut {NoStop}%
\bibitem [{\citenamefont {Kwon}\ \emph {et~al.}(2010)\citenamefont {Kwon},
  \citenamefont {Choi},\ and\ \citenamefont {Kim}}]{KwonPRE2010}%
  \BibitemOpen
  \bibfield  {author} {\bibinfo {author} {\bibfnamefont {S.}~\bibnamefont
  {Kwon}}, \bibinfo {author} {\bibfnamefont {W.}~\bibnamefont {Choi}}, \ and\
  \bibinfo {author} {\bibfnamefont {Y.}~\bibnamefont {Kim}},\ }\href {\doibase
  10.1103/PhysRevE.82.021108} {\bibfield  {journal} {\bibinfo  {journal} {Phys.
  Rev. E}\ }\textbf {\bibinfo {volume} {82}},\ \bibinfo {pages} {021108}
  (\bibinfo {year} {2010})}\BibitemShut {NoStop}%
\bibitem [{\citenamefont {Kishore}\ \emph {et~al.}(2012)\citenamefont
  {Kishore}, \citenamefont {Santhanam},\ and\ \citenamefont
  {Amritkar}}]{KishorePRE2012}%
  \BibitemOpen
  \bibfield  {author} {\bibinfo {author} {\bibfnamefont {V.}~\bibnamefont
  {Kishore}}, \bibinfo {author} {\bibfnamefont {M.~S.}\ \bibnamefont
  {Santhanam}}, \ and\ \bibinfo {author} {\bibfnamefont {R.~E.}\ \bibnamefont
  {Amritkar}},\ }\href {\doibase 10.1103/PhysRevE.85.056120} {\bibfield
  {journal} {\bibinfo  {journal} {Phys. Rev. E}\ }\textbf {\bibinfo {volume}
  {85}},\ \bibinfo {pages} {056120} (\bibinfo {year} {2012})}\BibitemShut
  {NoStop}%
\bibitem [{\citenamefont {Ling}\ \emph {et~al.}(2013)\citenamefont {Ling},
  \citenamefont {Hu}, \citenamefont {Ding}, \citenamefont {Shi},\ and\
  \citenamefont {Jiang}}]{LingEPJB2013}%
  \BibitemOpen
  \bibfield  {author} {\bibinfo {author} {\bibfnamefont {X.}~\bibnamefont
  {Ling}}, \bibinfo {author} {\bibfnamefont {M.-B.}\ \bibnamefont {Hu}},
  \bibinfo {author} {\bibfnamefont {J.-X.}\ \bibnamefont {Ding}}, \bibinfo
  {author} {\bibfnamefont {Q.}~\bibnamefont {Shi}}, \ and\ \bibinfo {author}
  {\bibfnamefont {R.}~\bibnamefont {Jiang}},\ }\href
  {https://doi.org/10.1140/epjb/e2013-30409-9} {\bibfield  {journal} {\bibinfo
  {journal} {Eur. Phys. J. B}\ }\textbf {\bibinfo {volume} {86}},\ \bibinfo
  {eid} {146} (\bibinfo {year} {2013})}\BibitemShut {NoStop}%
\bibitem [{\citenamefont {Lambiotte}\ \emph {et~al.}(2011)\citenamefont
  {Lambiotte}, \citenamefont {Sinatra}, \citenamefont {Delvenne}, \citenamefont
  {Evans}, \citenamefont {Barahona},\ and\ \citenamefont
  {Latora}}]{LambiottePRE2011}%
  \BibitemOpen
  \bibfield  {author} {\bibinfo {author} {\bibfnamefont {R.}~\bibnamefont
  {Lambiotte}}, \bibinfo {author} {\bibfnamefont {R.}~\bibnamefont {Sinatra}},
  \bibinfo {author} {\bibfnamefont {J.-C.}\ \bibnamefont {Delvenne}}, \bibinfo
  {author} {\bibfnamefont {T.~S.}\ \bibnamefont {Evans}}, \bibinfo {author}
  {\bibfnamefont {M.}~\bibnamefont {Barahona}}, \ and\ \bibinfo {author}
  {\bibfnamefont {V.}~\bibnamefont {Latora}},\ }\href {\doibase
  10.1103/PhysRevE.84.017102} {\bibfield  {journal} {\bibinfo  {journal} {Phys.
  Rev. E}\ }\textbf {\bibinfo {volume} {84}},\ \bibinfo {pages} {017102}
  (\bibinfo {year} {2011})}\BibitemShut {NoStop}%
\bibitem [{\citenamefont {Battiston}\ \emph {et~al.}(2016)\citenamefont
  {Battiston}, \citenamefont {Nicosia},\ and\ \citenamefont
  {Latora}}]{Battiston2016}%
  \BibitemOpen
  \bibfield  {author} {\bibinfo {author} {\bibfnamefont {F.}~\bibnamefont
  {Battiston}}, \bibinfo {author} {\bibfnamefont {V.}~\bibnamefont {Nicosia}},
  \ and\ \bibinfo {author} {\bibfnamefont {V.}~\bibnamefont {Latora}},\ }\href
  {http://stacks.iop.org/1367-2630/18/i=4/a=043035} {\bibfield  {journal}
  {\bibinfo  {journal} {New J. Phys.}\ }\textbf {\bibinfo {volume} {18}},\
  \bibinfo {pages} {043035} (\bibinfo {year} {2016})}\BibitemShut {NoStop}%
\bibitem [{\citenamefont {Zhang}\ \emph
  {et~al.}(2011{\natexlab{a}})\citenamefont {Zhang}, \citenamefont {Zhang},
  \citenamefont {Guan},\ and\ \citenamefont {Zhou}}]{ZhangJSMTE2011}%
  \BibitemOpen
  \bibfield  {author} {\bibinfo {author} {\bibfnamefont {Y.}~\bibnamefont
  {Zhang}}, \bibinfo {author} {\bibfnamefont {Z.}~\bibnamefont {Zhang}},
  \bibinfo {author} {\bibfnamefont {J.}~\bibnamefont {Guan}}, \ and\ \bibinfo
  {author} {\bibfnamefont {S.}~\bibnamefont {Zhou}},\ }\href
  {http://stacks.iop.org/1742-5468/2011/i=10/a=P10001} {\bibfield  {journal}
  {\bibinfo  {journal} {J. Stat. Mech.}\ }\textbf {\bibinfo {volume} {2011}},\
  \bibinfo {pages} {P10001} (\bibinfo {year} {2011}{\natexlab{a}})}\BibitemShut
  {NoStop}%
\bibitem [{\citenamefont {Burda}\ \emph {et~al.}(2009)\citenamefont {Burda},
  \citenamefont {Duda}, \citenamefont {Luck},\ and\ \citenamefont
  {Waclaw}}]{BurdaPRL2009}%
  \BibitemOpen
  \bibfield  {author} {\bibinfo {author} {\bibfnamefont {Z.}~\bibnamefont
  {Burda}}, \bibinfo {author} {\bibfnamefont {J.}~\bibnamefont {Duda}},
  \bibinfo {author} {\bibfnamefont {J.~M.}\ \bibnamefont {Luck}}, \ and\
  \bibinfo {author} {\bibfnamefont {B.}~\bibnamefont {Waclaw}},\ }\href
  {\doibase 10.1103/PhysRevLett.102.160602} {\bibfield  {journal} {\bibinfo
  {journal} {Phys. Rev. Lett.}\ }\textbf {\bibinfo {volume} {102}},\ \bibinfo
  {pages} {160602} (\bibinfo {year} {2009})}\BibitemShut {NoStop}%
\bibitem [{\citenamefont {Sinatra}\ \emph {et~al.}(2011)\citenamefont
  {Sinatra}, \citenamefont {G\'omez-Garde\~nes}, \citenamefont {Lambiotte},
  \citenamefont {Nicosia},\ and\ \citenamefont {Latora}}]{SinatraPRE2011}%
  \BibitemOpen
  \bibfield  {author} {\bibinfo {author} {\bibfnamefont {R.}~\bibnamefont
  {Sinatra}}, \bibinfo {author} {\bibfnamefont {J.}~\bibnamefont
  {G\'omez-Garde\~nes}}, \bibinfo {author} {\bibfnamefont {R.}~\bibnamefont
  {Lambiotte}}, \bibinfo {author} {\bibfnamefont {V.}~\bibnamefont {Nicosia}},
  \ and\ \bibinfo {author} {\bibfnamefont {V.}~\bibnamefont {Latora}},\ }\href
  {\doibase 10.1103/PhysRevE.83.030103} {\bibfield  {journal} {\bibinfo
  {journal} {Phys. Rev. E}\ }\textbf {\bibinfo {volume} {83}},\ \bibinfo
  {pages} {030103} (\bibinfo {year} {2011})}\BibitemShut {NoStop}%
\bibitem [{\citenamefont {Ochab}(2012)}]{OchabPRE2012}%
  \BibitemOpen
  \bibfield  {author} {\bibinfo {author} {\bibfnamefont {J.~K.}\ \bibnamefont
  {Ochab}},\ }\href {\doibase 10.1103/PhysRevE.86.066109} {\bibfield  {journal}
  {\bibinfo  {journal} {Phys. Rev. E}\ }\textbf {\bibinfo {volume} {86}},\
  \bibinfo {pages} {066109} (\bibinfo {year} {2012})}\BibitemShut {NoStop}%
\bibitem [{\citenamefont {Frank}\ and\ \citenamefont
  {Galinsky}(2014)}]{LawrencePRE2014}%
  \BibitemOpen
  \bibfield  {author} {\bibinfo {author} {\bibfnamefont {L.~R.}\ \bibnamefont
  {Frank}}\ and\ \bibinfo {author} {\bibfnamefont {V.~L.}\ \bibnamefont
  {Galinsky}},\ }\href {\doibase 10.1103/PhysRevE.89.032142} {\bibfield
  {journal} {\bibinfo  {journal} {Phys. Rev. E}\ }\textbf {\bibinfo {volume}
  {89}},\ \bibinfo {pages} {032142} (\bibinfo {year} {2014})}\BibitemShut
  {NoStop}%
\bibitem [{\citenamefont {Lin}\ and\ \citenamefont
  {Zhang}(2014)}]{YuanSciRep2014}%
  \BibitemOpen
  \bibfield  {author} {\bibinfo {author} {\bibfnamefont {Y.}~\bibnamefont
  {Lin}}\ and\ \bibinfo {author} {\bibfnamefont {Z.}~\bibnamefont {Zhang}},\
  }\href {http://dx.doi.org/10.1038/srep05365} {\bibfield  {journal} {\bibinfo
  {journal} {Sci. Rep.}\ }\textbf {\bibinfo {volume} {4}},\ \bibinfo {eid}
  {5365} (\bibinfo {year} {2014})}\BibitemShut {NoStop}%
\bibitem [{\citenamefont {Martin}\ \emph {et~al.}(2001)\citenamefont {Martin},
  \citenamefont {Fowlkes}, \citenamefont {Tal},\ and\ \citenamefont
  {Malik}}]{DataBase2001}%
  \BibitemOpen
  \bibfield  {author} {\bibinfo {author} {\bibfnamefont {D.}~\bibnamefont
  {Martin}}, \bibinfo {author} {\bibfnamefont {C.}~\bibnamefont {Fowlkes}},
  \bibinfo {author} {\bibfnamefont {D.}~\bibnamefont {Tal}}, \ and\ \bibinfo
  {author} {\bibfnamefont {J.}~\bibnamefont {Malik}},\ }in\ \href {\doibase
  10.1109/ICCV.2001.937655} {\emph {\bibinfo {booktitle} {Computer Vision,
  2001. ICCV 2001. Proceedings. Eighth IEEE International Conference on}}},\
  Vol.~\bibinfo {volume} {2}\ (\bibinfo {year} {2001})\ pp.\ \bibinfo {pages}
  {416--423 vol.2}\BibitemShut {NoStop}%
\bibitem [{\citenamefont {Grady}(2006)}]{GradyIEEE2006}%
  \BibitemOpen
  \bibfield  {author} {\bibinfo {author} {\bibfnamefont {L.}~\bibnamefont
  {Grady}},\ }\href {\doibase 10.1109/TPAMI.2006.233} {\bibfield  {journal}
  {\bibinfo  {journal} {IEEE Trans. Pattern Anal. Mach. Intell.}\ }\textbf
  {\bibinfo {volume} {28}},\ \bibinfo {pages} {1768} (\bibinfo {year}
  {2006})}\BibitemShut {NoStop}%
\bibitem [{\citenamefont {Zlati\ifmmode~\acute{c}\else \'{c}\fi{}}\ \emph
  {et~al.}(2010)\citenamefont {Zlati\ifmmode~\acute{c}\else \'{c}\fi{}},
  \citenamefont {Gabrielli},\ and\ \citenamefont {Caldarelli}}]{ZlaticPRE2010}%
  \BibitemOpen
  \bibfield  {author} {\bibinfo {author} {\bibfnamefont {V.}~\bibnamefont
  {Zlati\ifmmode~\acute{c}\else \'{c}\fi{}}}, \bibinfo {author} {\bibfnamefont
  {A.}~\bibnamefont {Gabrielli}}, \ and\ \bibinfo {author} {\bibfnamefont
  {G.}~\bibnamefont {Caldarelli}},\ }\href {\doibase
  10.1103/PhysRevE.82.066109} {\bibfield  {journal} {\bibinfo  {journal} {Phys.
  Rev. E}\ }\textbf {\bibinfo {volume} {82}},\ \bibinfo {pages} {066109}
  (\bibinfo {year} {2010})}\BibitemShut {NoStop}%
\bibitem [{\citenamefont {Sinop}\ and\ \citenamefont
  {Grady}(2007)}]{GradyIEEE2007}%
  \BibitemOpen
  \bibfield  {author} {\bibinfo {author} {\bibfnamefont {A.}~\bibnamefont
  {Sinop}}\ and\ \bibinfo {author} {\bibfnamefont {L.}~\bibnamefont {Grady}},\
  }in\ \href {\doibase 10.1109/ICCV.2007.4408927} {\emph {\bibinfo {booktitle}
  {Computer Vision, 2007. ICCV 2007. IEEE 11th International Conference on}}}\
  (\bibinfo {year} {2007})\ pp.\ \bibinfo {pages} {1--8}\BibitemShut {NoStop}%
\bibitem [{\citenamefont {Brin}\ and\ \citenamefont {Page}(1998)}]{Brin1998}%
  \BibitemOpen
  \bibfield  {author} {\bibinfo {author} {\bibfnamefont {S.}~\bibnamefont
  {Brin}}\ and\ \bibinfo {author} {\bibfnamefont {L.}~\bibnamefont {Page}},\
  }\href {\doibase 10.1016/S0169-7552(98)00110-X} {\bibfield  {journal}
  {\bibinfo  {journal} {Comput. Netw. ISDN Syst.}\ }\textbf {\bibinfo {volume}
  {30}},\ \bibinfo {pages} {107} (\bibinfo {year} {1998})}\BibitemShut
  {NoStop}%
\bibitem [{\citenamefont {Metzler}\ and\ \citenamefont
  {Klafter}(2004)}]{Ralf2004}%
  \BibitemOpen
  \bibfield  {author} {\bibinfo {author} {\bibfnamefont {R.}~\bibnamefont
  {Metzler}}\ and\ \bibinfo {author} {\bibfnamefont {J.}~\bibnamefont
  {Klafter}},\ }\href {http://stacks.iop.org/0305-4470/37/i=31/a=R01}
  {\bibfield  {journal} {\bibinfo  {journal} {J. Phys. A: Math. Gen.}\ }\textbf
  {\bibinfo {volume} {37}},\ \bibinfo {pages} {R161} (\bibinfo {year}
  {2004})}\BibitemShut {NoStop}%
\bibitem [{\citenamefont {Zaburdaev}\ \emph {et~al.}(2015)\citenamefont
  {Zaburdaev}, \citenamefont {Denisov},\ and\ \citenamefont
  {Klafter}}]{RevModPhysZaburdaev2015}%
  \BibitemOpen
  \bibfield  {author} {\bibinfo {author} {\bibfnamefont {V.}~\bibnamefont
  {Zaburdaev}}, \bibinfo {author} {\bibfnamefont {S.}~\bibnamefont {Denisov}},
  \ and\ \bibinfo {author} {\bibfnamefont {J.}~\bibnamefont {Klafter}},\ }\href
  {\doibase 10.1103/RevModPhys.87.483} {\bibfield  {journal} {\bibinfo
  {journal} {Rev. Mod. Phys.}\ }\textbf {\bibinfo {volume} {87}},\ \bibinfo
  {pages} {483} (\bibinfo {year} {2015})}\BibitemShut {NoStop}%
\bibitem [{\citenamefont {Ramos-Fern\'{a}ndez}\ \emph
  {et~al.}(2004)\citenamefont {Ramos-Fern\'{a}ndez}, \citenamefont {Mateos},
  \citenamefont {Miramontes}, \citenamefont {Cocho}, \citenamefont {Larralde},\
  and\ \citenamefont {Ayala-Orozco}}]{BES2004}%
  \BibitemOpen
  \bibfield  {author} {\bibinfo {author} {\bibfnamefont {G.}~\bibnamefont
  {Ramos-Fern\'{a}ndez}}, \bibinfo {author} {\bibfnamefont {J.~L.}\
  \bibnamefont {Mateos}}, \bibinfo {author} {\bibfnamefont {O.}~\bibnamefont
  {Miramontes}}, \bibinfo {author} {\bibfnamefont {G.}~\bibnamefont {Cocho}},
  \bibinfo {author} {\bibfnamefont {H.}~\bibnamefont {Larralde}}, \ and\
  \bibinfo {author} {\bibfnamefont {B.}~\bibnamefont {Ayala-Orozco}},\ }\href
  {http://dx.doi.org/10.1007/s00265-003-0700-6} {\bibfield  {journal} {\bibinfo
   {journal} {Behav. Ecol. Sociobiol.}\ }\textbf {\bibinfo {volume} {55}},\
  \bibinfo {pages} {223} (\bibinfo {year} {2004})}\BibitemShut {NoStop}%
\bibitem [{\citenamefont {Boyer}\ \emph {et~al.}(2006)\citenamefont {Boyer},
  \citenamefont {Ramos-Fern\'andez}, \citenamefont {Miramontes}, \citenamefont
  {Mateos}, \citenamefont {Cocho}, \citenamefont {Larralde}, \citenamefont
  {Ramos},\ and\ \citenamefont {Rojas}}]{Proc2006}%
  \BibitemOpen
  \bibfield  {author} {\bibinfo {author} {\bibfnamefont {D.}~\bibnamefont
  {Boyer}}, \bibinfo {author} {\bibfnamefont {G.}~\bibnamefont
  {Ramos-Fern\'andez}}, \bibinfo {author} {\bibfnamefont {O.}~\bibnamefont
  {Miramontes}}, \bibinfo {author} {\bibfnamefont {J.~L.}\ \bibnamefont
  {Mateos}}, \bibinfo {author} {\bibfnamefont {G.}~\bibnamefont {Cocho}},
  \bibinfo {author} {\bibfnamefont {H.}~\bibnamefont {Larralde}}, \bibinfo
  {author} {\bibfnamefont {H.}~\bibnamefont {Ramos}}, \ and\ \bibinfo {author}
  {\bibfnamefont {F.}~\bibnamefont {Rojas}},\ }\href {\doibase
  10.1098/rspb.2005.3462} {\bibfield  {journal} {\bibinfo  {journal} {Proc. R.
  Soc. B}\ }\textbf {\bibinfo {volume} {273}},\ \bibinfo {pages} {1743}
  (\bibinfo {year} {2006})}\BibitemShut {NoStop}%
\bibitem [{\citenamefont {Boyer}\ \emph {et~al.}(2012)\citenamefont {Boyer},
  \citenamefont {Crofoot},\ and\ \citenamefont {Walsh}}]{BoyerJRSI}%
  \BibitemOpen
  \bibfield  {author} {\bibinfo {author} {\bibfnamefont {D.}~\bibnamefont
  {Boyer}}, \bibinfo {author} {\bibfnamefont {M.~C.}\ \bibnamefont {Crofoot}},
  \ and\ \bibinfo {author} {\bibfnamefont {P.~D.}\ \bibnamefont {Walsh}},\
  }\href {\doibase 10.1098/rsif.2011.0582} {\bibfield  {journal} {\bibinfo
  {journal} {J. R. Soc. Interface}\ }\textbf {\bibinfo {volume} {9}},\ \bibinfo
  {pages} {842} (\bibinfo {year} {2012})}\BibitemShut {NoStop}%
\bibitem [{\citenamefont {Brockmann}\ \emph {et~al.}(2006)\citenamefont
  {Brockmann}, \citenamefont {Hufnagel},\ and\ \citenamefont
  {Geisel}}]{Brock2006}%
  \BibitemOpen
  \bibfield  {author} {\bibinfo {author} {\bibfnamefont {D.}~\bibnamefont
  {Brockmann}}, \bibinfo {author} {\bibfnamefont {L.}~\bibnamefont {Hufnagel}},
  \ and\ \bibinfo {author} {\bibfnamefont {T.}~\bibnamefont {Geisel}},\ }\href
  {\doibase 10.1038/nature04292} {\bibfield  {journal} {\bibinfo  {journal}
  {Nature (London)}\ }\textbf {\bibinfo {volume} {439}},\ \bibinfo {pages}
  {462} (\bibinfo {year} {2006})}\BibitemShut {NoStop}%
\bibitem [{\citenamefont {Brown}\ \emph {et~al.}(2007)\citenamefont {Brown},
  \citenamefont {Liebovitch},\ and\ \citenamefont {Glendon}}]{Brown}%
  \BibitemOpen
  \bibfield  {author} {\bibinfo {author} {\bibfnamefont {C.}~\bibnamefont
  {Brown}}, \bibinfo {author} {\bibfnamefont {L.}~\bibnamefont {Liebovitch}}, \
  and\ \bibinfo {author} {\bibfnamefont {R.}~\bibnamefont {Glendon}},\ }\href
  {\doibase 10.1007/s10745-006-9083-4} {\bibfield  {journal} {\bibinfo
  {journal} {Hum. Ecol.}\ }\textbf {\bibinfo {volume} {35}},\ \bibinfo {pages}
  {129} (\bibinfo {year} {2007})}\BibitemShut {NoStop}%
\bibitem [{\citenamefont {Rhee}\ \emph {et~al.}(2011)\citenamefont {Rhee},
  \citenamefont {Shin}, \citenamefont {Hong}, \citenamefont {Lee},
  \citenamefont {Kim},\ and\ \citenamefont {Chong}}]{Rhee}%
  \BibitemOpen
  \bibfield  {author} {\bibinfo {author} {\bibfnamefont {I.}~\bibnamefont
  {Rhee}}, \bibinfo {author} {\bibfnamefont {M.}~\bibnamefont {Shin}}, \bibinfo
  {author} {\bibfnamefont {S.}~\bibnamefont {Hong}}, \bibinfo {author}
  {\bibfnamefont {K.}~\bibnamefont {Lee}}, \bibinfo {author} {\bibfnamefont
  {S.~J.}\ \bibnamefont {Kim}}, \ and\ \bibinfo {author} {\bibfnamefont
  {S.}~\bibnamefont {Chong}},\ }\href {\doibase 10.1109/TNET.2011.2120618}
  {\bibfield  {journal} {\bibinfo  {journal} {IEEE/ACM Trans. Netw.}\ }\textbf
  {\bibinfo {volume} {19}},\ \bibinfo {pages} {630} (\bibinfo {year}
  {2011})}\BibitemShut {NoStop}%
\bibitem [{\citenamefont {Metzler}\ and\ \citenamefont
  {Klafter}(2000)}]{AnomalousDiff}%
  \BibitemOpen
  \bibfield  {author} {\bibinfo {author} {\bibfnamefont {R.}~\bibnamefont
  {Metzler}}\ and\ \bibinfo {author} {\bibfnamefont {J.}~\bibnamefont
  {Klafter}},\ }\href {\doibase 10.1016/S0370-1573(00)00070-3} {\bibfield
  {journal} {\bibinfo  {journal} {Phys. Rep.}\ }\textbf {\bibinfo {volume}
  {339}},\ \bibinfo {pages} {1} (\bibinfo {year} {2000})}\BibitemShut {NoStop}%
\bibitem [{\citenamefont {Barth\'elemy}(2011)}]{Barthelemy2011}%
  \BibitemOpen
  \bibfield  {author} {\bibinfo {author} {\bibfnamefont {M.}~\bibnamefont
  {Barth\'elemy}},\ }\href {\doibase 10.1016/j.physrep.2010.11.002} {\bibfield
  {journal} {\bibinfo  {journal} {Phys. Rep.}\ }\textbf {\bibinfo {volume}
  {499}},\ \bibinfo {pages} {1 } (\bibinfo {year} {2011})}\BibitemShut
  {NoStop}%
\bibitem [{\citenamefont {Barbosa}\ \emph {et~al.}(2018)\citenamefont
  {Barbosa}, \citenamefont {Barthelemy}, \citenamefont {Ghoshal}, \citenamefont
  {James}, \citenamefont {Lenormand}, \citenamefont {Louail}, \citenamefont
  {Menezes}, \citenamefont {Ramasco}, \citenamefont {Simini},\ and\
  \citenamefont {Tomasini}}]{Barbosa2018}%
  \BibitemOpen
  \bibfield  {author} {\bibinfo {author} {\bibfnamefont {H.}~\bibnamefont
  {Barbosa}}, \bibinfo {author} {\bibfnamefont {M.}~\bibnamefont {Barthelemy}},
  \bibinfo {author} {\bibfnamefont {G.}~\bibnamefont {Ghoshal}}, \bibinfo
  {author} {\bibfnamefont {C.~R.}\ \bibnamefont {James}}, \bibinfo {author}
  {\bibfnamefont {M.}~\bibnamefont {Lenormand}}, \bibinfo {author}
  {\bibfnamefont {T.}~\bibnamefont {Louail}}, \bibinfo {author} {\bibfnamefont
  {R.}~\bibnamefont {Menezes}}, \bibinfo {author} {\bibfnamefont {J.~J.}\
  \bibnamefont {Ramasco}}, \bibinfo {author} {\bibfnamefont {F.}~\bibnamefont
  {Simini}}, \ and\ \bibinfo {author} {\bibfnamefont {M.}~\bibnamefont
  {Tomasini}},\ }\href {\doibase 10.1016/j.physrep.2018.01.001} {\bibfield
  {journal} {\bibinfo  {journal} {Phys. Rep.}\ }\textbf {\bibinfo {volume}
  {734}},\ \bibinfo {pages} {1 } (\bibinfo {year} {2018})}\BibitemShut
  {NoStop}%
\bibitem [{\citenamefont {Liu}\ \emph {et~al.}(2014)\citenamefont {Liu},
  \citenamefont {Wang}, \citenamefont {Kang}, \citenamefont {Gao},\ and\
  \citenamefont {Lu}}]{YuGIS2014}%
  \BibitemOpen
  \bibfield  {author} {\bibinfo {author} {\bibfnamefont {Y.}~\bibnamefont
  {Liu}}, \bibinfo {author} {\bibfnamefont {F.}~\bibnamefont {Wang}}, \bibinfo
  {author} {\bibfnamefont {C.}~\bibnamefont {Kang}}, \bibinfo {author}
  {\bibfnamefont {Y.}~\bibnamefont {Gao}}, \ and\ \bibinfo {author}
  {\bibfnamefont {Y.}~\bibnamefont {Lu}},\ }\href {\doibase 10.1111/tgis.12023}
  {\bibfield  {journal} {\bibinfo  {journal} {Transactions in GIS}\ }\textbf
  {\bibinfo {volume} {18}},\ \bibinfo {pages} {89} (\bibinfo {year}
  {2014})}\BibitemShut {NoStop}%
\bibitem [{\citenamefont {Liben-Nowell}\ \emph {et~al.}(2005)\citenamefont
  {Liben-Nowell}, \citenamefont {Novak}, \citenamefont {Kumar}, \citenamefont
  {Raghavan},\ and\ \citenamefont {Tomkins}}]{Liben05}%
  \BibitemOpen
  \bibfield  {author} {\bibinfo {author} {\bibfnamefont {D.}~\bibnamefont
  {Liben-Nowell}}, \bibinfo {author} {\bibfnamefont {J.}~\bibnamefont {Novak}},
  \bibinfo {author} {\bibfnamefont {R.}~\bibnamefont {Kumar}}, \bibinfo
  {author} {\bibfnamefont {P.}~\bibnamefont {Raghavan}}, \ and\ \bibinfo
  {author} {\bibfnamefont {A.}~\bibnamefont {Tomkins}},\ }\href {\doibase
  10.1073/pnas.0503018102} {\bibfield  {journal} {\bibinfo  {journal} {Proc.
  Natl. Acad. Sci. USA}\ }\textbf {\bibinfo {volume} {102}},\ \bibinfo {pages}
  {11623} (\bibinfo {year} {2005})}\BibitemShut {NoStop}%
\bibitem [{\citenamefont {Noulas}\ \emph {et~al.}(2012)\citenamefont {Noulas},
  \citenamefont {Scellato}, \citenamefont {Lambiotte}, \citenamefont {Pontil},\
  and\ \citenamefont {Mascolo}}]{Noulas2012}%
  \BibitemOpen
  \bibfield  {author} {\bibinfo {author} {\bibfnamefont {A.}~\bibnamefont
  {Noulas}}, \bibinfo {author} {\bibfnamefont {S.}~\bibnamefont {Scellato}},
  \bibinfo {author} {\bibfnamefont {R.}~\bibnamefont {Lambiotte}}, \bibinfo
  {author} {\bibfnamefont {M.}~\bibnamefont {Pontil}}, \ and\ \bibinfo {author}
  {\bibfnamefont {C.}~\bibnamefont {Mascolo}},\ }\href {\doibase
  10.1371/journal.pone.0037027} {\bibfield  {journal} {\bibinfo  {journal}
  {PLoS ONE}\ }\textbf {\bibinfo {volume} {7}},\ \bibinfo {pages} {e37027}
  (\bibinfo {year} {2012})}\BibitemShut {NoStop}%
\bibitem [{\citenamefont {Pan}\ \emph {et~al.}(2013)\citenamefont {Pan},
  \citenamefont {Ghoshal}, \citenamefont {Krumme}, \citenamefont {Cebrian},\
  and\ \citenamefont {Pentland}}]{Pan2013}%
  \BibitemOpen
  \bibfield  {author} {\bibinfo {author} {\bibfnamefont {W.}~\bibnamefont
  {Pan}}, \bibinfo {author} {\bibfnamefont {G.}~\bibnamefont {Ghoshal}},
  \bibinfo {author} {\bibfnamefont {C.}~\bibnamefont {Krumme}}, \bibinfo
  {author} {\bibfnamefont {M.}~\bibnamefont {Cebrian}}, \ and\ \bibinfo
  {author} {\bibfnamefont {A.}~\bibnamefont {Pentland}},\ }\href {\doibase
  10.1038/ncomms2961} {\bibfield  {journal} {\bibinfo  {journal} {Nat.
  Commun.}\ }\textbf {\bibinfo {volume} {4}},\ \bibinfo {pages} {1961}
  (\bibinfo {year} {2013})}\BibitemShut {NoStop}%
\bibitem [{\citenamefont {Simini}\ \emph {et~al.}(2012)\citenamefont {Simini},
  \citenamefont {Gonz\'alez}, \citenamefont {Maritan},\ and\ \citenamefont
  {Barab\'{a}si}}]{Simini2012}%
  \BibitemOpen
  \bibfield  {author} {\bibinfo {author} {\bibfnamefont {F.}~\bibnamefont
  {Simini}}, \bibinfo {author} {\bibfnamefont {M.~C.}\ \bibnamefont
  {Gonz\'alez}}, \bibinfo {author} {\bibfnamefont {A.}~\bibnamefont {Maritan}},
  \ and\ \bibinfo {author} {\bibfnamefont {A.-L.}\ \bibnamefont
  {Barab\'{a}si}},\ }\href {\doibase 10.1038/nature10856} {\bibfield  {journal}
  {\bibinfo  {journal} {Nature (London)}\ }\textbf {\bibinfo {volume} {484}},\
  \bibinfo {pages} {96} (\bibinfo {year} {2012})}\BibitemShut {NoStop}%
\bibitem [{\citenamefont {Dall}\ and\ \citenamefont
  {Christensen}(2002)}]{Dall2002}%
  \BibitemOpen
  \bibfield  {author} {\bibinfo {author} {\bibfnamefont {J.}~\bibnamefont
  {Dall}}\ and\ \bibinfo {author} {\bibfnamefont {M.}~\bibnamefont
  {Christensen}},\ }\href {\doibase 10.1103/PhysRevE.66.016121} {\bibfield
  {journal} {\bibinfo  {journal} {Phys. Rev. E}\ }\textbf {\bibinfo {volume}
  {66}},\ \bibinfo {pages} {016121} (\bibinfo {year} {2002})}\BibitemShut
  {NoStop}%
\bibitem [{\citenamefont {Estrada}\ and\ \citenamefont
  {Sheerin}(2015)}]{Estrada2015}%
  \BibitemOpen
  \bibfield  {author} {\bibinfo {author} {\bibfnamefont {E.}~\bibnamefont
  {Estrada}}\ and\ \bibinfo {author} {\bibfnamefont {M.}~\bibnamefont
  {Sheerin}},\ }\href {\doibase 10.1103/PhysRevE.91.042805} {\bibfield
  {journal} {\bibinfo  {journal} {Phys. Rev. E}\ }\textbf {\bibinfo {volume}
  {91}},\ \bibinfo {pages} {042805} (\bibinfo {year} {2015})}\BibitemShut
  {NoStop}%
\bibitem [{\citenamefont {Arenas}\ \emph {et~al.}(2008)\citenamefont {Arenas},
  \citenamefont {D\'i­az-Guilera}, \citenamefont {Kurths}, \citenamefont
  {Moreno},\ and\ \citenamefont {Zhou}}]{ArenasPhysRep2008}%
  \BibitemOpen
  \bibfield  {author} {\bibinfo {author} {\bibfnamefont {A.}~\bibnamefont
  {Arenas}}, \bibinfo {author} {\bibfnamefont {A.}~\bibnamefont
  {D\'i­az-Guilera}}, \bibinfo {author} {\bibfnamefont {J.}~\bibnamefont
  {Kurths}}, \bibinfo {author} {\bibfnamefont {Y.}~\bibnamefont {Moreno}}, \
  and\ \bibinfo {author} {\bibfnamefont {C.}~\bibnamefont {Zhou}},\ }\href
  {\doibase 10.1016/j.physrep.2008.09.002} {\bibfield  {journal} {\bibinfo
  {journal} {Phys. Rep.}\ }\textbf {\bibinfo {volume} {469}},\ \bibinfo {pages}
  {93 } (\bibinfo {year} {2008})}\BibitemShut {NoStop}%
\bibitem [{\citenamefont {Estrada}(2015)}]{Estrada2015icn}%
  \BibitemOpen
  \bibfield  {author} {\bibinfo {author} {\bibfnamefont {E.}~\bibnamefont
  {Estrada}},\ }in\ \href {\doibase 10.1007/978-3-319-11322-7_3} {\emph
  {\bibinfo {booktitle} {Evolutionary Equations with Applications in Natural
  Sciences}}},\ \bibinfo {series} {Lecture Notes in Mathematics}, Vol.\
  \bibinfo {volume} {2126},\ \bibinfo {editor} {edited by\ \bibinfo {editor}
  {\bibfnamefont {J.}~\bibnamefont {Banasiak}}\ and\ \bibinfo {editor}
  {\bibfnamefont {M.}~\bibnamefont {Mokhtar-Kharroubi}}}\ (\bibinfo
  {publisher} {Springer International Publishing},\ \bibinfo {year} {2015})\
  pp.\ \bibinfo {pages} {93--131}\BibitemShut {NoStop}%
\bibitem [{\citenamefont {Mohar}(1991)}]{Mohar1991lsg}%
  \BibitemOpen
  \bibfield  {author} {\bibinfo {author} {\bibfnamefont {B.}~\bibnamefont
  {Mohar}},\ }\href@noop {} {\bibfield  {journal} {\bibinfo  {journal} {Graph
  Theory, Combinatorics, and Applications}\ }\textbf {\bibinfo {volume} {2}},\
  \bibinfo {pages} {871} (\bibinfo {year} {1991})}\BibitemShut {NoStop}%
\bibitem [{\citenamefont {Mohar}(1997)}]{Mohar1997sal}%
  \BibitemOpen
  \bibfield  {author} {\bibinfo {author} {\bibfnamefont {B.}~\bibnamefont
  {Mohar}},\ }\href@noop {} {\bibfield  {journal} {\bibinfo  {journal} {Graph
  Symmetry: Algebraic Methods and Applications}\ }\textbf {\bibinfo {volume}
  {497}},\ \bibinfo {pages} {227} (\bibinfo {year} {1997})}\BibitemShut
  {NoStop}%
\bibitem [{\citenamefont {McGraw}\ and\ \citenamefont
  {Menzinger}(2008)}]{LapSpectra}%
  \BibitemOpen
  \bibfield  {author} {\bibinfo {author} {\bibfnamefont {P.~N.}\ \bibnamefont
  {McGraw}}\ and\ \bibinfo {author} {\bibfnamefont {M.}~\bibnamefont
  {Menzinger}},\ }\href {\doibase 10.1103/PhysRevE.77.031102} {\bibfield
  {journal} {\bibinfo  {journal} {Phys. Rev. E}\ }\textbf {\bibinfo {volume}
  {77}},\ \bibinfo {pages} {031102} (\bibinfo {year} {2008})}\BibitemShut
  {NoStop}%
\bibitem [{\citenamefont {Estrada}\ \emph {et~al.}(2012)\citenamefont
  {Estrada}, \citenamefont {Hatano},\ and\ \citenamefont
  {Benzi}}]{Estrada2012}%
  \BibitemOpen
  \bibfield  {author} {\bibinfo {author} {\bibfnamefont {E.}~\bibnamefont
  {Estrada}}, \bibinfo {author} {\bibfnamefont {N.}~\bibnamefont {Hatano}}, \
  and\ \bibinfo {author} {\bibfnamefont {M.}~\bibnamefont {Benzi}},\ }\href
  {\doibase 10.1016/j.physrep.2012.01.006} {\bibfield  {journal} {\bibinfo
  {journal} {Phys. Rep.}\ }\textbf {\bibinfo {volume} {514}},\ \bibinfo {pages}
  {89 } (\bibinfo {year} {2012})}\BibitemShut {NoStop}%
\bibitem [{\citenamefont {Fouss}\ \emph {et~al.}(2016)\citenamefont {Fouss},
  \citenamefont {Saerens},\ and\ \citenamefont {Shimbo}}]{FoussBook2016}%
  \BibitemOpen
  \bibfield  {author} {\bibinfo {author} {\bibfnamefont {F.}~\bibnamefont
  {Fouss}}, \bibinfo {author} {\bibfnamefont {M.}~\bibnamefont {Saerens}}, \
  and\ \bibinfo {author} {\bibfnamefont {M.}~\bibnamefont {Shimbo}},\
  }\href@noop {} {\emph {\bibinfo {title} {Algorithms and Models for Network
  Data and Link Analysis}}}\ (\bibinfo  {publisher} {Cambridge University
  Press},\ \bibinfo {address} {New York},\ \bibinfo {year} {2016})\BibitemShut
  {NoStop}%
\bibitem [{\citenamefont {Bellman}(1960)}]{bellman1960}%
  \BibitemOpen
  \bibfield  {author} {\bibinfo {author} {\bibfnamefont {R.}~\bibnamefont
  {Bellman}},\ }\href@noop {} {\emph {\bibinfo {title} {{Introduction to Matrix
  Analysis}}}}\ (\bibinfo  {publisher} {McGraw-Hill, New York},\ \bibinfo
  {year} {1960})\BibitemShut {NoStop}%
\bibitem [{\citenamefont {Michelitsch}\ \emph
  {et~al.}(2016{\natexlab{b}})\citenamefont {Michelitsch}, \citenamefont
  {Collet}, \citenamefont {Nowakowski},\ and\ \citenamefont
  {Nicolleau}}]{Michelitsch2016}%
  \BibitemOpen
  \bibfield  {author} {\bibinfo {author} {\bibfnamefont {T.~M.}\ \bibnamefont
  {Michelitsch}}, \bibinfo {author} {\bibfnamefont {B.}~\bibnamefont {Collet}},
  \bibinfo {author} {\bibfnamefont {A.~F.}\ \bibnamefont {Nowakowski}}, \ and\
  \bibinfo {author} {\bibfnamefont {F.~C. G.~A.}\ \bibnamefont {Nicolleau}},\
  }\href {\doibase 10.1016/j.chaos.2015.10.035} {\bibfield  {journal} {\bibinfo
   {journal} {Chaos Solitons \& Fractals}\ }\textbf {\bibinfo {volume} {82}},\
  \bibinfo {pages} {38 } (\bibinfo {year} {2016}{\natexlab{b}})}\BibitemShut
  {NoStop}%
\bibitem [{\citenamefont {Michelitsch}\ \emph {et~al.}(2018)\citenamefont
  {Michelitsch}, \citenamefont {Collet}, \citenamefont {Riascos}, \citenamefont
  {Nowakowski},\ and\ \citenamefont {Nicolleau}}]{Michelitsch2018}%
  \BibitemOpen
  \bibfield  {author} {\bibinfo {author} {\bibfnamefont {T.}~\bibnamefont
  {Michelitsch}}, \bibinfo {author} {\bibfnamefont {B.}~\bibnamefont {Collet}},
  \bibinfo {author} {\bibfnamefont {A.~P.}\ \bibnamefont {Riascos}}, \bibinfo
  {author} {\bibfnamefont {A.}~\bibnamefont {Nowakowski}}, \ and\ \bibinfo
  {author} {\bibfnamefont {F.}~\bibnamefont {Nicolleau}},\ }\enquote {\bibinfo
  {title} {On recurrence and transience of fractional random walks in
  lattices},}\ in\ \href {\doibase 10.1007/978-3-319-72440-9_29} {\emph
  {\bibinfo {booktitle} {Generalized Models and Non-classical Approaches in
  Complex Materials 1}}},\ \bibinfo {editor} {edited by\ \bibinfo {editor}
  {\bibfnamefont {H.}~\bibnamefont {Altenbach}}, \bibinfo {editor}
  {\bibfnamefont {J.}~\bibnamefont {Pouget}}, \bibinfo {editor} {\bibfnamefont
  {M.}~\bibnamefont {Rousseau}}, \bibinfo {editor} {\bibfnamefont
  {B.}~\bibnamefont {Collet}}, \ and\ \bibinfo {editor} {\bibfnamefont
  {T.}~\bibnamefont {Michelitsch}}}\ (\bibinfo  {publisher} {Springer
  International Publishing},\ \bibinfo {address} {Cham},\ \bibinfo {year}
  {2018})\ pp.\ \bibinfo {pages} {555--580}\BibitemShut {NoStop}%
\bibitem [{\citenamefont {Abramowitz}\ and\ \citenamefont
  {Stegun}(1970)}]{HandbookAbramowitz}%
  \BibitemOpen
  \bibfield  {author} {\bibinfo {author} {\bibfnamefont {M.}~\bibnamefont
  {Abramowitz}}\ and\ \bibinfo {author} {\bibfnamefont {I.~A.}\ \bibnamefont
  {Stegun}},\ }\href@noop {} {\emph {\bibinfo {title} {{Handbook of
  Mathematical Functions}}}}\ (\bibinfo  {publisher} {Dover, New York},\
  \bibinfo {year} {1970})\BibitemShut {NoStop}%
\bibitem [{\citenamefont {Godsil}\ and\ \citenamefont
  {Royle}(2001)}]{GodsilBook}%
  \BibitemOpen
  \bibfield  {author} {\bibinfo {author} {\bibfnamefont {C.}~\bibnamefont
  {Godsil}}\ and\ \bibinfo {author} {\bibfnamefont {G.}~\bibnamefont {Royle}},\
  }\href@noop {} {\emph {\bibinfo {title} {Algebraic Graph Theory}}},\ \bibinfo
  {series} {Graduate Texts in Mathematics}, Vol.\ \bibinfo {volume} {207}\
  (\bibinfo  {publisher} {Springer},\ \bibinfo {address} {Berlin},\ \bibinfo
  {year} {2001})\BibitemShut {NoStop}%
\bibitem [{\citenamefont {de~Arruda}\ \emph {et~al.}(2014)\citenamefont
  {de~Arruda}, \citenamefont {Barbieri}, \citenamefont {Rodr\'{\i}guez},
  \citenamefont {Rodrigues}, \citenamefont {Moreno},\ and\ \citenamefont
  {Costa}}]{ArrudaPRE2014}%
  \BibitemOpen
  \bibfield  {author} {\bibinfo {author} {\bibfnamefont {G.~F.}\ \bibnamefont
  {de~Arruda}}, \bibinfo {author} {\bibfnamefont {A.~L.}\ \bibnamefont
  {Barbieri}}, \bibinfo {author} {\bibfnamefont {P.~M.}\ \bibnamefont
  {Rodr\'{\i}guez}}, \bibinfo {author} {\bibfnamefont {F.~A.}\ \bibnamefont
  {Rodrigues}}, \bibinfo {author} {\bibfnamefont {Y.}~\bibnamefont {Moreno}}, \
  and\ \bibinfo {author} {\bibfnamefont {L.~F.}\ \bibnamefont {Costa}},\ }\href
  {\doibase 10.1103/PhysRevE.90.032812} {\bibfield  {journal} {\bibinfo
  {journal} {Phys. Rev. E}\ }\textbf {\bibinfo {volume} {90}},\ \bibinfo
  {pages} {032812} (\bibinfo {year} {2014})}\BibitemShut {NoStop}%
\bibitem [{\citenamefont {Lin}\ and\ \citenamefont
  {Zhang}(2013)}]{YuanPRE2013}%
  \BibitemOpen
  \bibfield  {author} {\bibinfo {author} {\bibfnamefont {Y.}~\bibnamefont
  {Lin}}\ and\ \bibinfo {author} {\bibfnamefont {Z.}~\bibnamefont {Zhang}},\
  }\href {\doibase 10.1103/PhysRevE.87.062140} {\bibfield  {journal} {\bibinfo
  {journal} {Phys. Rev. E}\ }\textbf {\bibinfo {volume} {87}},\ \bibinfo
  {pages} {062140} (\bibinfo {year} {2013})}\BibitemShut {NoStop}%
\bibitem [{\citenamefont {Gonz\'alez}\ \emph {et~al.}(2008)\citenamefont
  {Gonz\'alez}, \citenamefont {Hidalgo},\ and\ \citenamefont
  {Barab\'{a}si}}]{Gonzalez2008}%
  \BibitemOpen
  \bibfield  {author} {\bibinfo {author} {\bibfnamefont {M.~C.}\ \bibnamefont
  {Gonz\'alez}}, \bibinfo {author} {\bibfnamefont {C.~A.}\ \bibnamefont
  {Hidalgo}}, \ and\ \bibinfo {author} {\bibfnamefont {A.-L.}\ \bibnamefont
  {Barab\'{a}si}},\ }\href {\doibase 10.1038/nature06958} {\bibfield  {journal}
  {\bibinfo  {journal} {Nature (London)}\ }\textbf {\bibinfo {volume} {453}},\
  \bibinfo {pages} {779} (\bibinfo {year} {2008})}\BibitemShut {NoStop}%
\bibitem [{\citenamefont {Estrada}\ \emph {et~al.}(2009)\citenamefont
  {Estrada}, \citenamefont {Higham},\ and\ \citenamefont
  {Hatano}}]{Estrada2009}%
  \BibitemOpen
  \bibfield  {author} {\bibinfo {author} {\bibfnamefont {E.}~\bibnamefont
  {Estrada}}, \bibinfo {author} {\bibfnamefont {D.~J.}\ \bibnamefont {Higham}},
  \ and\ \bibinfo {author} {\bibfnamefont {N.}~\bibnamefont {Hatano}},\ }\href
  {\doibase 10.1016/j.physa.2008.11.011} {\bibfield  {journal} {\bibinfo
  {journal} {Physica A: Stat. Mech. Appl.}\ }\textbf {\bibinfo {volume}
  {388}},\ \bibinfo {pages} {764 } (\bibinfo {year} {2009})}\BibitemShut
  {NoStop}%
\bibitem [{\citenamefont {Kemeny}\ and\ \citenamefont {Snell}(1960)}]{Kemeny}%
  \BibitemOpen
  \bibfield  {author} {\bibinfo {author} {\bibfnamefont {J.~G.}\ \bibnamefont
  {Kemeny}}\ and\ \bibinfo {author} {\bibfnamefont {J.~L.}\ \bibnamefont
  {Snell}},\ }\href@noop {} {\emph {\bibinfo {title} {{Finite Markov
  chains}}}}\ (\bibinfo  {publisher} {VanNostrand},\ \bibinfo {address}
  {Princeton},\ \bibinfo {year} {1960})\BibitemShut {NoStop}%
\bibitem [{\citenamefont {Zhang}\ \emph
  {et~al.}(2011{\natexlab{b}})\citenamefont {Zhang}, \citenamefont {Julaiti},
  \citenamefont {Hou}, \citenamefont {Zhang},\ and\ \citenamefont
  {Chen}}]{MFPTanalytic}%
  \BibitemOpen
  \bibfield  {author} {\bibinfo {author} {\bibfnamefont {Z.}~\bibnamefont
  {Zhang}}, \bibinfo {author} {\bibfnamefont {A.}~\bibnamefont {Julaiti}},
  \bibinfo {author} {\bibfnamefont {B.}~\bibnamefont {Hou}}, \bibinfo {author}
  {\bibfnamefont {H.}~\bibnamefont {Zhang}}, \ and\ \bibinfo {author}
  {\bibfnamefont {G.}~\bibnamefont {Chen}},\ }\href {\doibase
  10.1140/epjb/e2011-20834-1} {\bibfield  {journal} {\bibinfo  {journal} {Eur.
  Phys. J. B}\ }\textbf {\bibinfo {volume} {84}},\ \bibinfo {pages} {691}
  (\bibinfo {year} {2011}{\natexlab{b}})}\BibitemShut {NoStop}%
\bibitem [{\citenamefont {G\'omez-Garde\~nes}\ and\ \citenamefont
  {Latora}(2008)}]{LatoraPRE2008}%
  \BibitemOpen
  \bibfield  {author} {\bibinfo {author} {\bibfnamefont {J.}~\bibnamefont
  {G\'omez-Garde\~nes}}\ and\ \bibinfo {author} {\bibfnamefont
  {V.}~\bibnamefont {Latora}},\ }\href {\doibase 10.1103/PhysRevE.78.065102}
  {\bibfield  {journal} {\bibinfo  {journal} {Phys. Rev. E}\ }\textbf {\bibinfo
  {volume} {78}},\ \bibinfo {pages} {065102} (\bibinfo {year}
  {2008})}\BibitemShut {NoStop}%
\bibitem [{\citenamefont {Gray}(2006)}]{CirculantReview2006}%
  \BibitemOpen
  \bibfield  {author} {\bibinfo {author} {\bibfnamefont {R.~M.}\ \bibnamefont
  {Gray}},\ }\href {\doibase 10.1561/0100000006} {\bibfield  {journal}
  {\bibinfo  {journal} {Found. Trends Commun. Inf. Theory}\ }\textbf {\bibinfo
  {volume} {2}},\ \bibinfo {pages} {155} (\bibinfo {year} {2006})}\BibitemShut
  {NoStop}%
\bibitem [{\citenamefont {Watts}\ and\ \citenamefont
  {Strogatz}(1998)}]{WattsStrogatz}%
  \BibitemOpen
  \bibfield  {author} {\bibinfo {author} {\bibfnamefont {D.~J.}\ \bibnamefont
  {Watts}}\ and\ \bibinfo {author} {\bibfnamefont {S.~H.}\ \bibnamefont
  {Strogatz}},\ }\href {\doibase 10.1038/30918} {\bibfield  {journal} {\bibinfo
   {journal} {Nature (London)}\ }\textbf {\bibinfo {volume} {393}},\ \bibinfo
  {pages} {440} (\bibinfo {year} {1998})}\BibitemShut {NoStop}%
\bibitem [{\citenamefont {Erd\"{o}s}\ and\ \citenamefont
  {R\'{e}nyi}(1959)}]{ErdosRenyi}%
  \BibitemOpen
  \bibfield  {author} {\bibinfo {author} {\bibfnamefont {P.}~\bibnamefont
  {Erd\"{o}s}}\ and\ \bibinfo {author} {\bibfnamefont {A.}~\bibnamefont
  {R\'{e}nyi}},\ }\href@noop {} {\bibfield  {journal} {\bibinfo  {journal}
  {Publ. Math. (Debrecen)}\ }\textbf {\bibinfo {volume} {6}},\ \bibinfo {pages}
  {290} (\bibinfo {year} {1959})}\BibitemShut {NoStop}%
\bibitem [{\citenamefont {Barab\'asi}\ and\ \citenamefont
  {Albert}(1999)}]{BarabasiAlbert}%
  \BibitemOpen
  \bibfield  {author} {\bibinfo {author} {\bibfnamefont {A.-L.}\ \bibnamefont
  {Barab\'asi}}\ and\ \bibinfo {author} {\bibfnamefont {R.}~\bibnamefont
  {Albert}},\ }\href {\doibase 10.1126/science.286.5439.509} {\bibfield
  {journal} {\bibinfo  {journal} {Science}\ }\textbf {\bibinfo {volume}
  {286}},\ \bibinfo {pages} {509} (\bibinfo {year} {1999})}\BibitemShut
  {NoStop}%
\end{thebibliography}
\end{document}